\documentclass[11pt,onecolumn]{IEEEtran}

\usepackage[mathscr]{eucal}
\usepackage[cmex10]{amsmath}
\usepackage{epsfig,epsf,psfrag}
\usepackage{amssymb,amsmath,amsthm,amsfonts,latexsym}
\usepackage{amsmath,graphicx,bm,xcolor,url}
\usepackage[caption=false]{subfig} 
\usepackage{fixltx2e}
\usepackage{array}
\usepackage{verbatim}
\usepackage{bm}
\usepackage{algorithmic, cite}
\usepackage{algorithm}
\usepackage{verbatim}
\usepackage{textcomp}
\usepackage{mathrsfs}
\usepackage{epstopdf}

\catcode`~=11 \def\UrlSpecials{\do\~{\kern -.15em\lower .7ex\hbox{~}\kern .04em}} \catcode`~=13 

\allowdisplaybreaks[1]

\newcommand{\calA}{\mathcal{A}}
\newcommand{\calB}{\mathcal{B}}
\newcommand{\calC}{\mathcal{C}}
\newcommand{\calD}{\mathcal{D}}

\newcommand{\calI}{\mathcal{I}}
\newcommand{\calJ}{\mathcal{J}}

\newcommand{\calL}{\mathcal{L}}

\newcommand{\calN}{\mathcal{N}}

\newcommand{\calS}{\mathcal{S}}
\newcommand{\calT}{\mathcal{T}}

\newcommand{\rmd}{\mathrm{d}}

\newcommand{\rme}{\mathrm{e}}

\newcommand{\bbE}{\mathbb{E}}

\newcommand{\bbN}{\mathbb{N}}

\newcommand{\bbR}{\mathbb{R}}

\newcommand{\frakG}{\mathfrak{G}}

\newcommand{\frakP}{\mathfrak{P}}

\DeclareMathAlphabet{\mathbsf}{OT1}{cmss}{bx}{n}
\DeclareMathAlphabet{\mathssf}{OT1}{cmss}{m}{sl}

\newcommand{\rvC}{\mathsf{C}}

\newcommand{\rvG}{\mathsf{G}}

\newcommand{\rvQ}{\mathsf{Q}}

\newcommand{\rvV}{\mathsf{V}}

\newcommand{\rvW}{\mathsf{W}}

\newcommand{\rvZ}{\mathsf{Z}}

\DeclareSymbolFont{bsfletters}{OT1}{cmss}{bx}{n}  
\DeclareSymbolFont{ssfletters}{OT1}{cmss}{m}{n}
\DeclareMathSymbol{\bsfGamma}{0}{bsfletters}{'000}
\DeclareMathSymbol{\ssfGamma}{0}{ssfletters}{'000}
\DeclareMathSymbol{\bsfDelta}{0}{bsfletters}{'001}
\DeclareMathSymbol{\ssfDelta}{0}{ssfletters}{'001}
\DeclareMathSymbol{\bsfTheta}{0}{bsfletters}{'002}
\DeclareMathSymbol{\ssfTheta}{0}{ssfletters}{'002}
\DeclareMathSymbol{\bsfLambda}{0}{bsfletters}{'003}
\DeclareMathSymbol{\ssfLambda}{0}{ssfletters}{'003}
\DeclareMathSymbol{\bsfXi}{0}{bsfletters}{'004}
\DeclareMathSymbol{\ssfXi}{0}{ssfletters}{'004}
\DeclareMathSymbol{\bsfPi}{0}{bsfletters}{'005}
\DeclareMathSymbol{\ssfPi}{0}{ssfletters}{'005}
\DeclareMathSymbol{\bsfSigma}{0}{bsfletters}{'006}
\DeclareMathSymbol{\ssfSigma}{0}{ssfletters}{'006}
\DeclareMathSymbol{\bsfUpsilon}{0}{bsfletters}{'007}
\DeclareMathSymbol{\ssfUpsilon}{0}{ssfletters}{'007}
\DeclareMathSymbol{\bsfPhi}{0}{bsfletters}{'010}
\DeclareMathSymbol{\ssfPhi}{0}{ssfletters}{'010}
\DeclareMathSymbol{\bsfPsi}{0}{bsfletters}{'011}
\DeclareMathSymbol{\ssfPsi}{0}{ssfletters}{'011}
\DeclareMathSymbol{\bsfOmega}{0}{bsfletters}{'012}
\DeclareMathSymbol{\ssfOmega}{0}{ssfletters}{'012}

\newcommand{\tilq}{\tilde{q}}
\newcommand{\tilQ}{\tilde{Q}}

\newcommand{\hatW}{\hat{W}}

\newcommand{\hatx}{\hat{x}}

\newcommand{\tilx}{\tilde{x}}
\newcommand{\tilX}{\tilde{X}}

\newcommand{\tilY}{\tilde{Y}}

\newcommand{\tilZ}{\tilde{Z}}

\newcommand{\bard}{\bar{d}}

\newcommand{\barJ}{\bar{J}}

\newcommand{\barY}{\bar{Y}}

\newcommand{\eps}{\varepsilon}

\def\fndot{\, \cdot \,}

\DeclareMathOperator{\tr}{tr}

\DeclareMathOperator{\supp}{supp}

\newtheorem{theorem}{Theorem} 
\newtheorem{lemma}[theorem]{Lemma}

\newtheorem{proposition}[theorem]{Proposition}

\newtheorem{definition}{Definition}

\newtheorem{remark}{Remark}

\newcommand{\qednew}{\nobreak \ifvmode \relax \else
      \ifdim\lastskip<1.5em \hskip-\lastskip
      \hskip1.5em plus0em minus0.5em \fi \nobreak
      \vrule height0.75em width0.5em depth0.25em\fi}

\newcommand{\mmse}{\mathrm{mmse}}
\newcommand{\Jac}{\operatorname{Jac}}
\usepackage{epsfig} 
\usepackage{epstopdf,bbm} 
\usepackage{hyperref}
\usepackage{overpic}
\usepackage{booktabs, tabularx, array}
\usepackage{graphicx}
\usepackage{tikz}
\usetikzlibrary{arrows.meta,patterns}

\usepackage{pgfplots}
\usetikzlibrary{arrows.meta}
\pgfplotsset{compat=1.18}
\usetikzlibrary{arrows.meta,decorations.pathreplacing,shapes.geometric}

\hypersetup{hidelinks}

\usepackage{amsmath,amssymb}
\usepackage{xcolor}
\usetikzlibrary{arrows.meta,calc,positioning,decorations.pathreplacing}

\definecolor{lowerr}{HTML}{D9EEF5}
\definecolor{miderr}{HTML}{F8D49A}
\definecolor{higherr}{HTML}{D96B5F}
\definecolor{keepgreen}{HTML}{2F7F62}
\definecolor{trimgray}{HTML}{D8D8D8}
\definecolor{textgray}{HTML}{555555}

\tikzset{
  stagearrow/.style={-{Latex[length=3.2mm,width=2.2mm]}, very thick, draw=textgray},
  paneltitle/.style={font=\bfseries\small, align=center},
  annotation/.style={font=\scriptsize, align=center, text width=4.35cm},
  axislabel/.style={font=\scriptsize, text=textgray},
  matrixoutline/.style={draw=black!75, line width=0.55pt},
}

\begin{document}
\flushbottom
\title{Second-Order Asymptotics for the \\ Scalar Gaussian Broadcast Channel} 

\author{Vincent Y.~F.~Tan  %
\thanks{V.~Y.~F.~Tan is with the Department of Mathematics and 
the Department of Electrical and Computer Engineering, National University of Singapore
(email:\,vtan@nus.edu.sg).
}}
 
\IEEEpeerreviewmaketitle

\maketitle

\begin{abstract}
This paper characterizes the second-order coding region of the scalar Gaussian broadcast channel along the curved boundary of its capacity region. The characterization holds under both average and maximal error criteria  and allows correlation between the receiver noises. At each interior point of this boundary, the second-order region is a halfspace, as in the work of Scarlett and the author on Gaussian multiple-access channels with degraded message sets. Achievability follows from spherical superposition coding with a blocklength-dependent power allocation. The main innovation lies in the converse. Using Ornstein--Uhlenbeck entropy dissipation and Gaussian functional inequalities, we construct an auxiliary observation through an additive Gaussian noise channel and establish quantitative relative entropy bounds for the channel output distributions conditioned on this observation. The strong functional representation lemma of Li and El Gamal then allows us to select a realization of its independent random seed to obtain a deterministic partition of the retained receiver-1 message set. The component label plays a role analogous to the auxiliary random variable in the capacity characterization, while providing the distributional control needed at the second-order scale. These properties guide the construction of auxiliary output measures whose likelihood ratios admit the cancellations and normal approximation needed to evaluate an information spectrum converse.
\end{abstract}

 
\section{Introduction} 
The broadcast channel, introduced by Cover in 1972~\cite{cover72}, is a paradigmatic model in network information theory for noisy one-to-many communication. It provides a natural model for studying downlink communication in wireless systems, among other applications.  In this model, a single sender wishes to communicate a private message to each receiver. We consider only two receivers in this paper, and no common message is sent to both. The capacity region, defined as the closure of the set of achievable rate pairs, is unknown in general. For degraded broadcast channels, in which one receiver is ``stronger'' than the other, the capacity region is known. It is achieved via superposition coding~\cite{cover72,Bergmans1973}, in which the weaker receiver decodes its intended message, while the stronger receiver decodes both messages using successive cancellation. This scheme, however, is not optimal in general. Marton's inner bound~\cite{Marton79} is the best-known  achievability result, but it was recently shown to be   suboptimal for some broadcast channels~\cite{HuangLiuLiu2026}.

\begin{figure}[t]
  \centering
\begingroup
\def\BCSone{10}
\def\BCStwo{2}
\def\BCalpha{0.25}
\definecolor{BCcurve}{RGB}{0,83,138}
\definecolor{BCaxis}{RGB}{0,119,103}
\definecolor{BCendpoint}{RGB}{142,52,70}
\definecolor{BCsupport}{RGB}{169,93,13}
\pgfmathsetmacro{\BCcapone}{ln(1+\BCSone)/2}
\pgfmathsetmacro{\BCcaptwo}{ln(1+\BCStwo)/2}
\pgfmathsetmacro{\BCpointone}{ln(1+\BCalpha*\BCSone)/2}
\pgfmathsetmacro{\BCpointtwo}{ln((1+\BCStwo)/(1+\BCalpha*\BCStwo))/2}
\pgfmathsetmacro{\BCmu}{(\BCalpha+1/\BCStwo)/(\BCalpha+1/\BCSone)}
\pgfmathsetmacro{\BCweighted}{\BCpointone+\BCmu*\BCpointtwo}
\pgfmathsetmacro{\BCxintercept}{\BCweighted/\BCcapone}
\pgfmathsetmacro{\BCyintercept}{\BCweighted/(\BCmu*\BCcaptwo)}
\pgfmathsetmacro{\BCxaxisend}{max(1.32,\BCxintercept+.15)}
\pgfmathsetmacro{\BCyaxisend}{max(1.40,\BCyintercept+.13)}
\pgfmathsetmacro{\BCannotationheight}{max(1.24,\BCyintercept+.08)}
\begin{tikzpicture}[
  x=4.9cm,y=4.1cm,
  font=\footnotesize,
  line cap=round,line join=round,
  bc label/.style={inner sep=2pt},
  bc leader/.style={draw=black!65,line width=.4pt}
]
  \fill[black!4]
    (0,0)--
    plot[domain=0:1,samples=151,variable=\a]
      ({ln(1+\a*\BCSone)/ln(1+\BCSone)},
       {ln((1+\BCStwo)/(1+\a*\BCStwo))/ln(1+\BCStwo)})
    --cycle;

  \draw[->,line width=.5pt] (0,0)--(\BCxaxisend,0)
    node[anchor=west,bc label] {$R_1$};
  \draw[->,line width=.5pt] (0,0)--(0,\BCyaxisend)
    node[anchor=south,bc label] {$R_2$};
  \draw[BCaxis,line width=1.15pt,dash pattern=on 3pt off 2pt]
    (0,1)--(0,0)--(1,0);

  \draw[BCsupport,line width=1pt,dash pattern=on 5pt off 2.6pt]
    (0,\BCyintercept)--(\BCxintercept,0);
  \draw[BCsupport,line width=.7pt]
    (-.014,\BCyintercept)--(.014,\BCyintercept);
  \draw[BCsupport,line width=.7pt]
    (\BCxintercept,-.014)--(\BCxintercept,.014);
  \node[bc label,anchor=east,text=BCsupport]
    at (-.025,\BCyintercept) {$C_\alpha/\mu_\alpha$};
  \node[bc label,anchor=north west,text=BCsupport]
    at ({\BCxintercept+.008},-.025) {$C_\alpha$};
  \node[bc label,align=center,anchor=south,text=BCsupport]
    at (.80,\BCannotationheight)
    {\normalsize $R_1+\mu_\alpha R_2=C_\alpha$\\[4pt]
     $\displaystyle\frac{\rmd R_2}{\rmd R_1}=-\frac{1}{\mu_\alpha}$};

  \draw[BCcurve,line width=1.2pt]
    plot[domain=0:1,samples=151,variable=\a]
      ({ln(1+\a*\BCSone)/ln(1+\BCSone)},
       {ln((1+\BCStwo)/(1+\a*\BCStwo))/ln(1+\BCStwo)});
  \coordinate (BCinterior) at
    ({\BCpointone/\BCcapone},{\BCpointtwo/\BCcaptwo});
  \fill[BCcurve] (BCinterior) circle[radius=1.8pt];
  \node[bc label,align=center,anchor=south,text=BCcurve]
    (BCcurvelabel) at (.85,.84)
    {(i) $0<\alpha<1$\\[2pt]
     $\bigl(C_1(\alpha),C_2(\alpha)\bigr)$};
  \draw[bc leader] (BCcurvelabel.south)--(BCinterior);

  \node[rectangle,fill=BCendpoint,inner sep=0pt,minimum size=4.6pt]
    at (0,1) {};
  \node[rectangle,fill=BCendpoint,inner sep=0pt,minimum size=4.6pt]
    at (1,0) {};
  \node[bc label,anchor=east] at (-.025,1) {$\rvC(S_2)$};
  \node[bc label,anchor=north west,text=BCendpoint] at (.04,.90)
    {(ii)};
  \node[bc label,anchor=north east] at (.99,-.025) {$\rvC(S_1)$};
  \node[bc label,anchor=north,text=BCendpoint] at (1,-.14)
    {(iii) $\alpha=1$};


  \node[font=\LARGE,text=black!70] at (.35,.37) {$\calC$};
\end{tikzpicture}%
\endgroup
  \caption{The three boundary regimes of the Gaussian
  broadcast channel capacity region.}
  \label{fig:bc-capacity}
\end{figure}
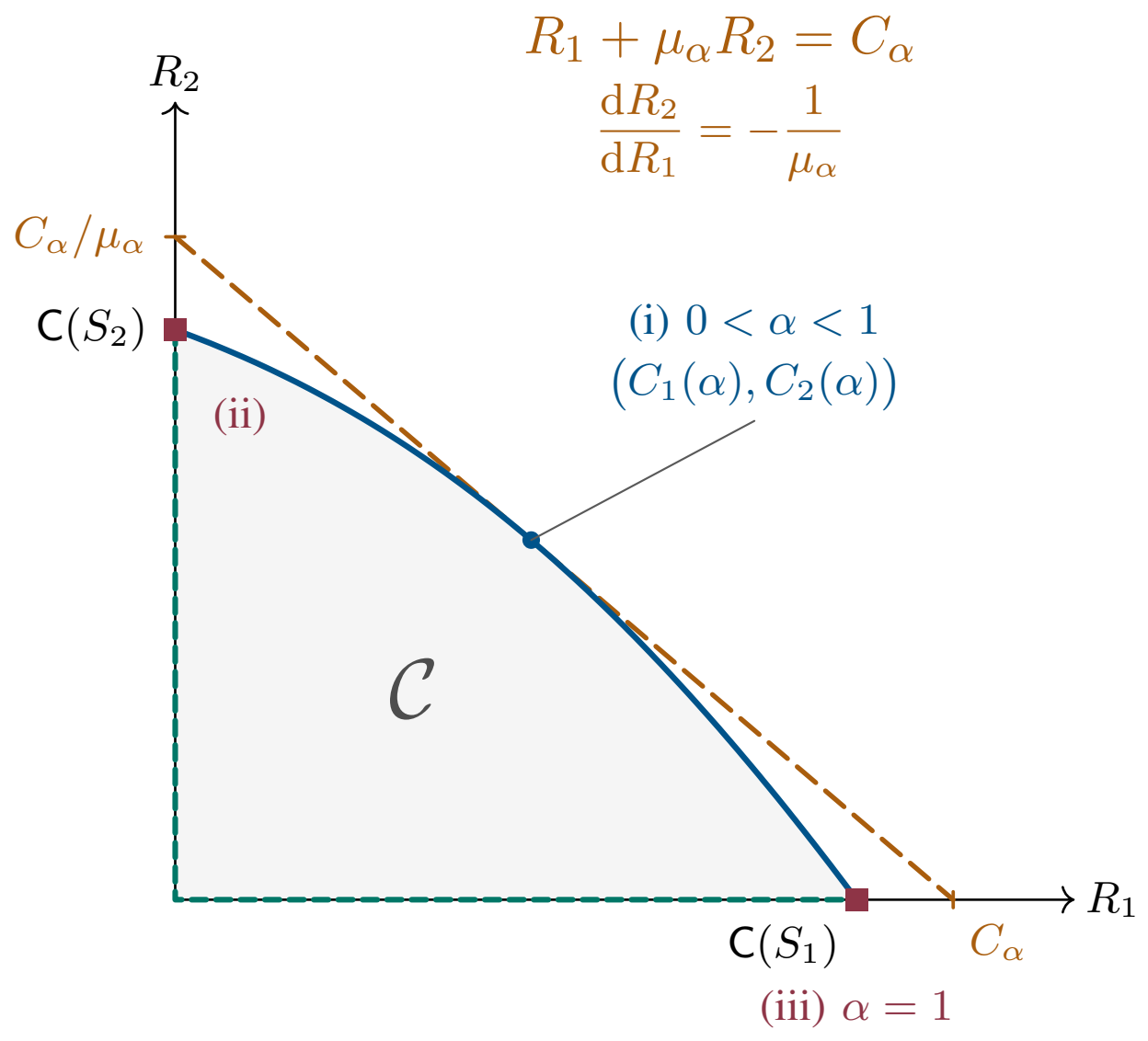

For the scalar {\em Gaussian} broadcast channel, which is stochastically degraded~\cite[Sec.~5.4]{elgamal}, the story is more complete, at least in the first-order sense. Cover~\cite{cover72} and Bergmans~\cite{Bergmans1974} showed that the capacity region consists of all nonnegative rate pairs $(R_1,R_2)$ such that
\begin{align}
    R_1\le \rvC(\alpha S_1),\qquad
    R_2\le \rvC\bigg(\frac{\bar{\alpha}S_2}{1+\alpha S_2}\bigg)
    \label{eqn:cap_region_intro}
\end{align}
for some $\alpha\in[0,1]$. Here, $\rvC(x):=\frac12\log(1+x)$ is the Gaussian capacity function, $S_1>S_2>0$ are the signal-to-noise ratios at the two receivers, and $\alpha$ and $\bar{\alpha}:=1-\alpha$ represent the fractions of transmit power allocated to their respective messages. The weak converse was proved by Bergmans using the entropy power inequality (EPI). See Fig.~\ref{fig:bc-capacity} for an illustration of the capacity region.

Our goal in this paper is to go beyond this first-order characterization and study the second-order asymptotics of the Gaussian broadcast channel. Second-order asymptotics \cite{Tan14} characterize how achievable rates approach the capacity boundary at the $n^{-1/2}$ scale for a fixed nonvanishing error probability. Nailing down the constants governing this approach at a fixed boundary point is the focus of our study. For the point-to-point Gaussian channel, the corresponding normal approximation and third-order refinement are well understood through the works of Hayashi~\cite{Hayashi09}, Polyanskiy, Poor, and Verd\'u~\cite{PPV10}, and Tan and Tomamichel~\cite{TanTom15}, among others.

To study the second-order behavior of the Gaussian broadcast channel around its   capacity boundary, it is imperative to understand whether the strong converse holds. That is, if we allow a fixed nonvanishing error probability strictly below one, is the capacity region still the one in~\eqref{eqn:cap_region_intro}? Under the per-codeword power constraint, Fong and Tan~\cite{FongTan2017GaussianBC} established the strong converse. Their converse also gives an \(O(n^{-1/2})\) outer approximation to the capacity region. This identifies the relevant scale but not the sharp second-order coefficients: for error probabilities below \(1/2\), it permits an {\em outward} displacement where an {\em inward} backoff is expected. 

 The main contribution of this paper is an exact characterization of the second-order coding region along the curved capacity boundary, stated in Theorem~\ref{thm:main}. The characterization applies under both the average and maximal error criteria, with an error occurring whenever either receiver fails to decode its intended message, and allows correlation between the receiver noises. At each interior point of this boundary, we show that the optimal second-order region is a {\em halfspace}. From the perspective of achievability, this form should not come as a surprise. The first-order region is achieved via superposition coding. Early second-order achievability bounds for discrete memoryless asymmetric broadcast channels were obtained by Tan and Kosut~\cite{TK12}. For the Gaussian multiple-access channel with degraded message sets, Scarlett and Tan~\cite{ST2015} subsequently characterized the second-order region using spherical superposition coding and normal approximations. A similar strategy is therefore natural here. Indeed, our achievability proof largely follows these Gaussian techniques; see also the Gaussian broadcast channel achievability results of Tuninetti et al.~\cite{TuninettiEtal2023}. Allowing the power allocation to vary at the $n^{-1/2}$ scale yields the claimed halfspaces.

The main innovation in this work lies in the converse. As noted in the author's monograph~\cite{Tan14}, establishing tight second-order converses has historically been challenging for network information theory problems whose standard first-order characterizations involve {\em auxiliary random variables}; recent works on various source coding problems provide notable examples of progress; see, e.g.,~\cite{Liu2021WAK}. The difficulty is that the information spectrum methods used for second-order analysis do not readily combine with the usual methods for identifying auxiliaries in weak and strong converses. In particular, identifying an auxiliary random variable that suffices for a first-order bound need not provide the distributional control required to retain the relevant second-order fluctuations. We explain this difficulty, and how the present proof addresses it, in the next two subsections. 

\subsection{Existing Converse Techniques and the Role of Auxiliaries}

At this point, it seems opportune to review existing converse techniques for broadcast channels, focusing on how auxiliaries are identified and what more is needed for a matching second-order converse.

\paragraph{Weak converses and the identification of auxiliaries}
Gallager's weak converse~\cite{Gallager1974} starts from Fano's inequality. For codes with vanishing error probability and rates converging to $(R_1,R_2)$,
\begin{align}
    nR_1\le I(W_1;Y_1^n)+o(n),\qquad
    nR_2\le I(W_2;Y_2^n)+o(n).
    \label{eqn:fano}
\end{align}
Here, $W_j$ is receiver $j$'s intended message and $Y_j^n$ its channel output. For this first-order argument, a stochastically degraded channel can be replaced by a physically degraded one with the same marginals. Gallager's ``nonintuitive'' identification of the auxiliary random variable, as described by El Gamal and Kim~\cite{elgamal}, is $U_i:=(W_2,Y_1^{i-1})$. The chain rule, degradedness, and time sharing then yield
\begin{align}
    R_1\le I(X;Y_1|U),\qquad R_2\le I(U;Y_2)
    \label{eqn:aux_intro}
\end{align}
for some $P_{U,X}$. Together with superposition coding, this characterizes the capacity region. Conditioning on the auxiliary random variable $U$ specifies the input distribution used in the characterization. In the Gaussian case, applying the EPI to these bounds introduces the power split $\alpha$ and yields~\eqref{eqn:cap_region_intro}.

Nair~\cite{Nair2013ConcaveEnvelopes} provides a complementary viewpoint by expressing the degraded broadcast channel capacity region through upper concave envelopes of mutual information functionals. This representation incorporates the optimization over mixtures of input distributions without displaying an auxiliary random variable explicitly. The resulting first-order characterization, however, does not by itself provide the quantitative control of joint information-density fluctuations required for a tight second-order converse.
\paragraph{Blowing-up and Gaussian concentration}
Ahlswede, G\'acs, and K\"orner~\cite{AhlswedeGacsKorner1976} enlarge decoding sets using the blowing-up lemma to obtain lists of subexponential size containing the correct messages with probability tending to one. For maximal error bounded away from one, a list version of Fano's inequality gives mutual information bounds with $o(n)$ remainders, after which the usual single-letterization steps apply. A related formulation uses the image-size characterizations developed by Csisz\'ar and K\"orner~\cite[Ch.~15]{Csi97}. For a fixed $\eta\in(0,1)$, an $\eta$-image of a set of input sequences is an output set having conditional probability at least $\eta$ for every input sequence in that set. These characterizations translate decoding requirements into entropy and mutual information constraints; their applications to network problems, including the asymmetric broadcast channel, are developed in~\cite[Ch.~16]{Csi97}.

For Gaussian channels, Fong and Tan~\cite{FongTan2017GaussianBC} instead use the Gaussian Poincar\'e inequality~\cite[Theorem~3.20]{BLM2013} to bound the conditional variances of certain information densities, given the transmitted message pair. After an average-to-maximal-error reduction, Chebyshev's inequality and averaging turn an information spectrum bound into mutual information bounds with $O(\sqrt n)$ remainders. Combining these with the EPI establishes the strong converse.

\paragraph{Reverse hypercontractivity}
Liu, van Handel, and Verd\'u~\cite{LiuVanHandelVerdu2019} smooth decoding functions using reverse hypercontractivity of Markov semigroups. This yields strengthened Fano-type inequalities and, in particular, converse bounds under the maximal error criterion with $O(\sqrt n)$ remainders for degraded discrete memoryless and Gaussian broadcast channels. In the Gaussian setting, the smoothing uses the Ornstein--Uhlenbeck semigroup~\cite[Sec.~2.7.1]{BakryGentilLedoux2014}, and the resulting mutual information bounds are evaluated using the EPI.  The resulting correction remains positive and therefore has the same directional limitation as Fong and Tan's bound when the error probability is below $1/2$.

\paragraph{Exponent methods and edge removal}
For discrete memoryless degraded broadcast channels, Oohama~\cite{Oohama2015DBC} combines information spectrum bounds, exponential moment estimates, and a recursive choice of auxiliary distributions. The resulting single-letter lower bound on the correct-decoding exponent is strictly positive outside the capacity region. A related method applies to broadcast channels with degraded message sets~\cite{OohamaAsymmetricBC}.

Kosut and Kliewer~\cite{KosutKliewer2019} relate strong converses to edge-removal properties, obtaining an exponential strong converse for general discrete memoryless broadcast channels. Their result does not require a known single-letter capacity characterization: proving a strong converse need not itself identify auxiliaries with which to evaluate the capacity region.

\paragraph{Change of measure and Markov constraints}

Tyagi and Watanabe~\cite{TyagiWatanabe2020} condition the original distribution on successful decoding. Since conditioning can alter both Markov relations and channel transition probabilities, they replace exact Markov constraints by information penalties in a variational characterization. Their applications include source coding and the wiretap channel. The same conditioning step applies to broadcast channels, although completing the converse requires a suitable penalized single-letter bound. This approach highlights another  difficulty of handling auxiliaries: the relations involving the auxiliaries must be controlled under the changed distribution before the weak converse argument can be recovered.

\subsection{The Challenge of a Tight Second-Order Converse and Our Approach}

For a tight second-order converse, identifying an auxiliary random variable that yields the correct mutual information bounds is not enough. These bounds concern {\em expectations}, whereas the second-order region depends on joint fluctuations of the {\em information densities}.  Since the second-order term in the log message sizes is itself of order $\sqrt n$, an unspecified $O(\sqrt n)$ remainder does not determine its coefficient. A Verd\'u--Han-type information spectrum bound~\cite{VH94} is therefore a natural starting point, but its evaluation requires additional distributional control.

Here is where our main contribution lies. We first construct an auxiliary additive Gaussian noise channel (see Fig.~\ref{fig:aux}) and establish quantitative relative entropy bounds for the broadcast channel output distributions conditioned on the resulting auxiliary observation.   We then apply the {\em strong functional representation lemma} (SFRL) of Li and El Gamal~\cite{LiElGamal2018} and select a suitable realization of its independent seed to obtain a deterministic partition of the retained receiver-1 message set. The component label plays a role analogous to the auxiliary $U$ in~\eqref{eqn:aux_intro}, while additionally providing the distributional control needed at the second-order scale.

The SFRL has found applications in universal compression representations~\cite{ZhangQianChenKhisti2021}, privacy-utility tradeoffs~\cite{ZamaniOechteringSkoglund2024}, and rate-distortion-perception theory~\cite{TheisWagner2021}. Here, it controls the conditional entropy of the component label given the receiver-2 output while retaining bounds on the average Gaussian relative entropy of the component outputs. Combined with Ornstein--Uhlenbeck entropy dissipation and Gaussian functional inequalities~\cite{BakryGentilLedoux2014}, these properties allow us to construct auxiliary output measures for which the codeword-dependent terms cancel in a weighted sum of the conditional means of the information densities. The remainder terms have zero conditional mean given the transmitted codeword and are controlled by variance bounds, while the principal joint fluctuations are evaluated by a normal approximation.  More details are provided in Section~\ref{sec:conv_prf}.  

\subsection{Notational Conventions}
Throughout the paper, logarithms are natural, so information quantities are measured in nats. We write $\rme^x=\exp(x)$. We use the usual information-theoretic notation~\cite{elgamal} for entropy $H(X)$, conditional entropy $H(X|Y)$, differential entropy $h(X)$, mutual information $I(X;Y)$, and relative entropy $D(P\|Q)$.

We write $X^n=(X_1,\ldots,X_n)$ for a length-$n$ random vector and $x^n=(x_1,\ldots,x_n)$ for its realizations. For $x^n,y^n\in\bbR^n$, let $\langle x^n,y^n\rangle=\sum_{i=1}^n x_i y_i$ denote their inner product and $\|x^n\|=\sqrt{\langle x^n,x^n\rangle}$ the Euclidean norm. For a real matrix $A$, its squared Hilbert--Schmidt norm is $\|A\|_{\mathrm{HS}}^2=\tr(A^\top A)$. The matrix $I_n$ denotes the $n\times n$ identity matrix. We use $\nabla h$ and $\nabla^2 h$ for the gradient and Hessian of a scalar-valued function $h$, respectively, and $\Jac F$ for the Jacobian matrix of a vector-valued function $F$. For a random vector $X^n$ with finite second moment, define the minimum mean-squared error given $Y$ as  $\mmse(X^n|Y):=\bbE\big[\|X^n-\bbE[X^n|Y]\|^2\big]$.

The Gaussian distribution with mean vector $\mu$ and covariance matrix $\Sigma$ is denoted by $\calN(\mu,\Sigma)$. The standard Gaussian density and cumulative distribution function are $\phi(x):=\frac{1}{\sqrt{2\pi}}\rme^{-x^2/2}$ and $\Phi(x):=\int_{-\infty}^x\phi(u)\,\rmd u$. In $n$ dimensions, we write $\phi_n :=\calN(0,I_n)$ for the standard Gaussian distribution and $\phi_n(x^n)$ for its density at $x^n$.

All asymptotics are as $n\to\infty$. For subsequent clarity, we adopt a slightly unconventional but intuitive sign convention for the $O(\cdot)$ notation: for a {\em positive} sequence $a_n$, we use $O(a_n)$ to denote a {\em nonnegative} sequence bounded above by a constant multiple of $a_n$ for all sufficiently large $n$. Accordingly, $-O(a_n)$ denotes the negative of such a sequence. The implicit constants may depend on   channel parameters, a   boundary point of the capacity region, and specified ranges of error probabilities, but not on $n$ or the codebook. Any additional uniformity is stated explicitly. Quantities such as relative entropies may depend on $n$, even when this dependence is not shown explicitly.

\section{Problem Setting and Main Result}
\subsection{Channel Model and Code Definition}
We consider the Gaussian broadcast channel
$\tilY_j^n=\tilX^n+\tilZ_j^n$, $j=1,2$, used without feedback
under the per-codeword power constraint $\|\tilde x^n(m_1,m_2)\|^2\le nP$, where $P>0$. The noise pairs $\{(\tilZ_{1,i}, \tilZ_{2,i})\}_{i=1}^n$ are zero-mean jointly Gaussian random vectors that are independent and identically distributed over channel uses and independent of the messages. They each have covariance matrix
\begin{align}
    \mathrm{Cov}\left( \begin{bmatrix}
        \tilZ_{1,i} \\ \tilZ_{2,i}
    \end{bmatrix} \right) = \begin{bmatrix}
         N_1 & \rho \sqrt{N_1 N_2}\\ \rho \sqrt{N_1 N_2} & N_2 
    \end{bmatrix}\quad\mbox{where}\quad  0<N_1< N_2\quad\mbox{and}\quad |\rho|\le1. 
\end{align}
Here $N_j$ is the noise variance from the transmitter to receiver $j$ and $\rho $ is the noise correlation coefficient. To reduce notation, we introduce the {\em signal-to-noise ratios}
\begin{align}
    S_j := \frac{P}{N_j},\quad j = 1,2,\qquad S_1>S_2>0.
\end{align}

We normalize the input, outputs, and noises by $\sqrt{P}$ and use $X^n := \tilX^n/\sqrt{P}$, $Y_j^n := \tilY_j^n/\sqrt{P}$ and $Z_j^n := \tilZ_j^n/\sqrt{P}$ in the rest of the paper. Thus, the normalized broadcast channel is 
\begin{align}
    Y_1^n = X^n+ Z_1^n,\qquad Y_2^n = X^n+ Z_2^n, \label{eqn:norm_channel}
\end{align}
with unit power constraint, i.e., $\|x^n(m_1,m_2)\|^2\le n$. Define the normalized noise variance $\nu_j := S_j^{-1}$. The pair of noises $(Z_{1,i}, Z_{2,i}) $ has joint distribution
\begin{align}
     \begin{bmatrix}
        Z_{1,i} \\ Z_{2,i}
    \end{bmatrix} \sim\calN \left( \begin{bmatrix}
        0\\ 0
    \end{bmatrix}, \begin{bmatrix}
        1/S_1 & \rho/\sqrt{S_1 S_2} \\ \rho/\sqrt{S_1 S_2} & 1/S_2
    \end{bmatrix}\right) .  \label{eqn:Z1Z2_noise}
\end{align}
Each normalized marginal channel's density is 
\begin{align}
    \rvW_j^n(y_j^n | x^n) = \Big(\frac{S_j}{2\pi}\Big)^{n/2}\exp\left\{ -\frac{S_j}{2}\|y_j^n-x^n\|^2 \right\}.
\end{align}

We formally define a code as follows. 
\begin{definition}[Gaussian  broadcast channel code]
    An $(n, M_{1,n}, M_{2,n})$-code for the (two-receiver) Gaussian  broadcast channel in~\eqref{eqn:norm_channel} and~\eqref{eqn:Z1Z2_noise} consists of an encoder $f_n : [M_{1,n}]\times [M_{2,n}]\to\bbR^n$ and  two (measurable) decoders $\varphi_{j,n}:\bbR^n\to [ M_{j,n} ], j = 1,2$ such that 
    \begin{align}
        \| f_n(m_1, m_2)\|^2\le n\qquad\forall\, (m_1,m_2) \in [M_{1,n}] \times [M_{2,n}] . \label{eqn:pow_constr}
    \end{align}
The messages $W_1$ and $W_2$ are independent and uniformly distributed on $[M_{1,n}]$ and $[M_{2,n}]$, respectively.  Message $W_j$ is intended for receiver $j$. When the randomly selected messages $W_1$ and $W_2$ are equal to $m_1$ and $m_2 $, respectively, the transmitted codeword is written as $x^n(m_1, m_2) = f_n(m_1, m_2)$. Let $\hatW_j = \varphi_{j,n}(Y_j^n)$ be the decoded message at receiver $j$. The {\em conditional error probability is}
\begin{align}
    \eps_n(m_1,m_2) := \Pr\big( (\hatW_1,\hatW_2)\ne  (m_1  ,m_2) \big| W_1=m_1, W_2 = m_2\big). \label{eqn:err_m1m2}
\end{align}
If an $(n, M_{1,n}, M_{2,n})$-code satisfies that its {\em average error probability}
\begin{align}
    \eps_{n, \mathrm{av}}:=\frac{1}{M_{1,n}M_{2,n}}\sum_{m_1, m_2}\eps_n(m_1,m_2) \le\eps_n, \label{eqn:ave_err}
\end{align}
we say that it is an  $(n, M_{1,n}, M_{2,n},\eps_n)_{\mathrm{av}}$-code. Similarly if an  $(n, M_{1,n}, M_{2,n})$-code satisfies that its {\em maximum error probability}
    \begin{align}
    \eps_{n, \mathrm{max}}:=\max_{m_1, m_2}\eps_n(m_1,m_2) \le\eps_n,
\end{align}
we say that it is an $(n, M_{1,n}, M_{2,n},\eps_n)_{\mathrm{max}}$-code.
 \end{definition}

For $x,y\ge0$, define the Gaussian {\em capacity},  {\em dispersion}  and {\em cross-dispersion} functions respectively as 
\begin{align}
    \rvC(x):=\frac{1}{2}\log(1+x),\quad \rvV(x):=\frac{x(x+2)}{2(x+1)^2}\quad\mbox{and}\quad
  \rvV(x,y):=\frac{x(y+2)}{2(x+1)(y+1)},
\end{align}
Note that $\rvV(x,x)=\rvV(x)$. 
The {\em capacity region} for the two-user Gaussian broadcast channel is defined as the closure of the set of all achievable (first-order) rates. Under the vanishing error probability criterion,   Cover~\cite{cover72} and  Bergmans~\cite{Bergmans1974} showed that the capacity region is
\begin{align}
    \calC = \bigcup_{\alpha \in [0,1]}\left\{ (R_1, R_2)\in\bbR_+^2: R_1\le C_1 (\alpha),~R_2 \le C_2(\alpha)\right\}
\end{align}
where
\begin{align}
    C_1(\alpha)=\rvC(\alpha S_1) \quad \mbox{and}\quad C_2(\alpha) = \rvC\bigg(\frac{\bar{\alpha}S_2}{1+\alpha S_2}\bigg)=\frac{1}{2}\log\frac{1+S_2}{1+\alpha S_2}. \label{eqn:C1C2}
\end{align}
The parameter $\alpha$  parameterizes an optimal Gaussian choice of the auxiliary
random variable (cf.~\eqref{eqn:aux_intro}) and can be regarded as the {\em power-splitting parameter}.
The Cover--Bergmans region \cite{cover72,Bergmans1974} applies to both the average and maximum error probability criteria; the strong converse was proved by Fong and Tan~\cite{FongTan2017GaussianBC}.  

\begin{remark}[Dependence on noise correlation]
The (first-order) capacity region depends only on the marginal channels~\cite[Lemma~5.1]{elgamal}. Under the joint-error criterion considered here (cf.~\eqref{eqn:err_m1m2}), however, the second-order coding region generally depends on the noise correlation coefficient~$\rho$ through the dispersion matrix $\Sigma_\alpha(\rho)$, defined below in~\eqref{eqn:Sigma_alpha}. For more details, see Subsection~\ref{sec:noise}.
\end{remark}

\begin{definition}[Second-order coding region]
Fix $0<\eps<1$, a first-order rate pair $R^*=(R_1^*, R_2^*)$, and $\dagger\in\{\mathrm{av},\mathrm{max}\}$. An offset pair $(L_1, L_2)\in\bbR^2$ is said to be {\em $(\eps,R^*)_\dagger$-second-order achievable} if there exists a sequence of $(n,M_{1,n}, M_{2,n},\eps_n)_\dagger$-codes such that 
\begin{align}
    \liminf_{n\to\infty}\frac{1}{\sqrt{n}}(\log M_{j,n}-nR_j^*)\ge L_j,\quad j = 1,2,\quad \mbox{and}\quad \limsup_{n\to\infty}\eps_n\le\eps. \label{eqn:second-order_def}
\end{align}
The closure of the set of all  $(\eps,R^*)_\dagger$-second-order achievable offset pairs is the (optimal) second-order region, denoted as $\calL_\dagger(\eps;R^*)$.
\end{definition}

In this paper, we restrict attention to first-order rate pairs $R^*=(C_1(\alpha),C_2(\alpha))$, $\alpha\in[0,1]$. These comprise the curved portion of the capacity boundary and its two endpoints, as shown in Fig.~\ref{fig:bc-capacity}.

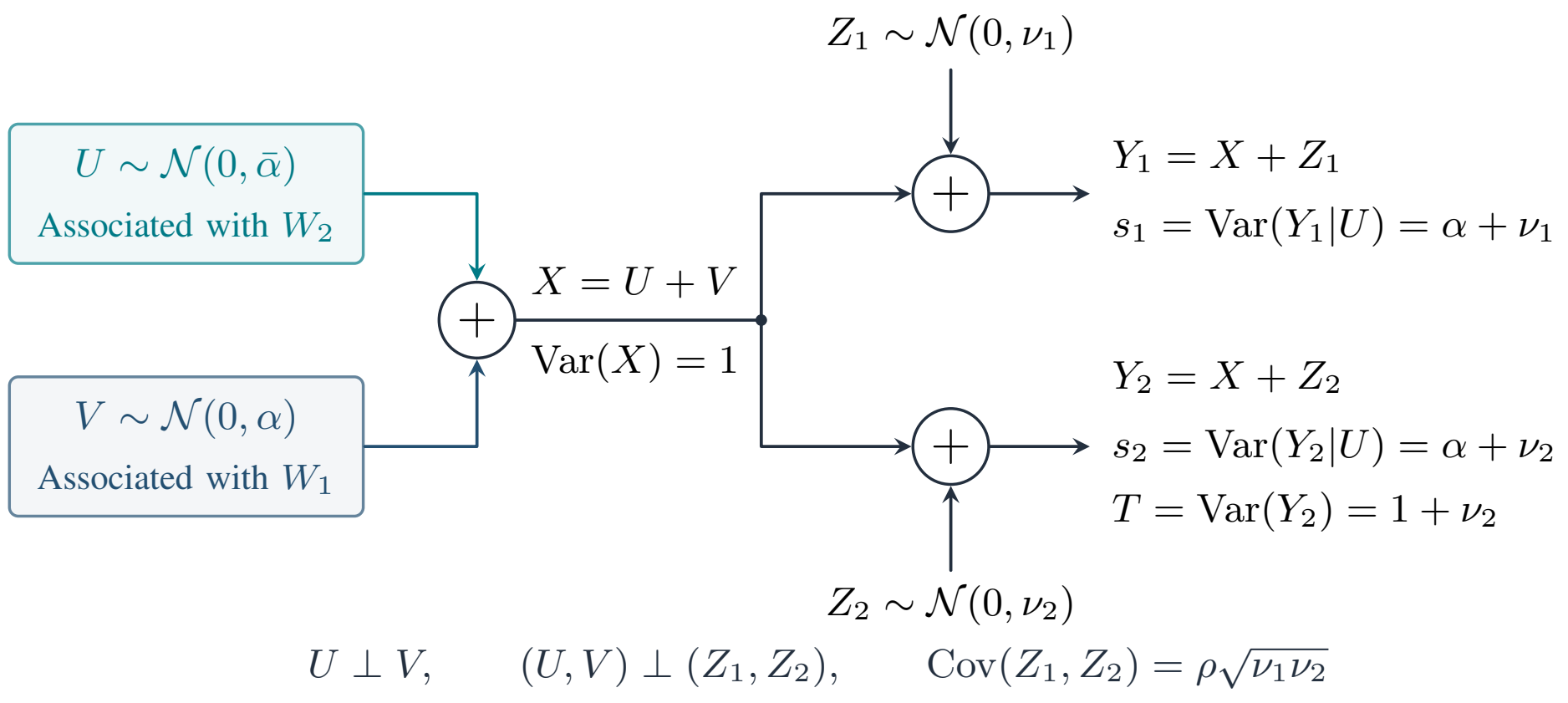
\begin{figure}
    \centering
\begingroup%
\definecolor{BCSupBlue}{RGB}{35,79,113}%
\definecolor{BCSupTeal}{RGB}{0,122,134}%
\definecolor{BCSupInk}{RGB}{35,47,62}%
\begin{tikzpicture}[
  x=1cm,y=1cm,
  font=\footnotesize,
  line cap=round,line join=round,
  bcsup arrow/.style={->,>=stealth,line width=.7pt,draw=BCSupInk},
  bcsup sum/.style={circle,draw=BCSupInk,fill=white,
    line width=.65pt,minimum size=6mm,inner sep=0pt,font=\large},
  bcsup source/.style={rounded corners=2pt,line width=.65pt,
    minimum width=2.8cm,minimum height=1.05cm,align=center,
    inner sep=5pt},
  bcsup output/.style={anchor=west,align=left,inner sep=0pt}
]
  \node[bcsup source,draw=BCSupTeal!70,fill=BCSupTeal!5,
    text=BCSupTeal] (U) at (1.4,1.0)
    {$U\sim\mathcal N(0,\bar\alpha)$\\[3pt]
     {\scriptsize Associated with $W_2$}};
  \node[bcsup source,draw=BCSupBlue!70,fill=BCSupBlue!5,
    text=BCSupBlue] (V) at (1.4,-1.0)
    {$V\sim\mathcal N(0,\alpha)$\\[3pt]
     {\scriptsize Associated with $W_1$}};
  \node[bcsup sum] (sumX) at (3.7,0) {$+$};
  \draw[bcsup arrow,draw=BCSupTeal] (U.east)--(3.7,1.0)--(sumX.north);
  \draw[bcsup arrow,draw=BCSupBlue] (V.east)--(3.7,-1.0)--(sumX.south);

  \coordinate (split) at (5.95,0);
  \draw[line width=.7pt,draw=BCSupInk] (sumX.east)--(split);
  \node[anchor=south,inner sep=2pt] at (4.95,.10) {$X=U+V$};
  \node[anchor=north,inner sep=2pt] at (4.95,-.10)
    {$\operatorname{Var}(X)=1$};
  \fill[BCSupInk] (split) circle[radius=1.3pt];

  \node[bcsup sum] (sumYone) at (7.45,1.0) {$+$};
  \node[bcsup sum] (sumYtwo) at (7.45,-1.0) {$+$};
  \draw[bcsup arrow] (split)|-(sumYone.west);
  \draw[bcsup arrow] (split)|-(sumYtwo.west);

  \node[anchor=south,inner sep=2pt] at (7.45,2.02)
    {$Z_1\sim\mathcal N(0,\nu_1)$};
  \draw[bcsup arrow] (7.45,1.98)--(sumYone.north);
  \node[anchor=north,inner sep=2pt] at (7.45,-2.02)
    {$Z_2\sim\mathcal N(0,\nu_2)$};
  \draw[bcsup arrow] (7.45,-1.98)--(sumYtwo.south);

  \draw[bcsup arrow] (sumYone.east)--(8.55,1.0);
  \draw[bcsup arrow] (sumYtwo.east)--(8.55,-1.0);
  \node[bcsup output] at (8.72,1.0)
    {$Y_1=X+Z_1$\\[5pt]
     $s_1=\operatorname{Var}(Y_1 | U)=\alpha+\nu_1$};
  \node[bcsup output] at (8.72,-1.0)
    {$Y_2=X+Z_2$\\[5pt]
     $s_2=\operatorname{Var}(Y_2| U)=\alpha+\nu_2$\\[4pt]
     $T=\operatorname{Var}(Y_2)=1+\nu_2$};

  \node[align=center,text=BCSupInk,inner sep=0pt] at (6.4,-2.75)
    {$U\perp V,\qquad (U,V)\perp(Z_1,Z_2),\qquad
      \operatorname{Cov}(Z_1,Z_2)=\rho\sqrt{\nu_1\nu_2}$};
\end{tikzpicture}%
\endgroup
    \caption{Illustration of the quantities in \eqref{eqn:sj_def}.}
    \label{fig:superposition}
\end{figure}
\subsection{The Main Result and an Alternative Characterization}
The parameter $\alpha\in[0,1]$ in \eqref{eqn:C1C2} represents the fraction of the transmit power that is allocated to message $W_1$ and $\bar{\alpha}$ is the fraction allocated to the weaker receiver's message $W_2$. Define
\begin{alignat}{2}
s_j &:= \alpha+\nu_j = \frac{1+\alpha S_j}{S_j},
&\qquad T &:= 1+\nu_2, \label{eqn:sj_def}\\
\mu_\alpha &:= \frac{s_2}{s_1}
= \frac{S_1(1+\alpha S_2)}{S_2(1+\alpha S_1)},
&\qquad C_\alpha &:= C_1(\alpha)+\mu_\alpha C_2(\alpha).  \label{eqn:mualpha_def}
\end{alignat}
Let us provide some intuition for these key quantities. Consider the Gaussian superposition model $X=U+V$ where $U\sim\calN(0,\bar{\alpha})$ and $V\sim \calN(0,\alpha)$  are independent of each other and of the noise pair. See Fig.~\ref{fig:superposition}. They represent the contributions associated with messages  \(W_2\) and \(W_1\), respectively. Conditional on $U=u$, $Y_j = u+V+Z_j \sim \calN(u,\alpha+\nu_j)$. Consequently $s_j =\mathrm{Var}(Y_j|U)$. Without conditioning on \(U\), both signal contributions remain. Under the same Gaussian model, $T=\mathrm{Var}(Y_2)=\alpha+\bar{\alpha}+\nu_2=1+\nu_2$. With $\mu_\alpha$ held fixed, the weighted sum $a\mapsto C_1(a)+\mu_\alpha C_2(a)$ is stationary at $a=\alpha$.  Indeed, $C_1'(\alpha)= 1/(2s_1)$ and $C_2'(\alpha)=-1/(2s_2)$ so $\mu_\alpha=-C_1'(\alpha)/C_2'(\alpha) = s_2/s_1$. Geometrically, the supporting line has slope $-1/\mu_\alpha$. The quantity $C_\alpha$ is the weighted sum capacity for the supporting weight $\mu_\alpha$ in the sense that 
\begin{align}
    C_\alpha=\max_{(R_1, R_2) \in \calC}\big\{R_1+\mu_\alpha R_2\big\}.
\end{align}
The corresponding first-order constraint is $R_1+\mu_\alpha R_2\le C_\alpha$. See Fig.~\ref{fig:bc-capacity}.

Define the two   dispersions and their cross-dispersion as
\begin{align}
V_1(\alpha)
&:=\rvV(\alpha S_1), \label{eqn:disp1}\\
V_2(\alpha)
&:=\rvV(S_2)+\rvV(\alpha S_2)
   -2\rvV(\alpha S_2,S_2),
\label{eq:dispersion-one-two}\\
V_{12}(\alpha,\rho)
&:=\rho\bigg(\rho-2\sqrt{\frac{S_2}{S_1}}\bigg)
   \bigl(
     \rvV(\alpha S_1,\alpha S_2)
     -\rvV(\alpha S_1,S_2)
   \bigr).
\label{eq:cross-dispersion}
\end{align}
These quantities are the conditional variances and covariance per channel use of the information densities analyzed in Subsection~\ref{sec:mmts}.  Now define the covariance (dispersion) matrix and bivariate Gaussian distribution function 
\begin{align}
    \Sigma_\alpha(\rho) &:= \begin{bmatrix}
        V_1(\alpha) & V_{12}(\alpha,\rho)\\ V_{12}(\alpha,\rho) & V_2(\alpha)
    \end{bmatrix} , \label{eqn:Sigma_alpha}\\
    \Psi_{\alpha,\rho}(z_1, z_2) &:= \Pr(\rvZ_1\le z_1, \rvZ_2\le z_2),\qquad (\rvZ_1, \rvZ_2)\sim\calN \big(0, \Sigma_\alpha(\rho) \big).
\end{align}
For $0<\alpha<1$,  $\Sigma_\alpha(\rho)$ is positive definite, as verified in Proposition \ref{prop:Fproperty}, even when the noise covariance in \eqref{eqn:Z1Z2_noise} is singular. For each $\alpha\in[0,1]$ and $\rho\in[-1,1]$, define the {\em  $(\alpha,\rho)$-modified Q-function}\footnote{The notation $\rvQ$ reflects the analogy with the usual Gaussian Q-function: here, a joint Gaussian upper-tail probability is maximized over threshold pairs with a prescribed weighted sum.}
\begin{align}
    \rvQ_{\alpha,\rho}(r) := \sup_{(z_1, z_2) \in\bbR^2:z_1+\mu_\alpha z_2 = r} \Pr(\rvZ_1\ge z_1, \rvZ_2\ge z_2). \label{eqn:Falpharho}
\end{align}
Let us provide some intuition for this quantity. The constraint $z_1+\mu_\alpha z_2=r$ has the same normal vector as the supporting line $R_1+\mu_\alpha R_2=C_\alpha$ of the capacity region at $(C_1(\alpha),C_2(\alpha))$ (cf.\ Fig.~\ref{fig:bc-capacity}). Thus, $\rvQ_{\alpha,\rho}(r)$  is the largest Gaussian upper-tail probability among threshold pairs with a prescribed weighted sum in this normal direction. At an interior boundary point, the optimization along this line corresponds to varying the second-order perturbation of the power allocation. Proposition~\ref{prop:Fproperty} shows that $\rvQ_{\alpha,\rho}\colon\mathbb R\to(0,1)$ is continuous and strictly decreasing, with limits~$1$ and~$0$ at
$-\infty$ and $+\infty$, respectively. Its inverse is therefore well defined, and we set
\begin{equation}
  \rvG_{\alpha,\rho}(\eps) :=\rvQ_{\alpha,\rho}^{-1}(1-\eps), \qquad \eps\in(0,1).
\end{equation}
We are now ready to state the main theorem, which is proved in Sections \ref{sec:ach_prf} (achievability) and \ref{sec:conv_prf} (converse).
\begin{theorem}\label{thm:main}
   For the Gaussian broadcast channel in~\eqref{eqn:norm_channel}
with $S_1>S_2>0$, $\rho\in[-1,1]$, and $\eps\in(0,1)$,
the second-order regions under the average and maximal
joint-error criteria coincide at every first-order rate pair
$R^*=(C_1(\alpha),C_2(\alpha))$, $\alpha\in[0,1]$.
Denote their common value by $\calL(\eps;R^*)$.  More precisely, 
    \begin{enumerate}
        \item[(i)] If $R^* = (C_1(\alpha), C_2(\alpha))$ for $0<\alpha<1$, the common region  is
        \begin{align}
            \calL (\eps;R^*)=\left\{(L_1, L_2)\in\bbR^2: L_1+\mu_\alpha L_2\le \rvG_{\alpha,\rho}(\eps) \right\}. \label{eqn:alpha_mid}
        \end{align}
        \item[(ii)] If $R^*=(0,C_2(0))=(0,\rvC(S_2))$, the common region  is
         \begin{align}
             \calL(\eps;R^*)= \left\{(L_1, L_2)\in\bbR^2: L_1^++ \frac{S_1}{S_2}L_2\le \frac{S_1}{S_2}\sqrt{\rvV(S_2})\Phi^{-1}(\eps)\right\}. \label{eqn:alpha0} 
         \end{align}
         \item[(iii)] If $R^*=(C_1(1),0)=(\rvC(S_1),0)$, the common region is 
         \begin{align}
             \calL (\eps;R^*)=\left\{(L_1, L_2)\in\bbR^2: L_1+ \frac{S_1(1+S_2)}{S_2(1+S_1)}L_2^+\le  \sqrt{\rvV(S_1})\Phi^{-1}(\eps)\right\}.  \label{eqn:alpha1} 
         \end{align}
    \end{enumerate}
\end{theorem}
The positive parts   in \eqref{eqn:alpha0}  and \eqref{eqn:alpha1}  (i.e., $L_1^+, L_2^+$) arise from the  convention in the definition of second-order achievability: when $R_j^*=0$, a negative target offset $L_j$ is already satisfied by $M_{j,n}=1$.

The halfspace in \eqref{eqn:alpha_mid} admits an alternative characterization analogous to those of second-order regions for multiple-access channels in~\cite{ScarlettMartinezGuillen2015,ST2015}. The union over $\beta$ in~\eqref{eqn:union_char} corresponds to allowing the power-splitting parameter to vary with the blocklength as
$\alpha_n=\alpha+\beta/\sqrt{n}$. These perturbations preserve the limiting first-order rate pair while changing the second-order offsets. Allowing $\beta$ to range over $\mathbb R$ enables the superposition  coding construction to attain the entire second-order region.

\begin{proposition}[Multivariate Gaussian characterization] \label{prop:Fproperty} 
For any fixed $\alpha \in [0,1]$ and $\rho \in [-1,1]$, the matrix $\Sigma_\alpha(\rho)$ in \eqref{eqn:Sigma_alpha} is positive semidefinite, and it is positive definite for every $\alpha \in (0,1)$.  The supremum that characterizes $\rvQ_{\alpha, \rho}$ is attained for every $r$. The function  $\rvQ_{\alpha, \rho}$  is continuous and strictly decreasing, with limits $1$ and $0$ at $-\infty$ and $+\infty$, respectively. Consequently $\rvG_{\alpha,\rho}$ is continuous on $(0,1)$ and it can be expressed as 
\begin{align}
    \rvG_{\alpha,\rho}(\eps) = \max_{ (z_1,z_2) \in\bbR^2 : \Pr(\rvZ_1 \ge z_1, \rvZ_2\ge z_2)\ge 1-\eps}\big(z_1 + \mu_\alpha z_2\big). \label{eqn:Gmax}
\end{align}
At the power endpoints,
\begin{alignat}{2}
\rvQ_{0,\rho}(r)
&= \Phi\bigg(-\frac{r}{\mu_0\sqrt{V_2(0)}}\bigg), &\qquad \rvQ_{1,\rho}(r)
&= \Phi\bigg(-\frac{r}{\sqrt{V_1(1)}}\bigg), \label{eqn:F_endpoints}\\
\rvG_{0,\rho}(\eps)
&= \mu_0\sqrt{V_2(0)}\,\Phi^{-1}(\eps),
&\qquad \rvG_{1,\rho}(\eps)
&= \sqrt{V_1(1)}\,\Phi^{-1}(\eps). \label{eqn:G_endpoints}
\end{alignat}
For $0<\alpha<1$, the halfspace in \eqref{eqn:alpha_mid} equals
\begin{align}
    \calL(\eps;R^*)=\bigcup_{\beta\in\bbR}\left\{ (L_1, L_2)\in\bbR^2: \Psi_{\alpha,\rho}\left( \frac{\beta S_1}{2(1+\alpha S_1)}-L_1, -\frac{\beta S_2}{2(1+\alpha S_2)}-L_2\right)\ge 1-\eps\right\}. \label{eqn:union_char}
\end{align}
For $0<\alpha<1$ and $\rho=0$ (i.e., independent noises $Z_1$ and $Z_2$), the bivariate Gaussian CDF factors into the two marginal Gaussian CDFs and 
\begin{align}
    \rvG_{\alpha,0}(\eps) = \max_{e_1, e_2\in (0,1): (1-e_1)(1-e_2)\ge 1-\eps}g_\alpha(e_1, e_2),  \label{eqn:error_alloc}
\end{align}
where for $e_1,e_2\in (0,1)$,
\begin{align}
    g_\alpha(e_1, e_2):=\sqrt{V_1(\alpha)}\Phi^{-1}(e_1)+\mu_\alpha \sqrt{V_2(\alpha)}\Phi^{-1}(e_2).
\end{align}    
\end{proposition}
\begin{proof}
First suppose that $0<\alpha<1$. Since
\begin{align}
 \big|\rho\big(\rho-2\sqrt{S_2/S_1}\big)\big| \le 1+2\sqrt{S_2/S_1},   
\end{align}
the absolute cross-dispersion is largest at $\rho=-1$ (a point to be reemphasized in Subsection \ref{sec:noise}).
Substituting the definitions of $\rvV$ gives
\begin{align}
 \det\Sigma_\alpha(\rho)
 \ge \det\Sigma_\alpha(-1)
 =\frac{\bar\alpha S_2}{(1+S_2)^2}
 \left[
   \rvV(\alpha S_1)
   +\frac{\alpha\bar\alpha S_1S_2
           (\alpha\sqrt{S_1S_2}-1)^2}
          {2(1+\alpha S_1)^2(1+\alpha S_2)^2}
 \right]
 >0.
\end{align}
Together with $V_1(\alpha)=\rvV(\alpha S_1)>0$, this proves
positive definiteness, including when $\rho=\pm1$.

For fixed $r$, the function $ 
 z_1\mapsto
 \Pr( 
   \rvZ_1\ge z_1,\,
   \rvZ_2\ge\frac{r-z_1}{\mu_\alpha})$
is continuous, positive, and tends to zero as $z_1\to\pm\infty$. Its maximum is therefore attained. These limits are uniform for $r$ in compact intervals, so the maximization can locally be restricted to a common compact set. Consequently, $\rvQ_{\alpha,\rho}$ is continuous.

To prove strict monotonicity, fix $r'<r$ and let
$(z_1^*,z_2^*)$ attain the maximum defining
$\rvQ_{\alpha,\rho}(r)$, so that
\begin{align}
 z_1^*+\mu_\alpha z_2^*=r,
 \qquad
 \Pr(\rvZ_1\ge z_1^*,\rvZ_2\ge z_2^*)
 =\rvQ_{\alpha,\rho}(r).   
\end{align}
Set $\delta=(r-r')/(1+\mu_\alpha)>0$.
The pair $(z_1^*-\delta,z_2^*-\delta)$ has weighted sum $r'$.
Consequently,
\begin{align}
 \rvQ_{\alpha,\rho}(r')
 \ge
 \Pr(\rvZ_1\ge z_1^*-\delta,\rvZ_2\ge z_2^*-\delta)
 >
 \Pr( \rvZ_1\ge z_1^*,\rvZ_2\ge z_2^*)
 =
 \rvQ_{\alpha,\rho}(r).   
\end{align}
The strict inequality holds because the Gaussian density is
strictly positive everywhere.

We next establish the limits at infinity. Put $t_r:=r/(1+\mu_\alpha)$. The pair $(t_r,t_r)$ is feasible in the definition of $\rvQ_{\alpha,\rho}(r)$. Hence, by the union bound,
\begin{align}
 \rvQ_{\alpha,\rho}(r)
  \ge \Pr\{\rvZ_1\ge t_r,\rvZ_2\ge t_r\}\ge 1-
 \Phi\bigg(\frac{t_r}{\sqrt{V_1(\alpha)}}\bigg) - \Phi\bigg(\frac{t_r}{\sqrt{V_2(\alpha)}}\bigg).
\end{align}
The last expression tends to one as $r\to-\infty$, proving
$\rvQ_{\alpha,\rho}(r)\to1$. Conversely, every pair satisfying
$z_1+\mu_\alpha z_2=r$ has at least one coordinate no smaller
than $t_r$. Hence
\begin{align}
 \rvQ_{\alpha,\rho}(r)
 \le
 \max_{j=1,2}
 \bigg[
  1-\Phi\bigg(\frac{t_r}{\sqrt{V_j(\alpha)}}\bigg)
 \bigg]
 \rightarrow0
 \quad\mbox{as} \quad r\to+\infty.
\end{align}
 
Thus $\rvQ_{\alpha,\rho}$ is a continuous, strictly decreasing
bijection from $\mathbb R$ onto $(0,1)$.
Its inverse is continuous, and hence so is
$\rvG_{\alpha,\rho}(\eps)=\rvQ_{\alpha,\rho}^{-1}(1-\eps)$. 
Moreover,
\begin{align}
 \rvQ_{\alpha,\rho}(r)\ge1-\eps \quad\Longleftrightarrow\quad r\le \rvG_{\alpha,\rho}(\eps).   
\end{align}
Attainment at $r=\rvG_{\alpha,\rho}(\eps)$ proves the asserted maximization formula for $\rvG_{\alpha,\rho}$.

At the endpoints, direct substitution yields $\Sigma_0(\rho)=\operatorname{diag}\bigl(0,\rvV(S_2)\bigr)$ and  $\Sigma_1(\rho) =\operatorname{diag}\bigl(\rvV(S_1),0\bigr).$
These matrices are positive semidefinite. The maximizing
thresholds are $(0,r/\mu_0)$ and $(r,0)$, respectively,
giving the stated endpoint formulas in \eqref{eqn:F_endpoints} and \eqref{eqn:G_endpoints} and all the remaining
claims about $\rvQ_{\alpha,\rho}$ and $\rvG_{\alpha,\rho}$ there.

For the union characterization in~\eqref{eqn:union_char}, fix $0<\alpha<1$. As $\beta$ ranges over $\bbR$, the thresholds
\begin{align}
 z_1=L_1-\frac{\beta S_1}{2(1+\alpha S_1)},
 \quad\mbox{and}\quad z_2=L_2+\frac{\beta S_2}{2(1+\alpha S_2)}   
\end{align}
range over the entire line
$z_1+\mu_\alpha z_2=L_1+\mu_\alpha L_2$.
By   symmetry around $(0,0)$, $ \Pr( \rvZ_1\ge z_1,\rvZ_2\ge z_2)=\Psi_{\alpha,\rho}(-z_1,-z_2)$. Hence the union over $\beta$ is equivalent to $\rvQ_{\alpha,\rho}(L_1+\mu_\alpha L_2)\ge1-\eps$,
which is   the asserted halfspace.

Finally, when $\rho=0$, the cross-dispersion vanishes, so the Gaussian random variables  $\rvZ_1$ and $\rvZ_2$ are independent. Substituting $z_j=\sqrt{V_j(\alpha)}\,\Phi^{-1}(e_j)$ into   $\rvG_{\alpha,0}$ in~\eqref{eqn:Gmax} gives the stated error allocation representation in \eqref{eqn:error_alloc}.
\end{proof}

\subsection{Effect of Noise Correlation} \label{sec:noise}

We illustrate the second-order region in Theorem~\ref{thm:main} and its equivalent characterization in Proposition~\ref{prop:Fproperty} by examining its dependence on the noise correlation coefficient $\rho$. Fix $S_1>S_2>0$, $\eps\in(0,1)$, and $\alpha\in(0,1)$. The corresponding first-order boundary point does not depend on $\rho$, whereas the second-order region is the halfspace $L_1+\mu_\alpha L_2\le \rvG_{\alpha,\rho}(\eps)$ (cf.\ \eqref{eqn:alpha_mid}). Since $\mu_\alpha$ is independent of $\rho$, these halfspaces have parallel boundaries. Thus, a larger value of $\rvG_{\alpha,\rho}(\eps)$ gives a larger second-order region.

For fixed $\alpha$, the marginal variances $V_1(\alpha)$ and $V_2(\alpha)$ do not depend on $\rho$. The Gaussian orthant comparison form of Slepian's inequality \cite[Theorem~2.1]{LiShao2002} states that the joint upper-tail probability is nondecreasing in the cross-covariance when the marginal variances are fixed. Taking the supremum over the same constraint $z_1+\mu_\alpha z_2=r$ (in~\eqref{eqn:Falpharho}) therefore gives
\begin{align}
    V_{12}(\alpha,\rho)\le V_{12}(\alpha,\rho')
    \quad\Longrightarrow\quad
    \rvQ_{\alpha,\rho}(r)\le \rvQ_{\alpha,\rho'}(r),
    \qquad r\in\bbR.
\end{align}
Since these functions are strictly decreasing and satisfy $\rvQ_{\alpha,\rho}(\rvG_{\alpha,\rho}(\eps))=1-\eps$, we obtain
\begin{align}
    \rvG_{\alpha,\rho}(\eps) \le \rvG_{\alpha,\rho'}(\eps).
\end{align}
Consequently, ordering the cross-dispersions also orders the second-order regions by inclusion.

To identify the extremal correlations, put $t:=\sqrt{S_2/S_1}\in(0,1)$ and
\begin{align}
    \Delta\rvV_\alpha
    &:=\rvV(\alpha S_1,\alpha S_2)
       -\rvV(\alpha S_1,S_2) =\frac{\alpha\bar\alpha S_1S_2}
       {2(1+\alpha S_1)(1+\alpha S_2)(1+S_2)}>0.
\end{align}
The cross-dispersion can then be written as
\begin{align}
    V_{12}(\alpha,\rho)
    =\Delta\rvV_\alpha\big[(\rho-t)^2-t^2\big].
\end{align}
Over $\rho\in[-1,1]$, this quantity is minimized at $\rho=t$ and maximized at $\rho=-1$. Hence, the smallest second-order region is attained at $\rho=\sqrt{S_2/S_1}$, while the largest is attained at $\rho=-1$.

The minimizing correlation has a useful interpretation. Recall that a broadcast channel is {\em physically degraded} if its joint channel distribution satisfies the Markov chain $X-Y_1-Y_2$~\cite[Sec.~5.4]{elgamal}. It is {\em stochastically degraded} if there exists a Markov kernel $K$ such that $\rvW_2(\rmd y_2|x)=\int_{\bbR}K(\rmd y_2|y_1)\rvW_1(y_1|x)\,\rmd y_1$. Thus, stochastic degradedness concerns only the marginal channels, whereas physical degradedness concerns their actual joint distribution. Every physically degraded broadcast channel is stochastically degraded, but the converse need not hold.

The Gaussian broadcast channel considered here is stochastically degraded for every $\rho$: the required kernel adds independent Gaussian noise of variance $\nu_2-\nu_1$ to the receiver-1 output. For the actual channel to be physically degraded, however, $Z_1$ and $Z_2-Z_1$ must be independent. By joint Gaussianity, this is equivalent to $\mathrm{Cov}(Z_1,Z_2-Z_1)=\rho\sqrt{\nu_1\nu_2}-\nu_1=0$, or $\rho=\sqrt{\nu_1/\nu_2}=\sqrt{S_2/S_1}$. Thus, among Gaussian broadcast channels with the same marginal channels, the physically degraded channel has the smallest second-order region under the joint-error criterion considered here. This is because physical degradedness makes the limiting information density fluctuations most negatively correlated (since $\bbE[Z_2 | Z_1=z]
=\frac{V_{12}(\alpha,\rho)}{V_1(\alpha)}\,z$ is most negative at $\rho=\sqrt{S_2/S_1}$), so favorable fluctuations at one receiver are least likely to coincide with favorable fluctuations at the other. The largest second-order region is attained at $\rho=-1$,
corresponding to perfectly negatively correlated receiver noises.

\begin{figure}
    \centering
    \input{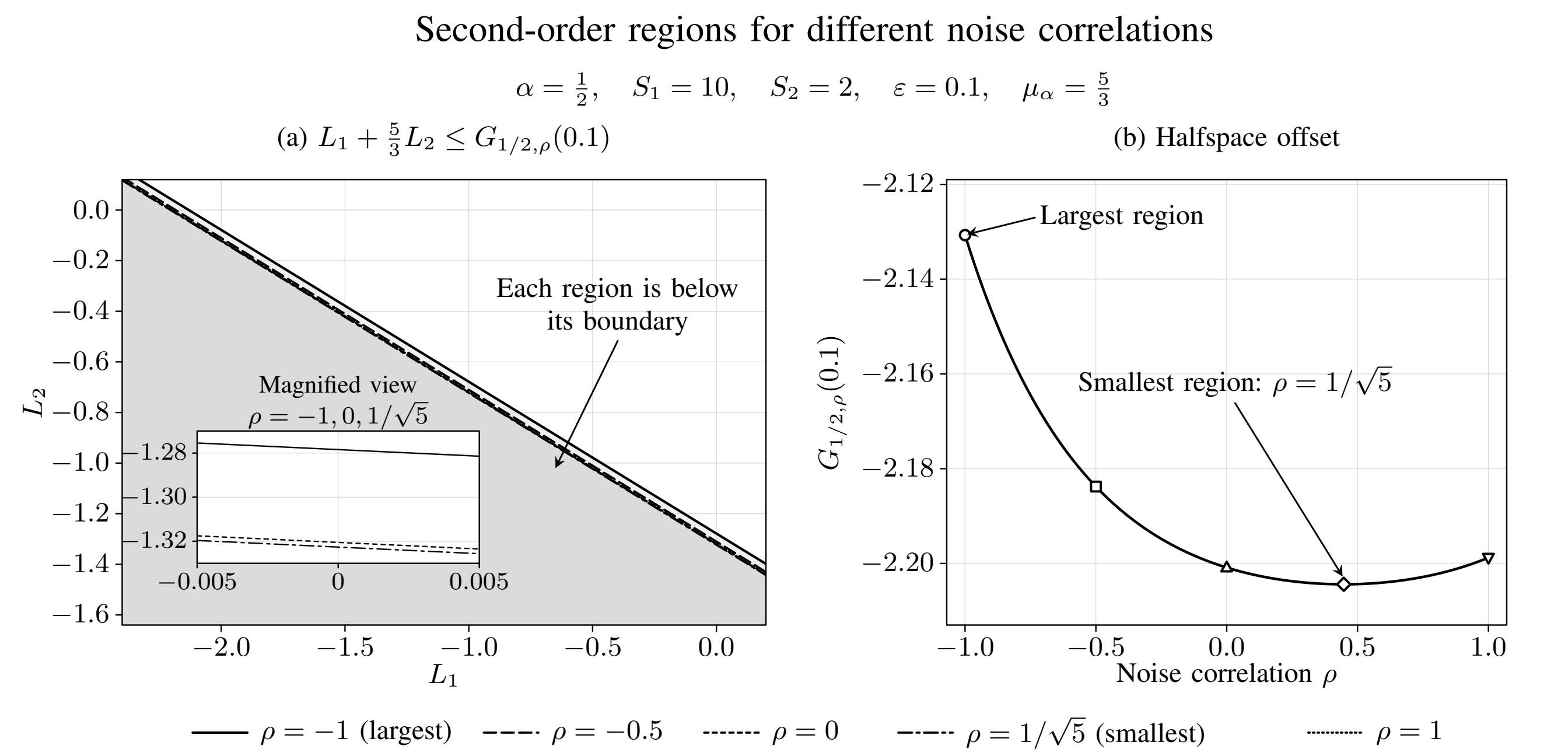}
    \caption{Effect of the noise correlation coefficient
    $\rho$ on the second-order region, with $S_1=10$,
    $S_2=2$, $\alpha=1/2$, and $\eps=0.1$.}
    \label{fig:vary_rho}
\end{figure}

Fig.~\ref{fig:vary_rho} illustrates these conclusions for $S_1=10$, $S_2=2$, $\alpha=1/2$, and $\eps=0.1$. Here, $\mu_{1/2}=5/3$, so all second-order boundaries have slope $-3/5$. Their positions depend on $\rvG_{1/2,\rho}(0.1)$, which is minimized at $\rho=1/\sqrt5$ and maximized at $\rho=-1$. The figure shows both the resulting halfspaces and the variation of $\rvG_{1/2,\rho}(0.1)$ with $\rho$.
 
\section{Achievability Proof of Theorem \ref{thm:main}} \label{sec:ach_prf}
Following the spherical superposition coding approach of Scarlett and Tan \cite{ST2015}, we derive a non-asymptotic bound and evaluate it by normal approximations. Blocklength-dependent power fractions (see also~\cite{ScarlettMartinezGuillen2015}) yield achievability in the three boundary regimes in Fig.~\ref{fig:bc-capacity}. Expurgation then gives the same regions under maximal error.


\subsection{Spherical Superposition Coding and the Finite Blocklength Bound} \label{sec:spherical}
For $n\ge3$ and $\alpha\in(0,1)$, let $u^n,v^n$ be any deterministic vectors satisfying 
\begin{align}
    \|u^n\|^2=n\bar{\alpha},\quad \|v^n\|^2=n\alpha, \quad \langle u^n,v^n\rangle =0. \label{eqn:uv}
\end{align}
Put $x^n=u^n+v^n$ and let $Y_j^n=x^n+Z_j^n$. Let $\phi_n(\fndot;t)$ denote the multivariate Gaussian density $\calN(0,tI_n)$ and $\rvW_j^n(\fndot|x^n)$ the density of $Y_j^n$. Define the ``output distributions'' $q_j(y^n|u^n)=\phi_n(y^n-u^n;s_j)$ and $q_{0,j}(y^n)=\phi_n(y^n;1+\nu_j)$ and the information densities
\begin{align}
    \iota_0 (u^n,Y_1^n)& :=\log\frac{q_1(Y_1^n|u^n) }{q_{0,1}(Y_1^n)} ,\quad
    \iota_1 (x^n,u^n,Y_1^n)  :=\log\frac{\rvW_1^n(Y_1^n|x^n) }{q_1(Y_1^n|u^n)} ,\quad
    \iota_2 (u^n,Y_2^n)  :=\log\frac{q_2(Y_2^n|u^n) }{q_{0,2}(Y_2^n) } . \label{eqn:info_dens}
\end{align}
All probabilities involving these statistics in the following lemma are taken over the correlated noise vectors $(Z_1^n, Z_2^n)$, with $u^n$ and $v^n$ held fixed.
\begin{lemma}[Finite blocklength upper bound] \label{lem:ach_fb} There are constants $K\ge 1$, $c>0$ and $n_0\ge 3$, depending only on $S_1$ and $S_2$ with the following properties. Fix $n\ge n_0$, $\alpha \in (0,1)$, $\rho \in [-1,1]$, integers $M_{1,n},M_{2,n}$  and $\gamma>0$. There is an $(n,M_{1,n},M_{2,n})_{\mathrm{av}}$-code with codeword energies $n$ such that receiver 1 estimates both messages $(W_1, W_2)$, receiver 2 estimates $W_2$, and the average probability $e_n$ that any of these three estimates is wrong satisfies
\begin{align}
    e_n&\le 1-\Pr(\iota_1>t_1, \iota_2>t_2)+ \Pr(\iota_0\le t_0)+ 3\rme^{-\gamma}+\rme^{-cn}, \label{eqn:ach_fb_bound} 
\end{align}
where $t_1 = \log M_{1,n}+\log K+\gamma$ and $t_0=t_2=\log M_{2,n}+\log K+\gamma$.  
\end{lemma}
We call $e_n$ the {\em augmented average error probability}. The quantity $\eps_{n,\mathrm{av}}$ in~\eqref{eqn:ave_err} is the average probability that at least one receiver incorrectly decodes its intended message. In defining $e_n$, we additionally count an error whenever receiver~1 incorrectly decodes $W_2$; thus, receiver~1 must decode {\em both} messages correctly. Consequently, $\eps_{n,\mathrm{av}}\le e_n$.
\begin{proof}
Generate independent ``cloud centers'' $U^n(m_2)$ uniformly on the sphere of radius $\sqrt{n\bar{\alpha}}$. Conditional on each center $u^n$, generate independent vectors $V^n(m_1,m_2)$ uniformly on the sphere of radius $\sqrt{n\alpha}$ in $(u^n)^\perp$. These vectors are the satellites associated with that cloud center. Transmit  $X^n(m_1, m_2)=U^n(m_2)+V^n(m_1, m_2)$, which has energy exactly $n$. This is the spherical superposition construction
used in \cite{ST2015} and \cite{TuninettiEtal2023}.

Let $ p_{Y_1^n | U^n}(y_1^n| u^n):= \int_{\mathbb R^n} \rvW_1^n(y_1^n| u^n+v^n)\, P_{V^n | U^n=u^n}(\mathrm d v^n)$ be the ensemble conditional output density. The standard spherical output comparisons give a ``typical'' set $\calT_{n,\alpha}$ and constants as in the statement such that
\begin{align}
    p_{Y_1^n|U^n}(y_1^n|u^n) &\le K  q_1(y_1^n|u^n)  \qquad \forall \, (u^n,y_1^n)\in \calT_{n,\alpha}, \label{eqn:output_compare}\\
    \Pr( (U^n, Y_1^n) \notin \calT_{n,\alpha})&\le \rme^{-cn}, \label{eqn:typ_set}\\
    \int q_j(y^n|u^n)\, \rmd P_{U^n}(u^n)&\le K q_{0,j}(y^n) \qquad j=1,2. \label{eqn:output_compare2}
\end{align}
For \eqref{eqn:output_compare} and \eqref{eqn:typ_set}, apply \cite[Lemma 8 and Appendix E]{ST2015}  after translation and scaling. For \eqref{eqn:output_compare2}, use \cite[Proposition~2]{MolavianJaziLaneman2015}
or \cite[Lemma 1, Eqn.~(37)]{yavas21}, at input power one and noise variance $\nu_j$.

Use the threshold decoding argument of \cite[Lemma~1]{bouch00}. Receiver 2 selects the unique cloud center whose information density  $\iota_2$ exceeds $t_2$. Receiver 1 selects the unique cloud center whose $\iota_0$ exceeds $t_0$, then the
unique satellite codeword whose $\iota_1$ exceeds $t_1$; the satellite test also requires $(u^n,y_1^n) \in \calT_{n,\alpha}$. 

Conditioned on a received vector, the expected
exponentials of the competing-cloud statistics $\iota_0$ and $\iota_2$   is at most $K$  by~\eqref{eqn:output_compare2}. Conditional on the true center and a received vector in $\calT_{n,\alpha}$, the same bound holds for a wrong
satellite by~\eqref{eqn:output_compare}.  By the union bound, the total contribution of competing cloud centers and satellite codewords to the decoding error probability is at most $3\rme^{-\gamma}$. Eqn.~\eqref{eqn:typ_set} bounds the additional error probability for the transmitted pair not being in the typical  set by $\rme^{-cn}$.  
\end{proof}

\subsection{Moments of Information Densities and Normal Approximation} \label{sec:mmts}
We now evaluate the moments of the information densities and derive the normal approximation for the probabilities in \eqref{eqn:ach_fb_bound}. For the additional
decoding test at receiver 1, set
\begin{align}
    C_0(\alpha)=\rvC(S_1) - \rvC(\alpha S_1)\quad\mbox{and}\quad V_0(\alpha)=\rvV(S_1) + \rvV(\alpha S_1) -2\rvV(\alpha S_1, S_1) .
\end{align}
\begin{lemma}[Moments and normal approximation] Fix $n\ge 3$, $\alpha \in (0,1)$, $\rho \in [-1,1]$, and deterministic vectors $u^n,v^n$ satisfying \eqref{eqn:uv}. For the information densities defined in \eqref{eqn:info_dens},
\begin{align}
    \bbE[\iota_j]&=n C_j(\alpha),\qquad\mathrm{Var}(\iota_j) = n V_j(\alpha)\qquad  j = 0,1, 2, \label{eqn:mmts1_id}\\
    \mathrm{Cov}(\iota_1, \iota_2) &= n V_{12}(\alpha,\rho). \label{eqn:mmts2_id}
\end{align}
There exists a constant $B<\infty$ (depending only on $S_1$ and $S_2$) such that for $j = 0,1,2$,
\begin{align}
    \sup_{t\in \bbR} \left| \Pr(\iota_j\le t)-\Phi \bigg(\frac{t-nC_j(\alpha)}{\sqrt{nV_j(\alpha)}} \bigg) \right|\le\frac{B}{\sqrt{n}}. \label{eqn:be_marg}
\end{align}
For every compact interval $I\subset (0,1)$, there is a constant $B_I<\infty$ (depending only on $I$, $S_1$ and $S_2$) such that for all $\alpha\in I$, 
\begin{align}
    \sup_{t_1,t_2\in \bbR} \left| \Pr(\iota_1> t_1, \iota_2>t_2)-\Psi_{\alpha,\rho} \bigg(\frac{nC_1(\alpha)-t_1}{\sqrt{n}} , \frac{nC_2(\alpha)-t_2}{\sqrt{n}}\bigg) \right|\le\frac{B_I}{\sqrt{n}}. \label{eqn:be_joint}
\end{align}
Both bounds hold uniformly over $\rho\in [-1,1]$ and all pairs in \eqref{eqn:uv}; \eqref{eqn:be_marg} is also uniform over $\alpha \in (0,1)$.
\end{lemma}
\begin{proof}
    Direct expansion of the information densities in \eqref{eqn:info_dens} gives
    \begin{align}
        \iota_1 - nC_1(\alpha) &= \frac{\langle v^n, Z_1^n\rangle}{s_1}-\frac{\alpha}{2\nu_1 s_1}( \|Z_1^n\|^2- n\nu_1 ) ,\label{eqn:inf_dens1}\\ 
        \iota_2 - nC_2(\alpha) &=  \left\langle \frac{u^n}{T}- \frac{\bar{\alpha}v^n}{Ts_2},Z_2^n \right\rangle - \frac{\bar{\alpha}}{2Ts_2} (\|Z_2^n\|^2-n\nu_2).\label{eqn:inf_dens2}
    \end{align}
    The expansion of $\iota_0 - nC_0(\alpha)$   is \eqref{eqn:inf_dens2}  with receiver~2 replaced by receiver~1 and \(T\) replaced by \(1+\nu_1\). The linear and quadratic terms in~\eqref{eqn:inf_dens1} and \eqref{eqn:inf_dens2} have zero mean and are uncorrelated. Using $\mathrm{Var}(\langle w^n, Z_j^n\rangle)= \nu_j\|w^n\|^2$  and $\mathrm{Var} \|Z_j^n\|^2=2n\nu_j^2$ proves~\eqref{eqn:mmts1_id}. For~\eqref{eqn:mmts2_id}, the linear terms contribute $-n\alpha\bar{\alpha}\rho\sqrt{\nu_1\nu_2}/(s_1s_2T)$. The identity $\mathrm{Cov}( \|Z_1^n\|^2, \|Z_2^n\|^2)=2n\rho^2\nu_1\nu_2$  gives the quadratic contribution $n\alpha\bar{\alpha}\rho^2\nu_2/(2s_1 s_2 T)$. Their sum is exactly $nV_{12}(\alpha, \rho) $ as defined in  \eqref{eq:cross-dispersion}.

    By rotational invariance and standard third moment bounds for Gaussian quadratic polynomials, the scalar and multivariate Berry--Esseen theorems \cite{Petrov,Ben05} yield \eqref{eqn:be_marg} and \eqref{eqn:be_joint}, respectively. For \eqref{eqn:be_joint}, uniformity follows also from Proposition~\ref{prop:Fproperty} and compactness of $I\times[-1,1]$.
\end{proof}

\subsection{Second-Order Achievability on the Boundary}
\begin{proposition}[Achievability with augmented error] \label{prop:ach_aug}
    For every second-order   rate pair $(L_1, L_2)$ in \eqref{eqn:alpha0}, \eqref{eqn:alpha1} and~\eqref{eqn:union_char}, there is a sequence of $(n, M_{1,n}, M_{2,n})_{\mathrm{av}}$-codes with rates at the corresponding first-order rate pair $R^*=(R_1^*, R_2^*)$ with~\eqref{eqn:second-order_def} satisfied.  Furthermore, the augmented error satisfies $\limsup_{n\to\infty}  e_n \le \eps$.
\end{proposition}
\begin{proof}
    Set $\gamma = \log n$ in Lemma \ref{lem:ach_fb} so that each threshold exceeds its log message size by $h_n = \log K +\log n$. For every test with $n C_j(\alpha_n)>t_j$,  Chebyshev's inequality gives (for $j = 0,1,2$),
    \begin{align}
        \Pr(\iota_j \le t_j)\le\frac{n V_j(\alpha_n)}{ (nC_j(\alpha_n)-t_j)^2}. \label{eqn:ach_cheby}
    \end{align}
    We use this bound for tests whose failure probabilities vanish and the Berry--Esseen bounds in \eqref{eqn:be_marg} or \eqref{eqn:be_joint} for the remaining tests.

    For case (i), fix $0<\alpha<1$ and choose $\beta \in \bbR$ such that the inequality  in \eqref{eqn:union_char} holds. Take $\alpha_n = \alpha+\beta/\sqrt{n}$ and $M_{j,n} = \lfloor \exp( n C_j(\alpha) + \sqrt{n }L_j)\rfloor$. By Taylor expansions, 
    \begin{align}
        n C_1 (\alpha_n) &=nC_1(\alpha) + \frac{\beta\sqrt{n}}{2s_1}-O(1), \label{eqn:taylor_alpha1}\\
        n C_2 (\alpha_n) &=nC_2(\alpha) - \frac{\beta\sqrt{n}}{2s_2}+O(1). \label{eqn:taylor_alpha2}
    \end{align}
    Since $C_0(\alpha)>C_2(\alpha)$, the failure probability for decoding the  cloud center at receiver~1 tends to zero by \eqref{eqn:ach_cheby}. The finite blocklength bound in \eqref{eqn:ach_fb_bound}, together with the multivariate Berry--Esseen statement in \eqref{eqn:be_joint} and the preceding Taylor expansions yields
    \begin{align}
        \limsup_{n\to\infty}e_n\le 1-\Psi_{\alpha,\rho}\bigg( \frac{\beta}{2s_1}-L_1,-\frac{\beta}{2s_2}-L_2 \bigg)\le\eps. 
    \end{align}
    Since $e_n$ is an upper bound on the actual decoding error, this proves the achievability in case (i) of the boundary of the capacity region. 

    We now handle case (ii) in which $R^* = (0 , C_2(0)) = (0, \rvC(S_2))$. Put $\beta = 2\nu_1L_1^+$ and choose  $\alpha_n = \beta/\sqrt{n} + n^{-2/3}$, $M_{1,n}=\lfloor \exp(\sqrt{n}L_1^+ )\rfloor$ and $M_{2,n} = \lfloor\exp( n \rvC(S_2) + \sqrt{n}L_2 )\rfloor$. Taylor expansions give
    \begin{align}
        nC_1(\alpha_n) &= \sqrt{n}L_1^+ + \frac{n^{1/3}}{2\nu_1}- O(1),\\
        nC_2(\alpha_n) &= n \rvC(S_2)- \frac{\beta\sqrt{n}}{2\nu_2} - \frac{n^{1/3}}{2\nu_2} +O(1).
    \end{align}
    The test at receiver 1 involving $\iota_1$ has a mean gap of $\Theta(n^{1/3})$ and variance $O(\sqrt n)$, so its failure probability tends to zero by \eqref{eqn:ach_cheby}. The same holds for the test involving $\iota_0$, since $C_0(\alpha_n)-C_2(0)\to\rvC(S_1)-\rvC(S_2)>0$.  Since $V_2(\alpha_n)\to \rvV(S_2)$, the finite blocklength bound in \eqref{eqn:ach_fb_bound}, together with the scalar Berry--Esseen statement in \eqref{eqn:be_marg} and the preceding Taylor expansions yields
     \begin{align}
         \limsup_{n\to\infty}e_n\le\Phi\bigg(\frac{L_2+\beta/(2\nu_2)}{\sqrt{\rvV(S_2)}} \bigg)\le\eps.
     \end{align}
     This proves case (ii) in the achievability statement in \eqref{eqn:alpha0}. 

     The argument for case (iii) is analogous. We simply put $\beta=2T L_2^+$ and
$\alpha_n=1-\beta/\sqrt n-n^{-2/3}$, and choose  $M_{1,n}= \lfloor \exp (n\rvC(S_1)+\sqrt nL_1 )\rfloor$, and  $M_{2,n}=\lfloor\exp(\sqrt nL_2^+)\rfloor$. Both tests for decoding the cloud center have vanishing failure probabilities by \eqref{eqn:ach_cheby}. The same Taylor expansion and
scalar Berry--Esseen argument give
\begin{align}
\limsup_{n\to\infty}e_n \le \Phi\bigg(\frac{L_1+\beta/(  2(1+\nu_1))}{\sqrt{\rvV(S_1)}} \bigg)\le\eps,
\end{align}
as required by \eqref{eqn:alpha1}.
\end{proof}

\subsection{Conversion to Maximal Error}
\begin{lemma}[Expurgation with cloud center-dependent relabeling] \label{lem:expurgate} Let an $(n,M_{1,n},M_{2,n})$-code have augmented average error probability at most $\eps+\xi/2$, where $\xi\in(0,1-\eps)$. Put $\theta=\xi/[2(\eps+\xi)]$. By selecting cloud centers and satellite codewords and relabeling the retained satellites separately within each selected center, we obtain a code whose message sizes and maximal error probability satisfy
\begin{align}
    M_{j,n}' = \left\lceil\frac{\theta M_{j,n}}{2} \right\rceil, \quad j = 1,2,\quad\mbox{and}
    \quad\eps_{n,\mathrm{max}}'\le\eps+\xi. \label{eqn:max_error}
\end{align}
    The subcode satisfies the original power constraint.
\end{lemma}
\begin{proof}
    Call a message pair  $(m_1, m_2)$ ``good'' if its conditional augmented error probability $e_n(m_1, m_2)$ is at most $\eps+\xi$. Markov's inequality shows that the fraction of good pairs is at least $\theta$. At least $\theta M_{2,n}/2$ cloud centers each contain at least $\theta M_{1,n}/2$ satellites that form good pairs. 

    Retain $\lceil \theta M_{2,n}/2 \rceil$ of these cloud centers and $\lceil \theta M_{1,n}/2 \rceil$   satellites such that all retained pairs are good. Relabel the centers, and relabel the retained satellites separately within each center. Receiver 1 uses its original estimate of $W_2$ to choose the appropriate relabeling of its estimate of $W_1$; receiver 2 relabels its original estimate of $W_2$. An estimate outside a retained set is mapped to an arbitrary valid new label. Whenever all original augmented estimates are correct, both new message estimates are correct. This proves the error bound in \eqref{eqn:max_error}. The selected codewords are unchanged, so the power constraint is preserved.
\end{proof}

\subsection{Completion of the Achievability Proof of Theorem \ref{thm:main}} 
Proposition \ref{prop:ach_aug} proves the average error achievability for \eqref{eqn:alpha_mid}--\eqref{eqn:alpha1}. For the maximal error case, write $e_n\le\eps+r_n$ where $r_n\ge 0$ tends to zero and choose $\xi_n=\max\{\sqrt{r_n}, 1/n\}$. Eventually, $r_n\le \xi_n/2$ and $\xi_n<1-\eps$. Apply Lemma \ref{lem:expurgate} with $\theta_n = \xi_n/[2(\eps+\xi_n)]$. Since $\theta_n\ge 1/(2n)$ for all sufficiently large $n$, \eqref{eqn:max_error} reduces each log message size by no more than $O(\log n)$. Since $\xi_n\to0$, $\limsup_{n\to\infty}\eps_{n,\mathrm{max}}'\le\eps$. Thus, every pair $(L_1, L_2)$ in~\eqref{eqn:alpha_mid}--\eqref{eqn:alpha1}  is also achievable under maximal error.

\section{Converse Proof of Theorem \ref{thm:main}} \label{sec:conv_prf}
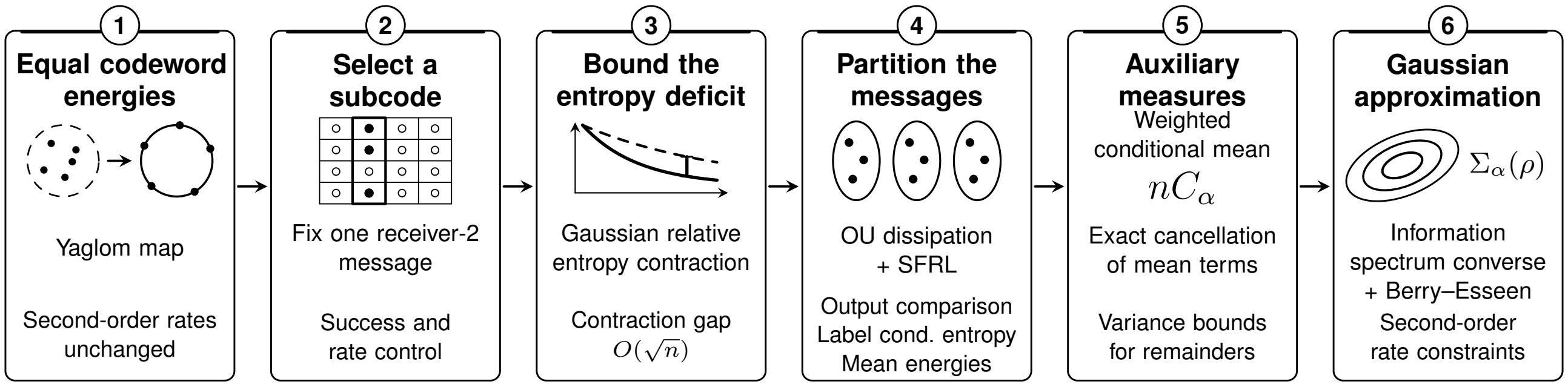
\begin{figure}
    \centering
\begingroup%
\begin{tikzpicture}[
  x=1cm,y=1cm,
  font=\sffamily\fontsize{8}{10}\selectfont,
  line cap=round,line join=round,
  bcmap heading/.style={align=center,text width=2.27cm,
    font=\sffamily\bfseries\fontsize{8.5}{10}\selectfont,
    inner sep=0pt},
  bcmap method/.style={align=center,text width=2.36cm,
    font=\sffamily\fontsize{7.2}{9}\selectfont,inner sep=0pt},
  bcmap result/.style={align=center,text width=2.36cm,
    font=\sffamily\fontsize{7}{9}\selectfont,inner sep=0pt},
  bcmap arrow/.style={->,>=stealth,draw=black,line width=.7pt}
]
  \def\BCMapCard#1#2#3{%
    \begin{scope}[shift={(#1,0)}]
      \draw[rounded corners=3pt,draw=black,fill=none,line width=.55pt]
        (0,0) rectangle (2.5,3.82);
      \draw[black,line width=1.1pt] (.18,3.81)--(2.32,3.81);
      \node[circle,draw=black,fill=white,text=black,line width=.55pt,minimum size=12pt,inner sep=0pt,
        font=\sffamily\bfseries\fontsize{7.5}{8}\selectfont]
        at (1.25,3.88) {#2};
      \node[bcmap heading,text=black,anchor=north] at (1.25,3.58) {#3};
    \end{scope}%
  }
  \BCMapCard{0}{1}{\mbox{Equal codeword}\\energies}
  \BCMapCard{2.9}{2}{Select a\\subcode}
  \BCMapCard{5.8}{3}{Bound the\\entropy deficit}
  \BCMapCard{8.7}{4}{Partition the\\messages}
  \BCMapCard{11.6}{5}{Auxiliary\\measures}
  \BCMapCard{14.5}{6}{Gaussian\\approximation}

  \foreach \a/\b in {2.54/2.86,5.44/5.76,8.34/8.66,11.24/11.56,14.14/14.46}
    \draw[bcmap arrow] (\a,2.12)--(\b,2.12);

  \begin{scope}[shift={(1.25,2.40)}]
    \draw[black,dashed,fill=none,line width=.5pt]
      (-.62,0) circle[radius=.39];
    \foreach \x/\y in {-.13/.19,.14/.12,-.21/-.10,.02/-.18,.10/-.01}
      \fill[black] ({-.62+\x},\y) circle[radius=1.15pt];
    \draw[bcmap arrow] (-.13,0)--(.13,0);
    \draw[black,fill=none,line width=.65pt]
      (.62,0) circle[radius=.39];
    \foreach \a in {20,85,155,225,300}
      \fill[black] ({.62+.39*cos(\a)},{.39*sin(\a)})
        circle[radius=1.25pt];
  \end{scope}
  \node[bcmap method,text=black] at (1.25,1.43) {Yaglom map};
  \node[bcmap result,text=black] at (1.25,.48)
    {Second-order rates\\unchanged};

  \begin{scope}[shift={(4.15,2.40)}]
    \foreach \x in {-.72,-.36,0,.36,.72}
      \draw[black,line width=.35pt] (\x,-.47)--(\x,.47);
    \foreach \y in {-.47,-.235,0,.235,.47}
      \draw[black,line width=.35pt] (-.72,\y)--(.72,\y);
    \foreach \x in {-.54,-.18,.18,.54}
      \foreach \y in {-.3525,-.1175,.1175,.3525}
        \draw[black,fill=white,line width=.35pt]
          (\x,\y) circle[radius=1.15pt];
    \foreach \y in {-.3525,.1175,.3525}
      \fill[black] (-.18,\y) circle[radius=1.45pt];
    \draw[black,line width=.8pt] (-.36,-.47) rectangle (0,.47);
  \end{scope}
  \node[bcmap method,text=black] at (4.15,1.43)
    {Fix one receiver-2\\message};
  \node[bcmap result,text=black] at (4.15,.48)
    {Success and\\rate control};

  \begin{scope}[shift={(7.05,2.40)}]
    \draw[->,>=stealth,black,line width=.4pt] (-.83,-.35)--(.84,-.35);
    \draw[->,>=stealth,black,line width=.4pt] (-.83,-.35)--(-.83,.42);
    \draw[black,dashed,line width=.8pt]
      plot[domain=-.75:.72,samples=41,variable=\t]
      ({\t},{.66*exp(-.65*(\t+.75))-.27});
    \draw[black,line width=1pt]
      plot[domain=-.75:.72,samples=41,variable=\t]
      ({\t},{.66*exp(-1.65*(\t+.75))-.27});
    \draw[black,line width=1pt]
      (.39,{.66*exp(-.65*1.14)-.27})--
      (.39,{.66*exp(-1.65*1.14)-.27});
    \foreach \z in {.65,1.65}
      \draw[black,line width=.6pt]
        (.33,{.66*exp(-\z*1.14)-.27})--(.45,{.66*exp(-\z*1.14)-.27});
  \end{scope}
  \node[bcmap method,text=black] at (7.05,1.43)
    {Gaussian relative\\entropy contraction};
  \node[bcmap result,text=black] at (7.05,.48)
    {Contraction gap\\$O(\sqrt n)$};

  \begin{scope}[shift={(9.95,2.40)}]
    \foreach \x in {-.66,0,.66}{
      \draw[black,fill=none,line width=.6pt]
        (\x,0) ellipse[x radius=.255,y radius=.43];
      \fill[black] ({\x-.08},.20) circle[radius=1.35pt];
      \fill[black] ({\x+.08},.01) circle[radius=1.35pt];
      \fill[black] ({\x-.04},-.20) circle[radius=1.35pt];
    }
  \end{scope}
  \node[bcmap method,text=black] at (9.95,1.43)
    {OU dissipation\\+ SFRL};
  \node[bcmap result,text=black] at (9.95,.48)
    {Output comparison\\Label cond.\ entropy\\Mean energies};

  \node[bcmap method,text=black] at (12.85,2.69)
    {Weighted\\conditional mean};
  \node[text=black,font=\fontsize{12}{14}\selectfont]
    at (12.85,2.10) {$nC_\alpha$};
  \node[bcmap method,text=black] at (12.85,1.43)
    {Exact cancellation\\of mean terms};
  \node[bcmap result,text=black] at (12.85,.48)
    {Variance bounds\\for remainders};

  \begin{scope}[shift={(15.25,2.35)},rotate=22,scale=.85]
    \draw[black,fill=none,line width=.55pt]
      (0,0) ellipse[x radius=.76,y radius=.39];
    \draw[black,fill=none,line width=.65pt]
      (0,0) ellipse[x radius=.53,y radius=.27];
    \draw[black,fill=none,line width=.75pt]
      (0,0) ellipse[x radius=.27,y radius=.14];
  \end{scope}
  \node[text=black,inner sep=1.5pt,
    font=\fontsize{9}{11}\selectfont] at (16.40,2.35)
    {$\Sigma_\alpha(\rho)$};
  \node[bcmap method,text=black] at (15.75,1.28)
    {Information\\spectrum converse\\+ Berry--Esseen};
  \node[bcmap result,text=black] at (15.75,.48)
    {Second-order\\rate constraints};

  \node[anchor=north,align=center,text=black,inner sep=0pt,
    font=\sffamily\fontsize{7}{9}\selectfont] at (8.5,-.25)
    {OU: Ornstein--Uhlenbeck.\quad
     SFRL: strong functional representation lemma.};
\end{tikzpicture}%
\endgroup
    \caption{Roadmap of the converse proof. Each panel above corresponds to a subsection in Section \ref{sec:conv_prf}. }
    \label{fig:roadmap}
\end{figure}

We first outline the converse proof before diving into details; see a roadmap in Fig.~\ref{fig:roadmap}. We consider an arbitrary sequence of Gaussian broadcast codes at a fixed boundary point, with second-order coding rates $(L_1,L_2)$ and asymptotic average error probability at most $\eps$. After reducing the code to one with equal codeword energies in Subsection~\ref{sec:yaglom}, in Subsection~\ref{sec:reduction} we fix a receiver-2 message $k$ and retain a subset of receiver-1 messages $\calI$. This selection preserves the required success probability up to a small slack. It also relates the original weighted sum of logarithmic message sizes to the number of retained messages and the Gaussian probability of their common receiver-2 decoding set, with only an $O(1)$ loss.

Subsection \ref{sec:ent_defi} bounds the entropy deficit, the gap in Gaussian relative entropy contraction, by \(O(\sqrt n)\).  The most important and novel contribution lies in Subsection~\ref{sec:partition}. It  combines this bound with Ornstein–Uhlenbeck entropy dissipation, Gaussian functional inequalities~\cite{BakryGentilLedoux2014}, and the SFRL~\cite{LiElGamal2018} to partition the retained message set. The component label is a deterministic function of the retained receiver-1 message and plays a role analogous to an auxiliary by specifying the conditional input distributions. The partition has three properties:
(i) The component output distributions have sublinear relative entropy, averaged over the components, from Gaussian distributions with the component means and prescribed covariance matrices. (ii) The component label has conditional entropy $o(\sqrt n)$ given the receiver-2 output. (iii) The normalized squared norms of the component means are close, on average, to the power fraction assigned to receiver~2 in the superposition construction in the achievability proof.

Subsection~\ref{sec:aux} constructs novel auxiliary output measures using the component distributions in the partition. Their choice ensures that the codeword-dependent terms cancel exactly in the weighted sum of the conditional means of the component log-likelihood ratios, leaving the constant $nC_\alpha$. Property~(i) provides variance bounds for the remaining centered terms and controls the deviations of the relevant norms and inner products from their values in the spherical superposition construction  used in the achievability proof. Together with the mean-energy bounds in property~(iii), these estimates provide the conditional covariance control needed for the Gaussian approximation.

Subsection~\ref{sec:combining} combines these estimates in a joint information-spectrum converse bound. Property~(ii) controls the posterior probability term introduced when passing from component output distributions to the output distribution of the entire selected subcode. Chebyshev's and Markov's inequalities control the remainder terms, while the covariance estimates obtained from properties~(i) and~(iii), together with Berry--Esseen bounds, yield the Gaussian approximation with dispersion matrix $\Sigma_\alpha(\rho)$. 

\subsection{Yaglom Map Reduction to Equal-Energy Codewords} \label{sec:yaglom}
The following dimension extension permits the equal-energy assumption throughout the converse; this follows the usual Gaussian channel reduction~\cite{PPV10, Tan14}.
\begin{lemma}[Reduction to equal-energy codewords] Every length-$n$ code satisfying  \eqref{eqn:pow_constr} induces a length-$N$ code, with $N=n+1$, with the same message sizes and every conditional error probability unchanged such that all codewords have squared norm $N$. For sequences of codes, the first- and second-order coding rates in \eqref{eqn:second-order_def} are unchanged. 
\end{lemma}
\begin{proof}
    For each message pair $(m_1, m_2)$ with corresponding codeword $x^n(m_1, m_2)$, define 
    \begin{align}
        \tilx^N(m_1, m_2 ) := \big(   x^n(m_1, m_2 ) , \sqrt{N-\|x^n(m_1, m_2 )\|^2  }\big).
    \end{align}
    Its squared norm is exactly $N$. Use the same Gaussian broadcast channel for the additional channel use, with a fresh noise pair having correlation $\rho$ and independent of the preceding noise pairs. Define the new decoders by $\tilde\varphi_{j,N}(y^N) :=\varphi_{j,n}(y_1,\ldots,y_n)$, $j=1,2$.  Both decoders ignore the new coordinate and thus the new joint error event coincides with the original event, conditional on every message pair. Since the blocklength only increases by~$1$, the first- and second-order coding rates are unaffected. 
\end{proof}
Henceforth, relabel the new blocklength $N$ as $n$ and suppress the tildes. Thus, we may assume that 
\begin{align}
    \|x^n(m_1, m_2 )\|^2=n\qquad\forall\, (m_1, m_2)\in [M_{1,n}]\times [M_{2,n}].
\end{align}
\subsection{Reduction from an Arbitrary Code} \label{sec:reduction}
Fix $\alpha\in [0,1]$ and $\eps\in (0,1)$. Consider any sequence of $(n,M_{1,n}, M_{2,n})_{\mathrm{av}}$-codes at $R^* = (C_1(\alpha), C_2(\alpha))$ with second-order coding rates $(L_1, L_2)$. The codebook is $\{x^n(m_1, m_2)\}$ and every codeword satisfies $\|x^n(m_1,m_2)\|^2 = n$. Write the decoding regions as 
\begin{align}
    \calA_{m_1} := \varphi_{1,n}^{-1}(m_1)\quad\mbox{and}\quad \calB_{m_2}:= \varphi_{2,n}^{-1}(m_2).
\end{align}
Let $\rvW_{1,2}^n(\fndot|x^n)$ be the joint channel distribution. For a transmitted pair of messages $(m_1, m_2)$, let the conditional probability of successful decoding be 
\begin{align}
    s(m_1, m_2):= \rvW_{1,2}^n( \calA_{m_1}\times \calB_{m_2}| x^n(m_1, m_2)).
\end{align}
Then the probability of successful decoding is
\begin{align}
    1-\eps_{n,\mathrm{av}}= \frac{1}{M_{1,n}M_{2,n}}\sum_{m_1, m_2}s(m_1, m_2).
\end{align}
Fix $\xi,\zeta>0$ with $\eps+\zeta<1$. For all sufficiently large $n$, 
\begin{align}
    \log M_{j,n}\ge n C_j(\alpha)+\sqrt{n} (L_j-\xi), \quad j = 1,2, \quad\mbox{and}\quad \eps_{n,\mathrm{av}}\le\eps+\zeta/2.
\end{align}
In particular,
\begin{align}
    \log M_{1,n}+\mu_\alpha\log M_{2,n}\ge n C_\alpha+\sqrt{n} \big(L_1 + \mu_\alpha L_2 -(1+\mu_\alpha)\xi \big). \label{eqn:rate_assump}
\end{align}

\begin{figure}
    \centering
\begingroup
\definecolor{BCselect}{RGB}{0,102,116}
\begin{tikzpicture}[
  x=1.8cm,y=1cm,font=\small,
  line cap=round,line join=round,
  bc text/.style={inner sep=2pt,align=center},
  bc code/.style={draw=BCselect!55,fill=BCselect!5,
    rounded corners=2pt,minimum width=4.4cm,minimum height=.43cm}
]
  \begin{scope}[xshift=.7065cm]
    \fill[BCselect!8] (1.90,1.25) rectangle (2.35,4.26);

    \foreach \c in {0,...,7}
      \draw[black!45,line width=.4pt]
        ({.55+.45*\c},1.25)--({.55+.45*\c},4.26);
    \foreach \r in {0,...,7}
      \draw[black!45,line width=.4pt]
        (.55,{1.25+.43*\r})--(3.70,{1.25+.43*\r});

    \foreach \c in {1,...,7}
      \foreach \r in {1,...,7}
        \draw[black!30,fill=white,line width=.35pt]
          ({.55+.45*(\c-.5)},{4.26-.43*(\r-.5)})
          circle[radius=1.7pt];

    \foreach \c/\r in {
      1/1,1/4,1/6,2/2,2/5,3/1,3/3,3/6,
      4/1,4/2,4/4,4/5,4/7,
      5/2,5/4,6/1,6/5,6/7,7/3,7/6}
      \fill[BCselect]
        ({.55+.45*(\c-.5)},{4.26-.43*(\r-.5)})
        circle[radius=1.9pt];

    \draw[BCselect,line width=.9pt]
      (1.90,1.25) rectangle (2.35,4.26);

    \node[bc text] at (2.125,4.57)
      {$m_2\in[M_{2,n}]$};
    \node[bc text,rotate=90] at (.16,2.755)
      {$m_1\in[M_{1,n}]$};
    \node[bc text,text=BCselect] at (2.125,1.03)
      {$k$};

    \node[bc text] (bcselLegend) at (2.125,.40)
      {$s(m_1,m_2)\ge1-\varepsilon-\zeta$};
    \fill[BCselect] ([xshift=-5pt]bcselLegend.west)
      circle[radius=1.9pt];
  \end{scope}

  \draw[->,>=stealth,line width=.8pt,BCselect]
    (4.4225,2.755)--(5.2075,2.755);

  \begin{scope}[xshift=-.7065cm]
    \node[bc text] at (7.18,4.57)
      {$\mathcal I=\{i_1,\ldots,i_M\}$};
    \node[bc code] at (7.18,4.00)
      {$x^n(i_1,k)$};
    \node[bc code] at (7.18,3.43)
      {$x^n(i_2,k)$};
    \node[bc text] at (7.18,2.89)
      {$\vdots$};
    \node[bc code] at (7.18,2.32)
      {$x^n(i_M,k)$};

    \node[bc text] at (7.18,1.43)
      {$\displaystyle M=|\mathcal I|
        \ge\frac{\theta M_{1,n}}2$};
    \node[bc text] at (7.18,.40)
      {$\displaystyle 0<q=Q_0(B_k)
        \le\frac{2}{\theta M_{2,n}}$};
  \end{scope}
\end{tikzpicture}%
\endgroup
    \caption{Illustration of the proof of Lemma \ref{lem:sel}}
    \label{fig:sel}
\end{figure}
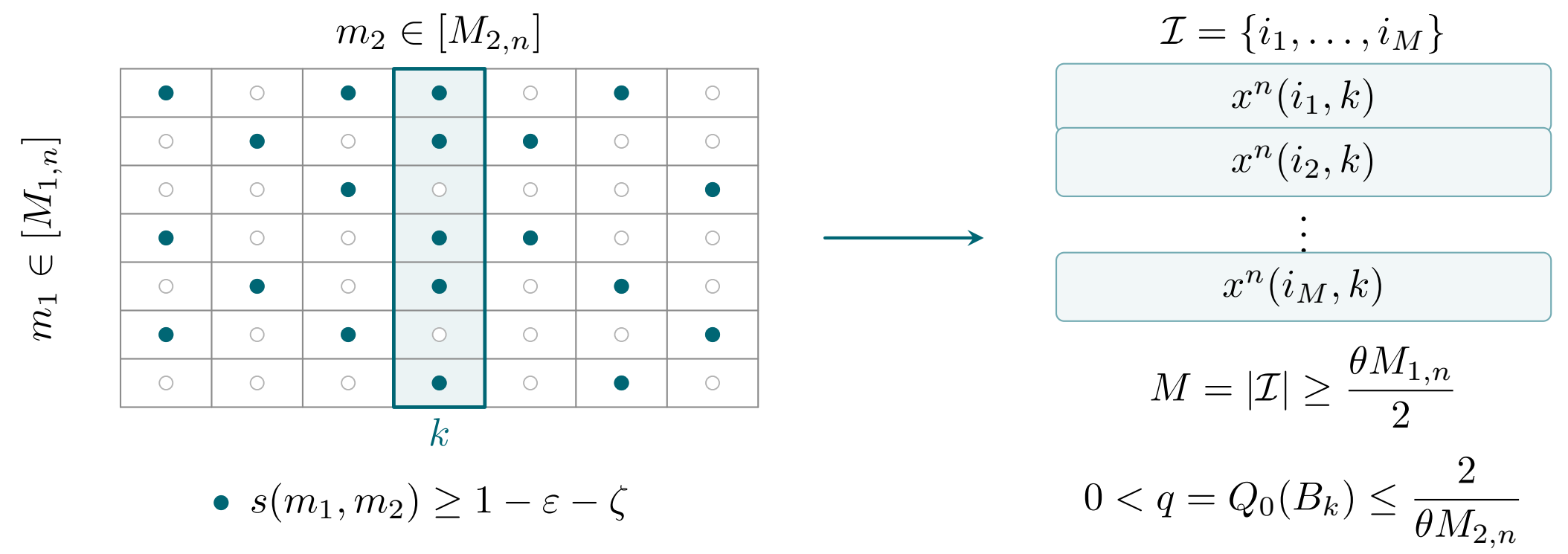

\begin{lemma}[Selection preserves success probability] \label{lem:sel}
Suppose that a code has average probability of successful decoding at least $1-\eps-\zeta/2$, where $\eps,\zeta>0$ and $\eps+\zeta<1$. Set $\theta = \zeta/[2(\eps+\zeta)]$. There is a message index $k \in [M_{2,n}]$ and a nonempty set $\calI \subset [M_{1,n}]$ such that, with $M= |\calI|$, $\calB=\calB_k$ and $Q_0 = \calN(0, TI_n)$,
    \begin{align}
        s(m_1, k) &\ge 1-\eps-\zeta \qquad\forall\, m_1 \in \calI ,\label{eqn:sel1}\\
        M &\ge \frac{\theta}{2}M_{1,n} ,\label{eqn:sel2}\\
        0< q&:=Q_0(\calB)\le \frac{2}{\theta M_{2,n}}. \label{eqn:sel3}
    \end{align}
    Using the original decoding regions, the average joint success probability $\frakP_{\mathrm{succ}}$ over the retained pairs satisfies
    \begin{align}
         \frakP_{\mathrm{succ}} =\frac{1}{M}\sum_{m_1 \in\calI}s(m_1, k)\ge 1-\eps-\zeta. \label{eqn:success_sel}
    \end{align}
    Consequently, \begin{align}
    \Xi:=\log M-\mu_\alpha\log q \ge \log M_{1,n}+\mu_\alpha\log M_{2,n}
    -(1+\mu_\alpha)\log\frac{2}{\theta}.
    \label{eqn:sel_final}
\end{align}
\end{lemma}

\begin{remark} \label{rmk:Xi}
The quantity $\Xi$ controls the weighted logarithmic sizes of the original message sets: by \eqref{eqn:sel_final}, an upper bound on $\Xi$ gives an upper bound on $\log M_{1,n}+\mu_\alpha\log M_{2,n}$ with only a constant loss, which is negligible at the second-order scale.
\end{remark}

The following proof is illustrated in Fig.~\ref{fig:sel}.
\begin{proof}[Proof of Lemma~\ref{lem:sel}]
    Let $\calS$  be the set of message pairs $(m_1, m_2)$ with $s(m_1,m_2)\ge 1-\eps-\zeta$ and put $p = |\calS|/(M_{1,n}M_{2,n})$. Since $1-\eps-\zeta/2\le p+(1-p)(1-\eps-\zeta)$, we have $p\ge\theta$. The decoding regions $\calB_k$ are disjoint so $\sum_k Q_0(\calB_k)\le 1$. Fewer than $\theta M_{2,n}/2$ receiver-2 messages $k$ have $Q_0(\calB_k)> 2/(\theta M_{2,n})$; these indices account for fewer than $\theta M_{1,n} M_{2,n}/2$ message pairs. Among the remaining receiver-2 message indices, one $k$ therefore has at least $\theta M_{1,n}/2$ pairs in $\calS$. Retain all these message 1 indices as $\calI\subset [M_{1,n}]$. This, together with the fact that  the marginal Gaussian densities are positive everywhere, proves \eqref{eqn:sel1}--\eqref{eqn:success_sel}.  Taking logs of the two size bounds in \eqref{eqn:sel2} and \eqref{eqn:sel3} gives~\eqref{eqn:sel_final}.
\end{proof}

We will apply this lemma for sufficiently large blocklengths. Put $e:=\eps+\zeta$. Since joint decoding success implies marginal decoding success, the selected subcode satisfies
\begin{align}
    \rvW_1^n(\calA_{m_1}|x^n(m_1, k))\ge 1-e\quad\mbox{and}\quad \rvW_2^n(\calB |x^n(m_1, k))\ge 1-e \qquad \forall\, m_1\in\calI. \label{eqn:images}
\end{align}
In other words, the decoding region $\calB$ is a common $(1-e)$-image \cite[Ch.~15]{Csi97} of the selected codewords $\{ x^n(m_1, k) : m_1\in\calI\}$. Additionally, the rate assumptions in \eqref{eqn:rate_assump} and the bound on $\Xi$  in~\eqref{eqn:sel_final} give 
\begin{align}
    \Xi &\ge nC_\alpha+\sqrt n\bigl(
        L_1+\mu_\alpha L_2-(1+\mu_\alpha)\xi
        \bigr)-O(1)
    \ge nC_\alpha-\kappa\sqrt n \label{eqn:lower_bd_Xi}
\end{align}
for some constant $0<\kappa<\infty$ independent of $n$.
Let $\Lambda_{1\rightarrow 2}$ denote the convolution operator with $\calN(0, (\nu_2-\nu_1)I_n)$, so  for bounded measurable $h$, define
\begin{align}
    (\Lambda_{1\rightarrow2}h)(y^n)
    :=\bbE\big[h(y^n+G^n)\big],
    \quad\mbox{where}\quad
    G^n\sim\calN\bigl(0,(\nu_2-\nu_1)I_n\bigr).
\end{align} 
Let $W$ be uniform on the extracted receiver~1 message set $\calI$ and put $X^n = x^n(W,k)$. Define the marginal output distributions
\begin{align}
    P_j := \frac{1}{M}\sum_{m_1\in\calI}\rvW_j^n(\fndot| x^n(m_1, k)) \qquad j = 1,2.  \label{eqn:marg_output}
\end{align}
With these definitions, $P_1 \Lambda_{1\rightarrow 2} = P_2$ is an identity between the two marginal output distributions.  

\subsection{Entropy Deficit} \label{sec:ent_defi}

We next show that the lower bound $\Xi\ge nC_\alpha-\kappa\sqrt n$ from \eqref{eqn:lower_bd_Xi} forces the entropy deficit (defined in \eqref{eqn:defDelta} below) to be $O(\sqrt n)$. The entropy deficit bound supplies the condition for the construction of the partition  in the next subsection. For this preliminary bound, it suffices to apply
change-of-measure and concentration arguments separately at each receiver, allowing additive $O(\sqrt n)$ errors. The sharp second-order term will be obtained later by analyzing the joint fluctuations at the two receivers.

\begin{lemma}[Bounded entropy deficit] \label{lem:defi} Fix $\tau\in(0,1/2)$. For a subcode satisfying~\eqref{eqn:images} with $e\in[0,1-\tau]$, define the {\em entropy deficit} by
\begin{align}
    \Delta:= D\big(P_1 \| \calN(0, s_1 I_n) \big) -\mu_\alpha D\big(P_2 \| \calN(0, s_2 I_n) \big) .  \label{eqn:defDelta} 
\end{align}
Then $\Delta \ge0$ and 
\begin{align}
    I(X^n; Y_1^n) + \mu_\alpha D(P_2 \| Q_0) = nC_\alpha-\Delta. \label{eqn:mi_plus_div}
\end{align}
Furthermore, under \eqref{eqn:lower_bd_Xi},
\begin{align}
    \Delta=O(\sqrt n). \label{eqn:ent_unused_power}
\end{align}
The implicit constant depends only on $\alpha$, $\nu_1$, $\nu_2$, $\tau$, and $\kappa$, and is uniform over $k$ and $\calI$.
\end{lemma}
\begin{proof}
    Let \(p_j\) denote the density of \(P_j\) (defined  in \eqref{eqn:marg_output}), and let \(q_0\) denote the density of $Q_0 = \calN(0, TI_n)$ defined in Lemma~\ref{lem:sel}. Define the log-likelihood ratios
    \begin{align}
        L_{1, m_1 }(y^n) := \log\frac{\rvW_1^n(y^n| x^n(m_1, k) )}{p_1(y^n)}\quad\mbox{and}\quad L_{2, m_1}(y^n) :=\log\frac{p_2(y^n)}{q_0(y^n)}.
    \end{align}
    The Gaussian score identity (e.g., \cite[Eqn.~(3)]{DytsoPoorShamai2023}) is 
    \begin{align}
        \nabla_{y^n} \log p_j (y^n) = \frac{\hatx_j^n(y^n)-y^n}{\nu_j} \quad\mbox{where}\quad \hatx_j^n(y^n) := \bbE[X^n|Y_j^n=y^n].
    \end{align}
    Since $\|\hatx_j^n(y^n)\|\le\sqrt{n}$, $\nabla_{y^n}L_{1,m_1}=(x^n(m_1,k)-\hatx_1^n(y^n))/\nu_1$ and  $\nabla_{y^n}L_{2,m_1}=\hatx_2^n(y^n)/\nu_2 - y^n/(\nu_2 T)$ have conditional mean squared norms of order $O(n)$. The conditional Gaussian Poincar\'e inequality \cite{BLM2013} then gives 
    \begin{align}
        \mathrm{Var}\big( L_{j,m_1}(Y_j^n)\mid W=m_1\big)= O(n) ,\qquad j=1,2. \label{eqn:unif_cond_var}
    \end{align}
    Let $\ell_{j,m_1}:=\bbE[L_{j,m_1}(Y_j^n)\mid W=m_1]$, $\delta:=\tau/2$,  and $\psi_{j,m_1}:=\ell_{j,m_1}+c\sqrt n$. By the uniform conditional variance bound in \eqref{eqn:unif_cond_var}, we may choose $c>0$, independently of $n$ and $m_1$, such that Chebyshev's inequality gives
    \begin{align}
    \Pr\bigl(
        L_{j,m_1}(Y_j^n)>\psi_{j,m_1}
        \mid W=m_1
    \bigr)\le\delta,
    \qquad j=1,2.    
    \end{align}
    From \eqref{eqn:images},  the conditional decoding success probability for each receiver is at least $1-e$. Hence, the intersection of its decoding success event with the event $\{L_{j,m_1}(Y_j^n)\le\psi_{j,m_1}\}$ has conditional probability at least $1-e-\delta\ge\tau-\delta=\delta$,  where we used $e\le1-\tau$.

For receiver~1, change of measure therefore gives
\begin{align}
    P_1(\calA_{m_1})
    &=\int_{\calA_{m_1}} \rme^{-L_{1,m_1}(y^n)} \, \rvW_1^n(\rmd y^n\mid x^n(m_1,k))\\
    &\ge \rme^{-\psi_{1,m_1}}\Pr\bigl(
            Y_1^n\in\calA_{m_1},\,
            L_{1,m_1}(Y_1^n)\le\psi_{1,m_1}\mid W=m_1 \bigr)\\
    &\ge\delta \rme^{-\psi_{1,m_1}}.
\end{align}
Since the regions $\calA_{m_1}$ are disjoint and $W$ is
uniform on $\calI$, summing over $m_1$ and applying
Jensen's inequality yields
\begin{align}
    1\ge\sum_{m_1\in\calI}P_1(\calA_{m_1})\ge\delta M\bbE \big[\rme^{-\psi_{1,W} } \big]\ge\delta M \rme^{-\bbE[\psi_{1,W}]}  . \label{eqn:com1}
\end{align}

For receiver~2, note that $\rme^{-L_{2,m_1}(y^n)}=q_0(y^n)/p_2(y^n)$ does not
depend on $m_1$. Using the mixture representation of $P_2$ and the same intersection bound, we obtain
\begin{align}
    q &=\int_{\calB}\frac{q_0(y^n)}{p_2(y^n)}\,P_2(\rmd y^n)\\
    &=\frac1M\sum_{m_1\in\calI}   \int_{\calB}\rme^{-L_{2,m_1}(y^n)}
        \, \rvW_2^n(\rmd y^n\mid x^n(m_1,k))\\
    &\ge\frac{\delta}{M}\sum_{m_1\in\calI}
        \rme^{-\psi_{2,m_1}}\\
    &=\delta\bbE \big[\rme^{-\psi_{2,W}} \big]
    \ge\delta \rme^{ -\bbE[\psi_{2,W}]}, \label{eqn:com2}
\end{align}
where the last step again follows from Jensen's inequality.

Taking logarithms of~\eqref{eqn:com1} and~\eqref{eqn:com2} and noting that $\bbE[\psi_{1,W}]=I(X^n;Y_1^n)+O(\sqrt{n})$ and $\bbE[\psi_{2,W}]=D(P_2\|Q_0)+O(\sqrt{n})$, we obtain
\begin{align}
    I(X^n;Y_1^n)\ge\log M - O(\sqrt{n})\quad\mbox{and}\quad D(P_2\|Q_0)\ge-\log q- O(\sqrt{n}). \label{eqn:mi_div_bds}
\end{align}

An application of the relative entropy contraction inequality under the Ornstein--Uhlenbeck semigroup \cite[Theorem~5.2.1]{BakryGentilLedoux2014}, after rescaling, gives
\begin{align}
    D\bigl(P_2\|\calN(0,s_2I_n)\bigr) \le \frac{1}{\mu_\alpha} D\bigl(P_1\|\calN(0,s_1I_n)\bigr). \label{eqn:contraction}
\end{align} 
Since $P_2 = P_1 \Lambda_{1\rightarrow 2} $,    $\Delta\ge0$ (recall its definition in~\eqref{eqn:defDelta}).

Expanding the densities of $\calN(0,s_1I_n)$, $\calN(0,s_2I_n)$, and $Q_0=\calN(0,TI_n)$, and using $\bbE[\|X^n\|^2]=n$, gives
\begin{align}
    I(X^n;Y_1^n) &= nC_1 (\alpha) +\frac{n\bar{\alpha}}{2s_1} - D(P_1 \| \calN(0,s_1I_n)), \label{eqn:mi}\\
    D(P_2 \| Q_0) &=   nC_2 (\alpha)  -\frac{n\bar{\alpha}}{2s_2} + D(P_2 \| \calN(0,s_2I_n)) . \label{eqn:div}
\end{align}
Since $\mu_\alpha/s_2=1/s_1$, adding \eqref{eqn:mi} to $\mu_\alpha$ times \eqref{eqn:div} proves \eqref{eqn:mi_plus_div}. By \eqref{eqn:mi_div_bds},
the left-hand side of \eqref{eqn:mi_plus_div} is at least $\Xi-O(\sqrt n)$. Together with \eqref{eqn:lower_bd_Xi}, this gives \eqref{eqn:ent_unused_power}, as desired.
\end{proof}
Since $\Delta \ge0$, it follows from \eqref{eqn:ent_unused_power} that  the entropy deficit $\Delta=O(\sqrt{n})$. This is  precisely the hypothesis used in the next subsection.

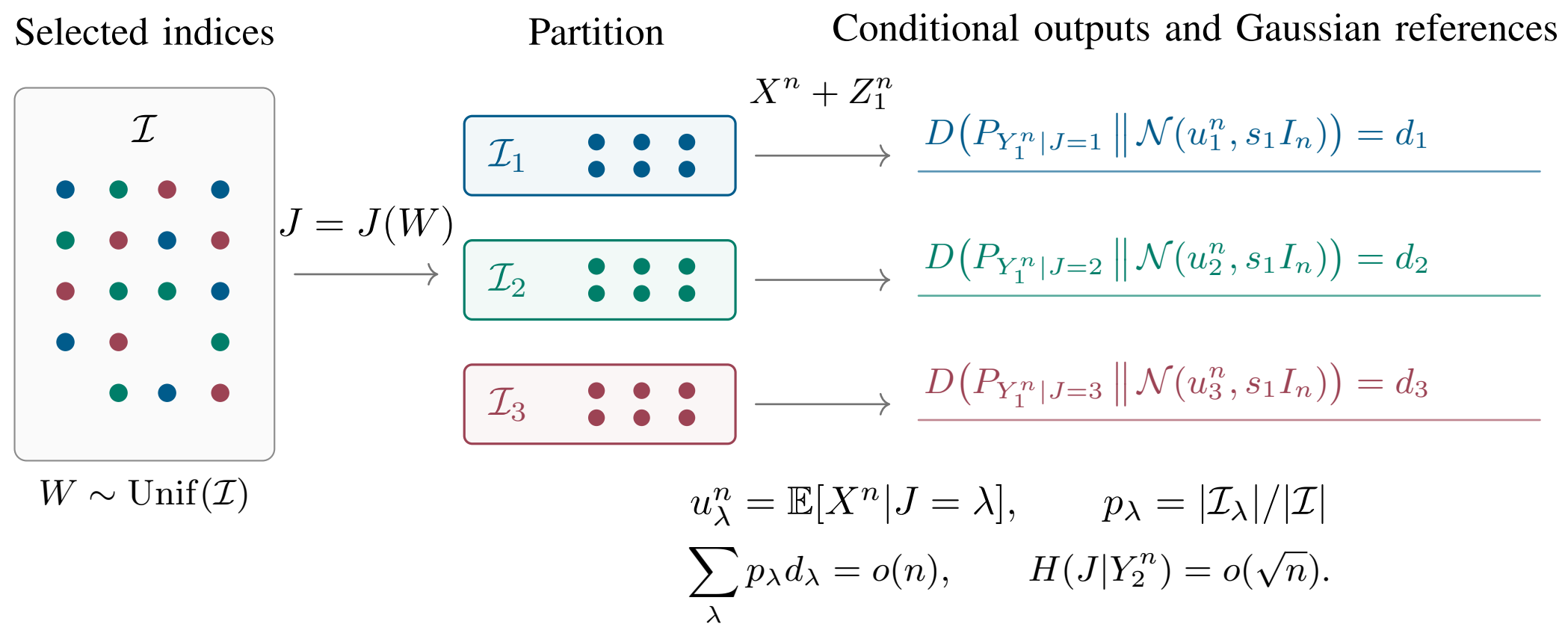
\begin{figure}
    \centering
\begingroup
\definecolor{BCpartone}{RGB}{0,91,139}
\definecolor{BCparttwo}{RGB}{0,126,105}
\definecolor{BCpartthree}{RGB}{156,67,84}
\begin{tikzpicture}[
  x=1cm,y=1cm,font=\small,
  line cap=round,line join=round,
  bc arrow/.style={->,draw=black!55,line width=.5pt},
  bc component/.style={rounded corners=2pt,line width=.6pt},
  bc label/.style={inner sep=2pt,align=center}
]
  \node[bc label] at (1.35,4.15) {Selected indices};
  \node[bc label] at (5.35,4.15) {Partition};
  \node[bc label] at (10.65,4.15)
    {Conditional outputs and Gaussian references};

  \draw[rounded corners=3pt,black!45,fill=black!2]
    (.2,.35) rectangle (2.5,3.65);
  \node[bc label] at (1.35,3.29) {$\mathcal I$};
  \foreach \xx/\yy in {.65/2.75,1.55/2.3,2.02/1.85,.65/1.4,1.55/.95,2.02/2.75}
    \fill[BCpartone] (\xx,\yy) circle[radius=2.3pt];
  \foreach \xx/\yy in {1.12/2.75,.65/2.3,1.55/1.85,2.02/1.4,1.12/.95,1.12/1.85}
    \fill[BCparttwo] (\xx,\yy) circle[radius=2.3pt];
  \foreach \xx/\yy in {1.55/2.75,2.02/2.3,.65/1.85,1.12/1.4,2.02/.95,1.12/2.3}
    \fill[BCpartthree] (\xx,\yy) circle[radius=2.3pt];
  \node[bc label,font=\footnotesize] at (1.35,.04)
    {$W\sim\mathrm{Unif}(\mathcal I)$};

  \draw[bc arrow] (2.68,2)--(3.95,2)
    node[midway,above=5pt,bc label] {$J=J(W)$};
  \foreach \yy/\col/\lab in {
    3.05/BCpartone/1,1.95/BCparttwo/2,.85/BCpartthree/3}{
    \draw[bc component,draw=\col,fill=\col!5]
      (4.18,\yy-.35) rectangle (6.58,\yy+.35);
    \node[text=\col,anchor=west,bc label] at (4.32,\yy)
      {$\mathcal I_{\lab}$};
    \foreach \xx in {5.35,5.75,6.15}
      \foreach \dy in {-.12,.12}
        \fill[\col] (\xx,\yy+\dy) circle[radius=2.1pt];
    \draw[bc arrow] (6.76,\yy)--(7.95,\yy);

    \node[anchor=west,bc label,text=\col,font=\footnotesize]
      at (8.17,\yy+.17)
      {$D\!\left(
          P_{Y_1^n\mid J=\lab}
          \,\middle\|\,
          \mathcal N(u_{\lab}^n,s_1I_n)
        \right)=d_{\lab}$};
    \draw[draw=\col!65,line width=.5pt]
      (8.2,\yy-.14)--(13.7,\yy-.14);
  }
  \node[bc label,font=\footnotesize] at (7.35,3.60)
    {$X^n+Z_1^n$};

  \node[bc label] at (9.0,-.04)
    {$u_\lambda^n=\mathbb E[X^n | J=\lambda],\qquad
      p_\lambda=|\mathcal I_\lambda|/|\mathcal I|$};

  \node[bc label,font=\footnotesize] at (9.0,-.75)
    {$\displaystyle
       \sum_\lambda p_\lambda d_\lambda = o(n),\qquad H(J|Y_2^n)= o(\sqrt{n})$.};
\end{tikzpicture}%
\endgroup
    \caption{Illustration of the partition constructed in Proposition~\ref{prop:partition}.}
    \label{fig:partition}
\end{figure}

\subsection{Partition with Approximately Gaussian Outputs} \label{sec:partition}
We continue with the selected subcode $\{x^n(m_1, k):m_1\in\calI\}$ and $X^n = x^n(W,k)$ where $W$ is uniform on $\calI$. We make the dependence of the nonnegative entropy deficit $\Delta$ on $\alpha$ and the distribution $P_{X^n}$ of $X^n$ explicit in the following, so we write
\begin{align}
    \Delta_\alpha(P_{X^n}) = D(P_1 \| \calN(0,s_1 I_n))-\mu_\alpha D(P_2 \| \calN(0,s_2 I_n)),
\end{align}
which is of order $O(\sqrt{n})$ by Lemma \ref{lem:defi}. We now use this entropy deficit bound to construct a deterministic partition of the retained message set. The following proposition controls the component output distributions, the conditional entropy of the component label given $Y_2^n$, and the energies of the component means. Fig.~\ref{fig:partition} illustrates the constructed partition and the quantities involved.


\begin{proposition}[Partition with Gaussian output approximation] \label{prop:partition}
   Fix $\alpha\in[0,1]$ and $0\le\kappa<\infty$. Suppose that
the subcode satisfies
$\Delta_\alpha(P_{X^n})\le\kappa\sqrt n$.
This hypothesis holds for a suitable finite $\kappa$
by Lemma~\ref{lem:defi}.

There is a partition of its index set
$\calI=\sqcup_\lambda\calI_\lambda$ into nonempty components
with the following properties. Let $J$ be the component
label, defined by $J=\lambda$ when $W\in\calI_\lambda$,
and write $M_\lambda=|\calI_\lambda|$,
$p_\lambda=M_\lambda/M$, and
    \begin{align}
        u_\lambda^n &:= \frac{1}{M_\lambda}\sum_{m_1\in\calI_\lambda}x^n(m_1, k) ,\\
        d_\lambda &:= D\big( P_{Y_1^n|J=\lambda}\big\|\calN(u_\lambda^n,s_1 I_n)  \big)\qquad
        \bard := \sum_{\lambda  } p_\lambda d_\lambda. \label{eqn:d_lambda}
    \end{align}
    Then the entropy of the component label given $Y_2^n$ and the component output distributions $P_{Y_1^n | J=\lambda}$ satisfy
    \begin{align}
        H(J| Y_2^n) & = O(n^{1/3}), \label{eqn:small_ce}\\
        \bard & = O(n^{5/6}). \label{eqn:sublinear_div}
    \end{align}
    The component means $u_\lambda^n$ also satisfy 
    \begin{align}
        \sum_{\lambda} p_\lambda\left|  \frac{\|u_\lambda^n\|^2}{n} - \bar{\alpha} \right|\le 2 (1+\nu_1) \sqrt{ \frac{\bard}{n}  } = O(n^{-1/12}). \label{eqn:comp_means}
    \end{align}
    All implicit constants depend only on $\alpha$, $\nu_1$, $\nu_2$, and $\kappa$, and are independent of the subcode, $n$, and $\rho\in[-1,1]$.
\end{proposition}
The bounds $H(J|Y_2^n)=o(\sqrt n)$ and $\bard=o(n)$ serve different purposes in the converse. In the proof of Proposition~\ref{prop:conv_normal}, Bayes' rule relates the component output distributions to the output distribution of the entire selected subcode. The conditional entropy bound makes the negative logarithm of the posterior probability of the true component negligible in probability at the $\sqrt n$ scale.

After averaging over the components, $\bard=o(n)$ gives $o(n)$ bounds on the conditional variances of the centered remainder terms in Lemma~\ref{lem:ll_ratio}. This refines the variance-control approach of Fong and Tan~\cite{FongTan2017GaussianBC} by separating negligible remainder terms from the principal fluctuations that determine the dispersion. Together with the bound on the component means and the geometric estimates in Lemma~\ref{lem:geom}, it also yields the covariance convergence established in Lemma~\ref{lem:ll_ratio}.

To prove this proposition, we first construct a function of the codeword that receiver 2 can estimate with small mean squared error. 

\begin{lemma}[A consequence of Ornstein--Uhlenbeck entropy dissipation] \label{lem:ou}
    Let $\Delta:=\Delta_{\alpha}(P_{X^n})$. There exists a function $F : \supp(X^n)\to\bbR^n$ and a measurable function $V : \bbR^n\to\bbR^n$ such that 
    \begin{align}
        \bbE \big[ \| F(X^n) - V(Y_2^n)\|^2 \big] & = O(\Delta), \label{eqn:FminusV}\\
        \bbE \big[ \| X^n  - F(X^n)\|^2 \big] & \le  n\alpha +O( \sqrt{n  \Delta}). \label{eqn:XminusF}
    \end{align}
    Furthermore, $\| F(x^n(m_1, k)) \|=O(\sqrt{n})$ for every $m_1\in\calI$. 
\end{lemma}
\begin{figure}
    \centering
\begingroup
\definecolor{OUblue}{RGB}{0,83,138}
\definecolor{OUteal}{RGB}{0,119,103}
\definecolor{OUorange}{RGB}{179,83,25}
\begin{tikzpicture}[
  x=1cm,y=1cm,font=\small,
  line cap=round,line join=round,
  >=Stealth,
  ou label/.style={inner sep=2pt,align=center},
  ou guide/.style={draw=black!22,densely dotted,line width=.4pt}
]
  \def\OUtau{4.0}
  \def\OUstar{7.8}
  \def\OUend{11.6}

  \node[ou label] at (5.45,2.30)
    {$\displaystyle
      \barY_t^n=\frac{X^n+Z_{\omega(t)}^n}{\sqrt{s(t)}},
      \qquad s(t)=s_1\rme^{2t},\qquad \omega(t)=s(t)-\alpha$};

  \node[ou label,text=OUblue] at (0,1.64) {Receiver 1};
  \node[ou label,text=OUblue] at (\OUtau,1.64) {Receiver 2};
  \node[ou label,text=OUorange] at (\OUstar,1.64) {Selected law};
  \node[ou label,text=OUteal] at (\OUend,1.64) {Further smoothing};
  \node[ou label] at (0,.88)
    {$\displaystyle \frac{Y_1^n}{\sqrt{s_1}}$};
  \node[ou label] at (\OUtau,.88)
    {$\displaystyle \frac{Y_2^n}{\sqrt{s_2}}$};
  \node[ou label,text=OUorange] at (\OUstar,.88)
    {$\displaystyle \frac{X^n+Z_\omega^n}{\sqrt{s}}$};
  \node[ou label] at (\OUend,.88)
    {$\displaystyle \frac{X^n+Z_{2\nu_2}^n}{\sqrt{\alpha+2\nu_2}}$};

  \fill[OUteal!13,rounded corners=3pt]
    (\OUtau,-.17) rectangle (\OUend,.17);
  \draw[->,black!65,line width=.65pt] (-.35,0)--(12.45,0)
    node[right,ou label] {$t$};
  \draw[OUblue,line width=1.4pt] (0,0)--(\OUtau,0);
  \draw[OUteal,line width=1.4pt] (\OUtau,0)--(\OUend,0);

  \foreach \x in {0,\OUtau,\OUend}{
    \draw[ou guide] (\x,.37)--(\x,.12);
    \draw[ou guide] (\x,-.68)--(\x,-2.08);
  }
  \draw[OUorange!60,densely dotted,line width=.5pt]
    (\OUstar,.40)--(\OUstar,.13);
  \draw[OUorange!45,densely dotted,line width=.5pt]
    (\OUstar,-.68)--(\OUstar,-2.08);
  \fill[OUblue] (0,0) circle[radius=2.3pt];
  \fill[OUblue] (\OUtau,0) circle[radius=2.3pt];
  \fill[OUteal] (\OUend,0) circle[radius=2.3pt];
  \node[draw=OUorange,fill=OUorange,diamond,
    inner sep=0pt,minimum size=6pt] at (\OUstar,0) {};

  \node[ou label] at (0,-.43) {$0$};
  \node[ou label] at (\OUtau,-.43) {$\tau$};
  \node[ou label,text=OUorange] at (\OUstar,-.43) {$t_*$};
  \node[ou label] at (\OUend,-.43) {$\tau+h$};

  \node[ou label,anchor=east] at (-.65,-1.07) {$\omega(t)$};
  \node[ou label,anchor=east] at (-.65,-1.72) {$s(t)$};
  \foreach \x/\a/\b in {0/\nu_1/s_1,\OUtau/\nu_2/s_2,\OUend/2\nu_2/{\alpha+2\nu_2}}{
    \node[ou label,fill=white] at (\x,-1.07) {$\a$};
    \node[ou label,fill=white] at (\x,-1.72) {$\b$};
  }
  \node[ou label,text=OUorange,fill=white] at (\OUstar,-1.07)
    {$\omega=\omega(t_*)$};
  \node[ou label,text=OUorange,fill=white] at (\OUstar,-1.72)
    {$s=\alpha+\omega$};

  \draw[OUteal,decorate,
    decoration={brace,mirror,amplitude=4pt},line width=.55pt]
    (\OUtau,-2.23)--(\OUend,-2.23)
    node[midway,below=7pt,ou label,text=OUteal]
    {$\displaystyle t_*\in[\tau,\tau+h],\qquad
       \mathcal J(t_*)\le\frac{\Delta}{h(\mu_\alpha-1)}$};

\end{tikzpicture}%
\endgroup
    \caption{Illustration of the Ornstein--Uhlenbeck evolution in the proof of Lemma \ref{lem:ou}}
    \label{fig:ou}
\end{figure}
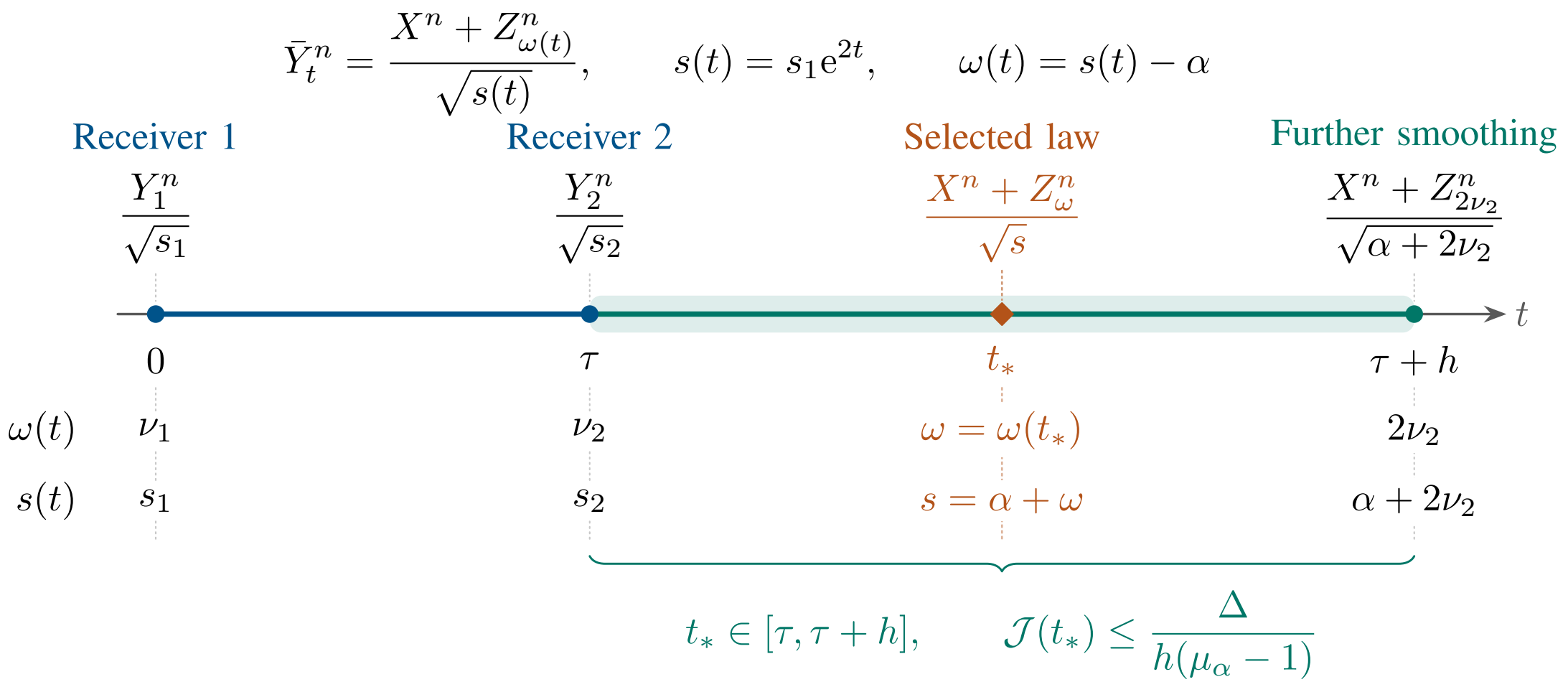

\begin{proof}[Proof of Lemma \ref{lem:ou}]
    Put $\tau := \frac{1}{2}\log \mu_\alpha$. For $t\ge0$, define 
    \begin{align}
        s(t) := s_1\rme^{ 2t } ,\qquad \omega(t) := s(t)-\alpha,\qquad\barY_t^n:= \frac{X^n+ Z_{\omega(t)}^n}{\sqrt{s(t)}}.
    \end{align}
    Here $Z_v^n \sim \calN(0,vI_n)$ is independent of $X^n$. The marginal distributions of $\barY_t^n$ form the standard Ornstein--Uhlenbeck (OU) evolution \cite[Sec.~2.7.1, Eqn.~(2.7.3)]{BakryGentilLedoux2014}.  Indeed, for $t \ge r\ge 0$ their Gaussian transition kernel is (in distribution),
    \begin{align}
        \barY_t^n \stackrel{\rmd}{=} \rme^{-(t-r)}\barY_r^n + \sqrt{ 1-\rme^{-2(t-r)} }G^n,\quad\mbox{where}\quad G^n\sim\phi_n=\calN(0,I_n),\quad G^n\perp \barY_r^n .
    \end{align}
    This process and the derivations below are illustrated schematically in Fig.~\ref{fig:ou}. 
    
    Let $f_t = \rmd P_{ \barY_t^n}/\rmd \phi_n$ be the density of $\barY_t^n$ relative to the standard Gaussian measure. Define the relative entropy of $\barY_t^n$ with respect to $\phi_n$ (along the OU evolution), the relative Fisher information, and the expected squared Hilbert–Schmidt norm of the Hessian of the relative log density  respectively as
    \begin{align}
        \calD(t) &:= \int f_t\, \log f_t \, \rmd \phi_n ,\\
        \calI(t) &:= \int \|\nabla \log f_t\|^2 \, f_t \, \rmd \phi_n ,\\
        \calJ(t) &:= \int \|\nabla^2 \log f_t\|_{\mathrm{HS}}^2\, f_t \, \rmd \phi_n .
    \end{align}
    The entropy and Fisher information dissipation identities \cite[Prop.~5.2.2 and Sec.~5.7]{BakryGentilLedoux2014}, specialized to the Ornstein--Uhlenbeck semigroup, give
    \begin{align}
        \calD'(t) = -\calI(t),\qquad \calI'(t) = -2\calI(t)-2 \calJ(t). \label{eqn:diss_ide}
    \end{align}
     Define $\delta(t) := \frac{1}{2}\calI(t) - \calD(t)$. The Gaussian log-Sobolev inequality \cite[Prop.~5.5.1]{BakryGentilLedoux2014} and the dissipation identities in~\eqref{eqn:diss_ide} give, respectively,
    \begin{align}
    \delta(t)\ge0\quad\mbox{and}\quad \delta'(t)=-\calJ(t). \label{eqn:gauss_log_sobo}
    \end{align}
    Since $\omega(0) = \nu_1$ and $\omega(\tau)=\nu_2$, 
    \begin{align}
        \Delta = \calD(0) - \rme^{2\tau}\calD(\tau) = 2\int_0^\tau \rme^{2t}\delta(t)\, \rmd t.
    \end{align}
    Since $\calJ(t)\ge0$, the function $\delta$ is nonincreasing. Consequently,
\begin{align}
    \Delta \ge 2\delta(\tau)\int_0^\tau \rme^{2t}\,\rmd t =(\rme^{2\tau}-1)\delta(\tau) =(\mu_\alpha-1)\delta(\tau),
\end{align}
which gives $\delta(\tau)\le\Delta/(\mu_\alpha-1)$. Let 
\begin{align}
    h:=\frac{1}{2}\log\frac{\alpha+2\nu_2}{\alpha+\nu_2}>0.
\end{align}
Integrating the second identity in \eqref{eqn:gauss_log_sobo}, we obtain
\begin{align}
    \int_\tau^{\tau+h}\calJ(t)\, \rmd t \le \delta(\tau)\le \frac{\Delta}{\mu_\alpha-1}.
\end{align}
Hence, by the mean value theorem for integrals, there exists
$t_*\in[\tau,\tau+h]$ such that $\calJ(t_*)\le \Delta/[h(\mu_\alpha-1)]$.

Put $\omega:=\omega(t_*)\in[\nu_2,2\nu_2]$ and $s:=\alpha+\omega$. Let $p_\omega$ denote the density of $Y_\omega^n=X^n+Z_\omega^n$, and define
\begin{align}
    \eta(y^n):=y^n+s\nabla_{y^n}\log p_\omega(y^n),\qquad y^n\in\bbR^n. \label{eqn:def_eta}
\end{align}
Under the change of variables $z^n= y^n/\sqrt{s}$, the Hessian of the relative logarithmic density is $\nabla_{z^n}^2\log f_{t_*}(z^n) = I_n + s\nabla_{y^n}^2\log p_{\omega}(y^n) = \Jac \eta(y^n)$, where $\Jac \eta$ is the Jacobian matrix of  the map $\eta$. Thus, by the choice of $t_*$,
\begin{align}
    J_* :=\bbE\big[\|\Jac \eta(Y_\omega^n)\|_{\mathrm{HS}}^2\big] =\calJ(t_*)
    \le \frac{\Delta}{h(\mu_\alpha-1)}
    =O(\Delta). \label{eqn:boundJ_star}
\end{align}
The desired functions are defined as follows:
\begin{align}
    F(x^n):=\bbE\big[\eta(x^n+Z_\omega^n) \big]\quad\mbox{and}\quad V(y^n):=\bbE\big[\eta(y^n+Z_{\omega-\nu_2}^n) \big]. \label{eqn:defFV}
\end{align}
For the following calculation, let
$G^n\sim\calN(0,(\omega-\nu_2)I_n)$ be independent of
$(X^n,Y_2^n)$ and construct
$Y_\omega^n:=Y_2^n+G^n$.
Since $Z_2^n+G^n\sim\calN(0,\omega I_n)$ and is independent
of $X^n$, this construction preserves the required joint
distribution of $(X^n,Y_\omega^n)$. 
When $\omega=\nu_2$, we have $G^n=0$ almost surely and
$V(y^n)=\eta(y^n)$.
By the definitions of $F$ and $V$ and the tower property,
$\bbE[V(Y_2^n)\mid X^n]=F(X^n)$.
Applying the Gaussian Poincar\'e inequality coordinatewise,
conditionally on $X^n$, and then averaging over $X^n$, gives
\begin{align}
    \bbE\big[\|F(X^n)-V(Y_2^n)\|^2\big]
    \le \nu_2\bbE\big[
        \|\Jac V(Y_2^n)\|_{\mathrm{HS}}^2
    \big].
    \label{eqn:gauss_poincare}
\end{align}
By the definition of $V$ and differentiation under the expectation,
\begin{align}
    \Jac V(Y_2^n)
    =\bbE\big[\Jac\eta(Y_\omega^n)\mid Y_2^n\big]
    \qquad\text{a.s.}
\end{align}
Since the squared Hilbert--Schmidt norm is convex, conditional
Jensen's inequality gives
\begin{align}
    \big\|\Jac V(Y_2^n)\big\|_{\mathrm{HS}}^2
    &\le
    \bbE\big[
        \|\Jac\eta(Y_\omega^n)\|_{\mathrm{HS}}^2
        \mid Y_2^n
    \big]
    \qquad\text{a.s.} 
\end{align}
Taking expectations yields
\begin{align}
    \bbE\big[\|\Jac V(Y_2^n)\|_{\mathrm{HS}}^2\big]
    &\le
    \bbE\big[\|\Jac\eta(Y_\omega^n)\|_{\mathrm{HS}}^2\big]
    =O( \Delta). \label{eqn:after_Jens}
\end{align}
Combining \eqref{eqn:gauss_poincare}--\eqref{eqn:after_Jens} gives \eqref{eqn:FminusV}.

Next we prove the energy assertion in \eqref{eqn:XminusF}. Put 
\begin{align}
\chi:=\bbE\big[ \tr( \Jac \eta (Y_\omega^n) ) \big]    \quad \mbox{and} \quad v:= \bbE\big[  \|\eta(Y_\omega^n) - F(X^n) \|^2\big].
\end{align}
Then $|\chi|\le \sqrt{n J_*}$, while $v\ge0$. Writing $\psi(y^n):=\nabla_{y^n}\log p_\omega (y^n)$, integration by parts yields
\begin{align}
    \bbE \big[ (Y_\omega^n)^\top \psi(Y_\omega^n) \big]=-n\quad\mbox{and}\quad \chi=n-s\bbE\big[ \| \psi(Y_\omega^n)\|^2  \big]. \label{eqn:chi_byparts}
\end{align}
We provide details of the second calculation involving $\chi$. Writing $\partial_i:=\partial/\partial y_i$, integration by parts gives
\begin{align}
    \bbE \big[\partial_i\psi_i(Y_\omega^n)\big]
    &=\int_{\bbR^n}\partial_i\psi_i(y^n)\,
        p_\omega(y^n)\,\rmd y^n
    =-\int_{\bbR^n}\psi_i(y^n)\,
        \partial_i p_\omega(y^n)\,\rmd y^n
    =-\bbE\big[\psi_i(Y_\omega^n)^2\big],
\end{align}
where we used $\partial_i p_\omega=\psi_i p_\omega$. The boundary terms vanish because $p_\omega$ is a finite Gaussian mixture and $\psi$ grows at most linearly. Since $\Jac\eta=I_n+s\Jac\psi$, summing over $i$ yields
\begin{align}
    \chi
    =n+s\sum_{i=1}^n\bbE[\partial_i\psi_i(Y_\omega^n)]
    =n-s\bbE[\|\psi(Y_\omega^n)\|^2],
\end{align}
which proves the second identity in \eqref{eqn:chi_byparts}. 
Expansion of the square of \eqref{eqn:def_eta} then yields
\begin{align}
    \bbE\big[ \|\eta (Y_\omega^n)\|^2 \big] = \bbE\big[ \|X^n\|^2\big]-n\alpha-s\chi. \label{eqn:sqare_eta}
\end{align}
Also, $\bbE [ (Y_\omega^n)^\top\eta(Y_\omega^n) ]=\bbE\big[ \|X^n\|^2\big]-n\alpha$, and conditional Gaussian integration by parts gives $\bbE\big[ (Z_\omega^n)^\top \eta (Y_\omega^n)\big]=\omega\chi$. Hence,
\begin{align}
    \bbE\big[ (X^n)^\top\eta( Y_\omega^n)\big] = \bbE\big[ \|X^n\|^2\big]-n\alpha-\omega\chi. \label{eqn:XetaY}
\end{align}
Since $F(X^n)=\bbE[\eta(Y_\omega^n)\mid X^n]$, we have
$\bbE[\|F(X^n)\|^2]=\bbE[\|\eta(Y_\omega^n)\|^2]-v$
and
$\bbE[(X^n)^\top F(X^n)]
=\bbE[(X^n)^\top\eta(Y_\omega^n)]$.
Expanding $\bbE[\|X^n-F(X^n)\|^2]$ and using
\eqref{eqn:sqare_eta} and \eqref{eqn:XetaY} therefore gives
\begin{align}
    \bbE\big[ \|X^n- F(X^n)\|^2 \big]= n\alpha+(\omega-\alpha)\chi-v.
\end{align}
Since $v\ge0$, $|\chi|\le\sqrt{nJ_*}$, and $J_* =O(\Delta)$ by \eqref{eqn:boundJ_star}, we obtain
\begin{align}
    \bbE\big[\|X^n-F(X^n)\|^2\big]
    &\le n\alpha+|\omega-\alpha|\sqrt{nJ_*}\\
    &= n\alpha+O(\sqrt{n\Delta}),
\end{align}
where the last step uses $\alpha\in[0,1]$ and
$\omega\in[\nu_2,2\nu_2]$. This proves \eqref{eqn:XminusF}.

Finally, the Gaussian posterior mean formula gives
\begin{align}
    \eta(y^n) = \frac{s}{\omega}\bbE[X^n \mid Y_\omega^n=y^n]-\frac{\alpha}{\omega}y^n.
\end{align}
Taking expectations with respect to $y^n$ conditionally on $X^n = x^n(m_1, k)$ and using the peak power constraint shows that 
\begin{align}
    \|F(x^n(m_1, k))\|\le (1+ 2\alpha/\omega)\sqrt{n}
\end{align}
as desired. 
\end{proof}

\begin{proof}[Proof of Proposition~\ref{prop:partition}]
The proof has three steps. First, we construct an auxiliary channel such that \(P_{Y_1^n|U^n}\) has sublinear expected relative entropy from \(\mathcal N(\mathbb E[X^n|U^n],s_1I_n)\) and \(I(X^n;U^n|Y_2^n)=O(n^{1/3})\). Second, we apply the SFRL and select a seed to obtain the partition satisfying \eqref{eqn:small_ce}--\eqref{eqn:sublinear_div}. Finally, Gaussian maximum entropy and equal codeword energies yield \eqref{eqn:comp_means}.


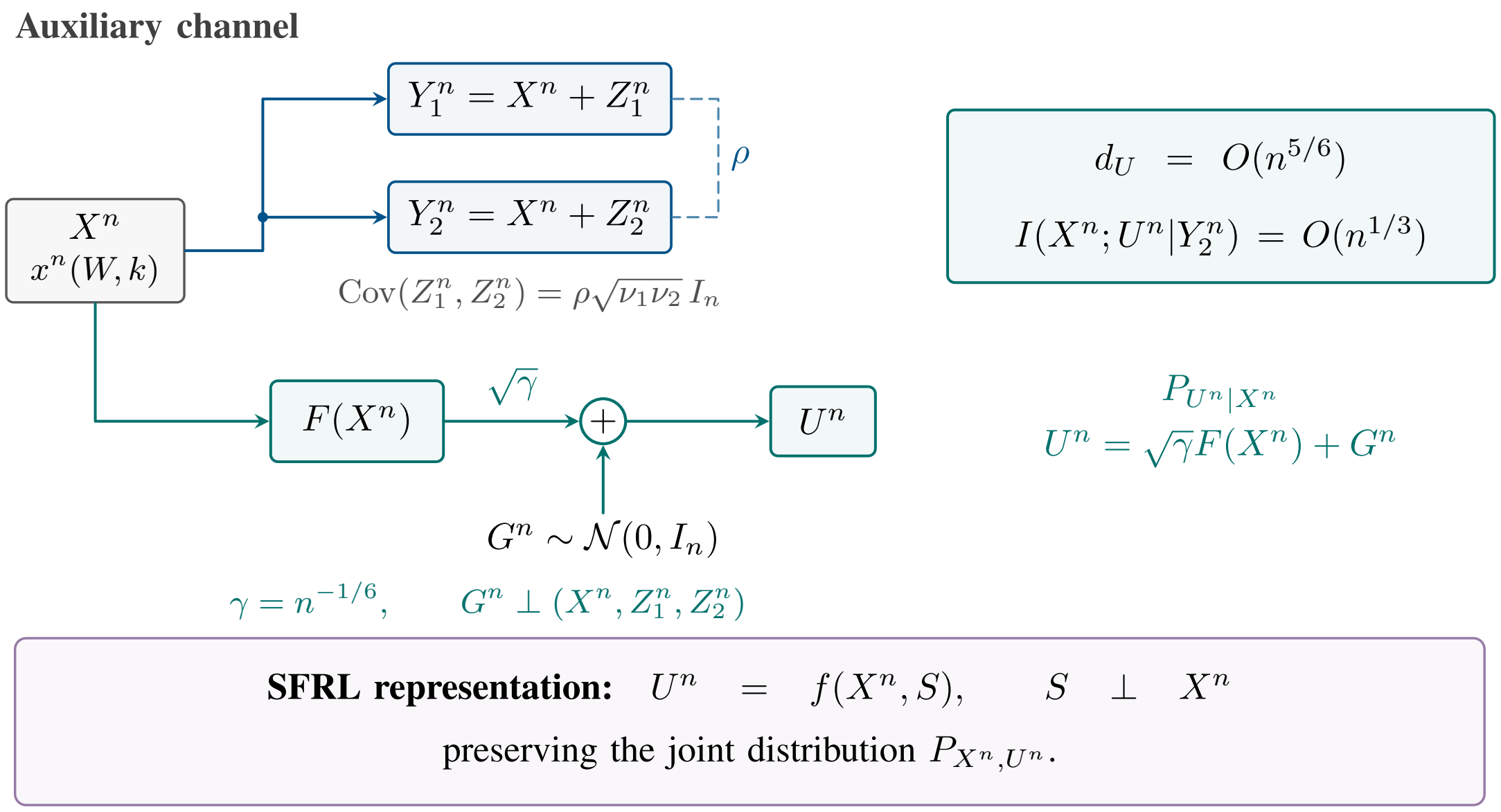
\begin{figure}
    \centering
%
\begingroup
\definecolor{BCauxblue}{RGB}{0,83,138}
\definecolor{BCauxteal}{RGB}{0,113,108}
\definecolor{BCauxpurple}{RGB}{106,65,126}
\begin{tikzpicture}[
  x=1cm,y=1cm,font=\small,
  line cap=round,line join=round,>=stealth,
  bc input/.style={draw=black!65,fill=black!3,rounded corners=2pt,
    line width=.65pt,minimum width=1.7cm,minimum height=.8cm,
    inner sep=4pt,align=center},
  bc output/.style={draw=BCauxblue,fill=BCauxblue!5,
    rounded corners=2pt,line width=.7pt,minimum width=2.7cm,
    minimum height=.58cm,inner sep=5pt,align=center},
  bc aux/.style={draw=BCauxteal,fill=BCauxteal!5,
    rounded corners=2pt,line width=.75pt,minimum height=.6cm,
    inner sep=6pt,align=center},
  bc proof/.style={draw=BCauxpurple!70,fill=BCauxpurple!4,
    rounded corners=3pt,line width=.65pt,inner sep=9pt,align=center},
  bc arrow/.style={->,line width=.75pt},
  bc label/.style={inner sep=2pt,align=center}
]
  \node[anchor=west,font=\small\bfseries,text=black!75]
    at (.15,4.72) {Auxiliary channel};

  \node[bc input] (auxX) at (1.05,2.6)
    {$X^n$\\[-1pt]{\footnotesize $x^n(W,k)$}};
  \node[bc output] (auxYone) at (5.2,4.05)
    {$Y_1^n=X^n+Z_1^n$};
  \node[bc output] (auxYtwo) at (5.2,2.92)
    {$Y_2^n=X^n+Z_2^n$};
  \draw[bc arrow,BCauxblue]
    (auxX.east)--(2.65,2.6)--(2.65,4.05)--(auxYone.west);
  \draw[bc arrow,BCauxblue]
    (2.65,2.92)--(auxYtwo.west);
  \fill[BCauxblue] (2.65,2.92) circle[radius=1.4pt];

  \draw[BCauxblue!70,densely dashed,line width=.6pt]
    (auxYone.east)--(7.0,4.05)--(7.0,2.92)--(auxYtwo.east);
  \node[bc label,anchor=west,text=BCauxblue] at (7.04,3.48)
    {$\rho$};
  \node[bc label,font=\footnotesize,text=black!70] at (5.2,2.18)
    {$\operatorname{Cov}(Z_1^n,Z_2^n)
       =\rho\sqrt{\nu_1\nu_2}\,I_n$};

  \node[bc aux,minimum width=1.65cm] (auxF) at (3.55,.97)
    {$F(X^n)$};
  \node[circle,draw=BCauxteal,fill=white,line width=.8pt,
    inner sep=0pt,minimum size=.43cm] (auxSum) at (5.9,.97) {$+$};
  \node[bc aux,minimum width=1.0cm] (auxU) at (8.0,.97) {$U^n$};
  \draw[bc arrow,BCauxteal]
    (auxX.south)--(1.05,.97)--(auxF.west);
  \draw[bc arrow,BCauxteal] (auxF.east)--(auxSum.west)
    node[midway,above=2pt,bc label] {$\sqrt{\gamma}$};
  \draw[bc arrow,BCauxteal] (auxSum.east)--(auxU.west);
  \node[bc label] (auxG) at (5.9,-.16)
    {$G^n\sim\mathcal N(0,I_n)$};
  \draw[bc arrow,BCauxteal] (auxG.north)--(auxSum.south);
  \node[bc label,font=\footnotesize,text=BCauxteal] at (4.8,-.75)
    {$\gamma=n^{-1/6},\qquad
       G^n\perp (X^n,Z_1^n,Z_2^n)$};

  \node[bc aux,text width=4.8cm,minimum height=1.65cm]
    at (11.8,3.12)
    {$d_U=O(n^{5/6})$\\[9pt]
     $I(X^n;U^n  | Y_2^n)=O(n^{1/3})$};
  \node[bc label,text=BCauxteal] at (11.8,.97)
    {$P_{U^n | X^n}$\\[3pt]
     $U^n=\sqrt{\gamma}F(X^n)+G^n$};

  \node[bc proof,text width=13.4cm,minimum height=1.05cm]
    at (7.3,-1.9)
    {\textbf{SFRL representation:}\quad
     $U^n=f(X^n,S),\qquad S\perp X^n$\\[5pt]
     preserving the joint distribution $P_{X^n,U^n}$.};
\end{tikzpicture}%
\endgroup
    \caption{Illustration of the auxiliary channel in~\eqref{eqn:Uchannel} and its consequences. }
    \label{fig:aux}
\end{figure}
{\em 1. Construction of an auxiliary channel and bounding the conditional relative entropy:}

Recall that  $\Delta:=\Delta_\alpha(P_{X^n})$ is the (unconditional)
entropy deficit. For any random variable $A$ satisfying $A-X^n-(Y_1^n,Y_2^n)$, the chain rule for relative entropy gives
\begin{align}
    \bbE_A\big[
        D (P_{Y_j^n|A}\|\calN(0,s_jI_n) )
    \big]
    =
    D (P_{Y_j^n} \|\calN(0,s_jI_n))
    +I(A;Y_j^n),\qquad j=1,2.
\end{align}
Subtracting $\mu_\alpha$ times the identity for $j=2$
from that for $j=1$, and applying  relative entropy contraction along the Ornstein--Uhlenbeck semigroup \cite[Thm.~5.2.1 and Prop.~5.5.1]{BakryGentilLedoux2014}  to each conditional input distribution (as in~\eqref{eqn:contraction}), yields
\begin{align}
    0\le  \bbE_A[\Delta_\alpha(P_{X^n|A})] =\Delta+I(A;Y_1^n)-\mu_\alpha I(A;Y_2^n).
\end{align}
Substitute $I(A;Y_j^n) = I(X^n;A) - I(X^n;A|Y_j^n)$ to obtain
\begin{align}
    (\mu_\alpha-1)I(X^n;A)\le \Delta+ \mu_\alpha I(X^n;A|Y_2^n)- I(X^n;A|Y_1^n). \label{eqn:substitute_IAY}
\end{align}
Now take the constructed functions $F $ and $V$ from Lemma \ref{lem:ou}, and set 
\begin{align}
    \gamma = n^{-1/6},\qquad U^n:= \sqrt{\gamma}F(X^n) + G^n,\qquad G^n\sim\calN(0, I_n) \label{eqn:Uchannel}
\end{align}
with $G^n$ independent of the input and the channel noises. In \eqref{eqn:Uchannel}, we have constructed an auxiliary channel from $F(X^n)$ to $U^n$; see Fig.~\ref{fig:aux}. Comparing the conditional distribution of $U^n$ given $(X^n,Y_2^n)$ to $\calN( \sqrt{\gamma}V(Y_2^n), I_n) $ gives
\begin{align}
    I(X^n;U^n|Y_2^n) \le\frac{\gamma}{2} \bbE\big[ \| F(X^n)- V(Y_2^n)\|^2 \big]=O(\gamma\Delta), \label{eqn:cond_mi}
\end{align}
where the last equality follows from \eqref{eqn:FminusV}. Using  \eqref{eqn:substitute_IAY} with $A:=U^n$ and dropping the final nonpositive term gives
\begin{align}
    I(X^n;U^n)=O(\Delta).\label{eqn:IXU}
\end{align}
We conclude from \eqref{eqn:cond_mi} that $I(X^n;U^n|Y_2^n)=O(n^{1/3})$ and from \eqref{eqn:IXU} that $I(X^n;U^n)=O(\sqrt{n})$.

 Since $F(X^n)$ is a function of $X^n$ (cf.~\eqref{eqn:Uchannel}) and $X^n-F(X^n)-U^n$, we have $I(F(X^n);U^n)=I(X^n;U^n)$. The vector I--MMSE relation and the fact that MMSE is nonincreasing in the Gaussian channel parameter \cite[Thm.~2 and Cor.~2]{GuoShamaiVerdu2005} yield
\begin{align}
    \mmse(F(X^n)\mid U^n) \le \frac{2I(X^n;U^n)}{\gamma} =O\bigg(\frac{\Delta}{\gamma}\bigg) =O(n^{2/3}).
\end{align}
Use $\bbE[F(X^n) \mid U^n]$ as an estimator of $X^n$. Expanding its squared error and applying Cauchy--Schwarz to the cross term yields
\begin{align}
    \mmse(X^n\mid U^n)&\le \bbE\big[ \|X^n- F(X^n)\|^2 \big]+ \mmse (F(X^n)\mid U^n)\notag \\
    &\qquad+ 2\sqrt{  \bbE\big[ \|X^n- F(X^n)\|^2 \big]  \mmse (F(X^n)\mid U^n) } \label{eqn:estimator1}\\
    &\le n\alpha+O(n^{5/6}). \label{eqn:estimator2}
\end{align}
Here, \eqref{eqn:XminusF} bounds the first term  in~\eqref{eqn:estimator1} by $n\alpha+ O(n^{3/4})$. The second term in \eqref{eqn:estimator1} is $O(n^{2/3})$, and the cross term is $O(n^{5/6})$. The last exponent dominates the preceding terms. 

We next prove the entropy estimate needed to convert the MMSE bound in \eqref{eqn:estimator2} into a bound on the conditional relative entropy in \eqref{eqn:sublinear_div}. Write $I_j := I(X^n;Y_j^n)$ for $j = 1,2$ and define
\begin{align}
    v_* := \nu_1(\rme^{2I_1/n}-1). \label{eqn:v_star}
\end{align}
Nonnegativity of mutual information and Gaussian maximum entropy under $\bbE[\|X^n\|^2]\le n$ imply $0\le v_*\le1$.  For   $v\ge0$, put $\xi_v := (v+\nu_2)/(v+\nu_1)$. Recalling the Gaussian mixture output distributions $P_j$ defined in~\eqref{eqn:marg_output}, we expand their relative entropies with respect to $\calN(0,(v+\nu_j)I_n)$, $j=1,2$. The second moment terms cancel because $\xi_v/(v+\nu_2)=1/(v+\nu_1)$, yielding
\begin{align}
    D(P_1 \| \calN( 0, (v+\nu_1)I_n))-\xi_v D(P_2 \| \calN( 0, (v+\nu_2)I_n)) =\frac{n}{2}\log \frac{v+\nu_1}{\nu_1}-I_1-\xi_v\left(\frac{n}{2}\log \frac{v+\nu_2}{\nu_2}-I_2\right). \label{eqn:re_expansion}
\end{align}
Another application of  Gaussian entropy contraction  \cite[Thm.~5.2.1 and Prop.~5.5.1]{BakryGentilLedoux2014} ensures that the left-hand side is nonnegative. Furthermore, at $v=v_*$ (cf.~\eqref{eqn:v_star}), the first two terms on the right-hand side cancel and so 
\begin{align}
    I_2\ge\frac{n}{2}\log\frac{v_*+\nu_2}{\nu_2}. \label{eqn:I2_lower_bound}
\end{align}
Setting $v=\alpha$ in \eqref{eqn:re_expansion}, substituting the defining expression for $I_1$ in terms of $v_*$, and using \eqref{eqn:I2_lower_bound}, we obtain
\begin{align}
    \Delta\ge nd_\alpha(v_*)\quad\mbox{where}\quad d_\alpha(v) := \frac{1}{2}\left[  \mu_\alpha\log\frac{v+\nu_2}{s_2}-\log \frac{v+\nu_1}{s_1}\right].
\end{align}
Since $d_\alpha(\alpha)=0$ and $d_\alpha'(v)=\frac{(\mu_\alpha-1)(v-\alpha) }{2(v+\nu_1)(v+\nu_2)}$, integration on the interval between $\alpha$ and $v\in [0,1]$ gives the quadratic lower bound
\begin{align}
    d_\alpha(v)\ge \frac{(\mu_\alpha-1)(v-\alpha)^2}{4(1+\nu_1)(1+\nu_2)}. \label{eqn:lb_dalpha}
\end{align}
The logarithm is Lipschitz on $[\nu_1,1+\nu_1]$. Eqns.~\eqref{eqn:v_star} and~\eqref{eqn:lb_dalpha} thus give
\begin{align}
    \left|  I(X^n;Y_1^n)- \frac{n}{2}\log\frac{s_1}{\nu_1}\right| &=  \frac{n}{2}\big|\log (v_*+\nu_1)- \log (\alpha+\nu_1)\big|\\ 
    &\le\frac{n}{2\nu_1} |v_*-\alpha| \label{eqn:use_Lip}\\
    &\le\frac{1}{ \nu_1} \sqrt{\frac{(1+\nu_1)(1+\nu_2)}{\mu_\alpha-1} }\sqrt{n\Delta}=O(n^{3/4}),\label{eqn:use_quad_lb}
\end{align}
where \eqref{eqn:use_Lip} follows from the fact that the derivative of $\log x $ on the interval $[\nu_1,1+\nu_1] $ (Lipschitz constant) is at most $1/\nu_1$ and \eqref{eqn:use_quad_lb} uses the quadratic lower bound in \eqref{eqn:lb_dalpha}.

Define the conditional relative entropy 
\begin{align}
    d_U:=\bbE_{U^n}\Big[D\big(P_{Y_1^n|U^n} \big\| \calN( \bbE[X^n|U^n], s_1 I_n)\big)\Big].
\end{align}
Expanding the logarithm of the density of \(\mathcal N(\mathbb E[X^n\mid U^n],s_1I_n)\) gives
\begin{align}
    d_U=\frac{n}{2}\log\frac{s_1}{\nu_1}- I(X^n;Y_1^n)+ I(U^n;Y_1^n) + \frac{1}{2s_1}(\mmse (X^n\mid U^n)-n\alpha). \label{eqn:du_ub}
\end{align}
Here the mutual information terms follow from \(I(X^n;Y_1^n\mid U^n)=I(X^n;Y_1^n)-I(U^n;Y_1^n)\), using \(U^n-X^n-Y_1^n\). Now use \eqref{eqn:use_quad_lb}, $I(U^n;Y_1^n)\le I(U^n;X^n)$, \eqref{eqn:IXU} and  \eqref{eqn:estimator2} to conclude that  \eqref{eqn:du_ub} can be upper bounded as
\begin{align}
    d_U=O(n^{5/6}). \label{eqn:bound_dU}
\end{align}

{\em 2. A deterministic partition:} The  SFRL  of Li and El Gamal \cite{LiElGamal2018} converts the auxiliary observation $U^n$ into a partition of message indices. Its additional information cost is only $O(\log n)$; see \eqref{eqn:overhead}.

Apply the SFRL \cite{LiElGamal2018} to $(X^n,U^n)$ just constructed. It provides a random variable $S$, which we call the {\em seed}, independent of $X^n$ and a measurable function $f$ such that $U^n=f(X^n,S)$ has the prescribed joint distribution with $X^n$, and
\begin{align}
    I(X^n;S|U^n)\le\log\left(\frac{I(X^n;U^n)}{\log 2}+1 \right)+4\log 2= O(\log n), \label{eqn:overhead}
\end{align}
where the final bound follows from \eqref{eqn:IXU}.

Sample $S$ independently of $W$, and set $X^n=x^n(W,k)$ and $U^n=f(X^n,S)$. Generate $(Y_1^n,Y_2^n)$ conditionally on $X^n$ according
to the joint channel distribution $\rvW_{1,2}^n(\fndot|X^n)$, independently of $S$. This preserves the prescribed joint distribution of $(W,X^n,U^n,Y_1^n,Y_2^n)$ and gives $S\perp(W,X^n,Y_1^n,Y_2^n)$.

\begin{figure}
    \centering
\begingroup
\definecolor{BCseedblue}{RGB}{0,83,138}
\definecolor{BCseedgreen}{RGB}{0,119,103}
\definecolor{BCseedpurple}{RGB}{116,73,131}
\begin{tikzpicture}[
  x=1cm,y=1cm,font=\small,
  >=stealth,line cap=round,line join=round,
  seed cell/.style={rounded corners=5pt,line width=.7pt,
    minimum width=3.1cm,minimum height=.68cm,inner sep=4pt},
  seed label/.style={rounded corners=4pt,line width=.7pt,
    minimum width=1.25cm,minimum height=.68cm,inner sep=4pt},
  seed value/.style={rounded corners=3pt,line width=.7pt,
    minimum width=3.25cm,minimum height=.68cm,inner sep=4pt}
]
  \node at (5.25,3.55) {$S=s\quad (k\text{ fixed})$};
  \node at (1.8,2.92)
    {$\displaystyle\mathcal I=\bigsqcup_{\lambda}\mathcal I_\lambda$};
  \node at (5.25,2.82) {$J_s=\lambda$};
  \node at (8.7,2.82) {$U^n=f(X^n,s)$};

  \draw[black!35,rounded corners=7pt,line width=.5pt]
    (0,-.48) rectangle (3.6,2.48);

  \foreach \yy/\idx/\members/\col in {
    2/1/{1,4,6}/BCseedblue,
    1/2/{2,5}/BCseedgreen,
    0/3/{3,7}/BCseedpurple}{
    \node[seed cell,draw=\col,fill=\col!6] (cell\idx)
      at (1.8,\yy) {$\mathcal I_{\idx}=\{\members\}$};
    \node[seed label,draw=\col,fill=\col!6] (label\idx)
      at (5.25,\yy) {$\lambda=\idx$};
    \node[seed value,draw=\col,fill=\col!6] (value\idx)
      at (8.7,\yy) {$f(x^n(\idx,k),s)$};
    \draw[->,draw=\col,line width=.75pt]
      (cell\idx.east)--(label\idx.west);
    \draw[<->,draw=\col,line width=.85pt]
      (label\idx.east)--(value\idx.west);
  }
  \node at (5.25,-1.02)
    {$m_1\in\mathcal I_\lambda
      \quad\Longleftrightarrow\quad J_s(m_1)=\lambda$};
\end{tikzpicture}%
\endgroup
    \caption{Partition of the selected message set $\calI=\sqcup_\lambda\calI_\lambda$ for a fixed seed $S=s$
and cloud index $k$. Each cell $\calI_\lambda$ is labeled
by its first message index $\lambda$, and
$f(x^n(m_1,k),s)=f(x^n(\lambda,k),s)$ for every
$m_1\in\calI_\lambda$. Distinct cells correspond to distinct values of $U^n$. Hence, given $S=s$, the component label $J_s=\lambda$ and the value of $U^n$ determine one another.}
    \label{fig:seeds}
\end{figure}
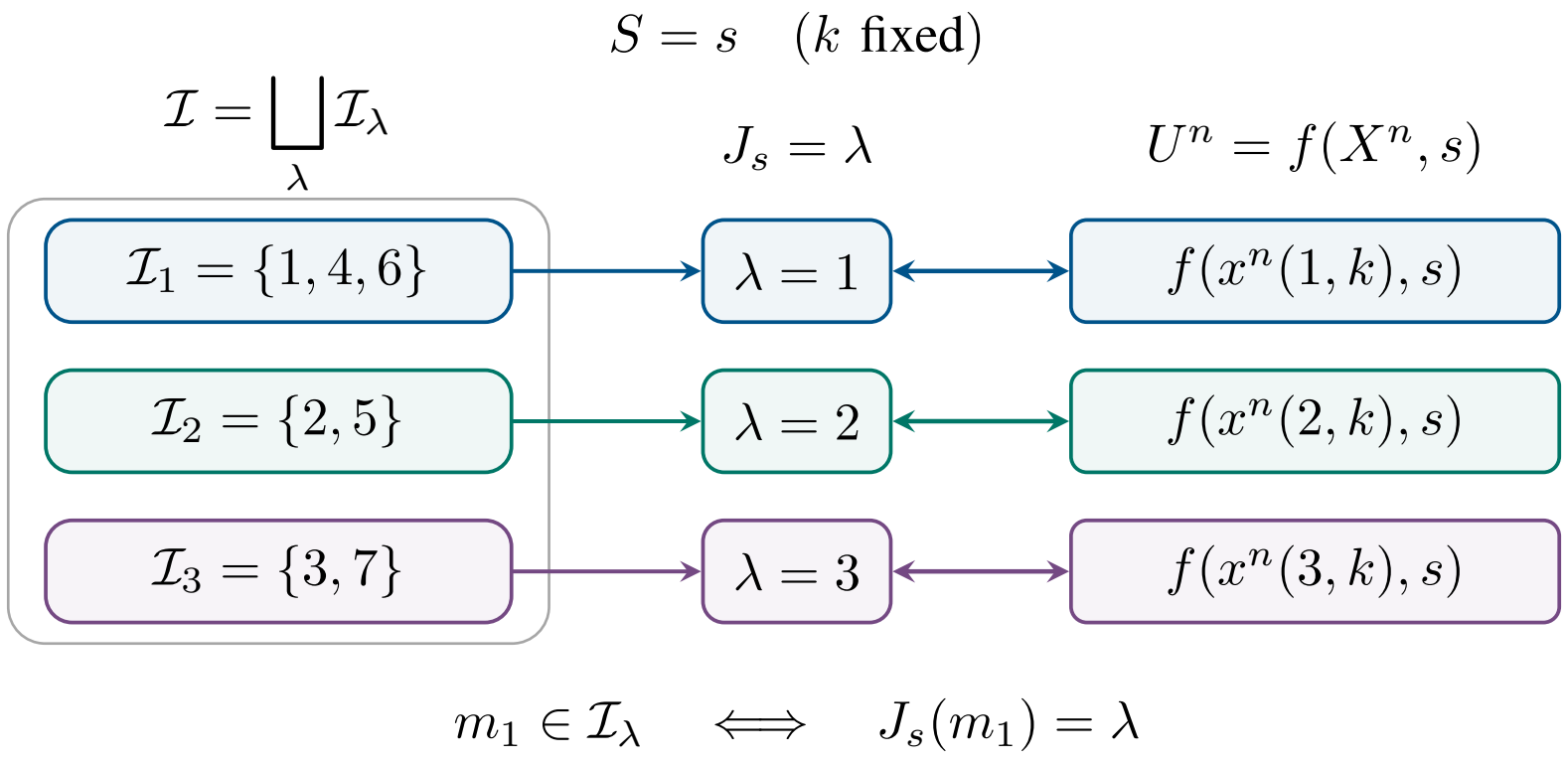

For a fixed seed $s$, the map $x^n(m_1, k)\mapsto f( x^n(m_1, k), s)$ has a finite range. Its preimages form a partition of the selected subcode; let $J_s$ be its component label. Fix an ordering of $\calI$ and label each
cell by its first index. Given $S=s$, the labels $J_s$ and the values of $U^n$ determine one another; see Fig.~\ref{fig:seeds}. Consequently, 
\begin{align}
    H(J_S|Y_2^n, S) &= I(X^n;U^n|Y_2^n,S) \\
    &= I(X^n;U^n|Y_2^n) + I(X^n;S|U^n,Y_2^n) ,\label{eqn:cond_indep} \\
    &\le I(X^n;U^n|Y_2^n) + I(X^n;S|U^n)  \label{eqn:markov_chain}\\
    &=O(n^{1/3}),  \label{eqn:use_overhead}
\end{align}
where 
\begin{itemize}
    \item \eqref{eqn:cond_indep} follows from $I(X^n;S|Y_2^n)=0$ ($S$ is independent of $(X^n,Y_2^n)$);
    \item \eqref{eqn:markov_chain} follows because \(U^n=f(X^n,S)\) and the channel noise is independent of \((X^n,S)\). Thus, even after conditioning on \(U^n\) and \(S\), the conditional distribution of \(Y_2^n\) given \(X^n\) remains the channel distribution $\rvW_2^n (\fndot|X^n)$. Thus,  $S-X^n-Y_2^n$ forms a Markov chain conditionally on $U^n$ so $I(S;Y_2^n|X^n, U^n)=0$;
    \item  \eqref{eqn:use_overhead} follows from $I(X^n;U^n|Y_2^n)=O(n^{1/3})$ below \eqref{eqn:cond_mi} and the overhead calculation in~\eqref{eqn:overhead}.
\end{itemize}

We also need an averaged divergence bound after revealing the seed. For brevity, write $m_{U^n}:=\bbE[X^n\mid U^n]$ and  $m_{U^n,S}:=\bbE[X^n\mid U^n,S]$. Then
\begin{align}
     \bbE_{U^n,S}\Big[D\big( P_{Y_1^n|U^n,S }\|\calN(m_{U^n,S}, s_1I_n) \big) \Big]  &\le \bbE_{U^n,S}\Big[D\big( P_{Y_1^n|U^n,S }\|\calN(m_{U^n}, s_1I_n) \big)\Big] \label{eqn:match_means}\\
    &= d_U+ I(Y_1^n;S|U^n)  \label{eqn:chain_rule}\\
    &\le d_U+ I(X^n;S|U^n) \label{eqn:conditional_data_processing} \\
    & =O(n^{5/6}),\label{eqn:bounddI}
\end{align}
where 
\begin{itemize}
    \item \eqref{eqn:match_means} follows from the fact that among Gaussian distributions with covariance $s_1I_n$, choosing the mean to match the distribution being approximated minimizes relative entropy;
    \item \eqref{eqn:chain_rule} follows from the chain rule for relative entropy;
    \item \eqref{eqn:conditional_data_processing} follows from the (conditional) data processing inequality for mutual information;
    \item \eqref{eqn:bounddI} follows from the bound on the   relative entropy in \eqref{eqn:bound_dU} and the overhead in \eqref{eqn:overhead}.
 \end{itemize}

The left-hand sides of the critical inequalities in~\eqref{eqn:use_overhead} and~\eqref{eqn:bounddI} are expectations over $S$ of nonnegative quantities. By Markov's inequality, each exceeds $3$ times its upper bound with probability no more than $1/3$. Choose a single seed $s_*$ for which both bounds hold and put $J:=J_{s_*}$. For this fixed seed, conditioning on $U^n$ is equivalent
to conditioning on $J$. Moreover, $W$ conditional on
$J=\lambda$ is uniform on $\calI_\lambda$, so the
corresponding conditional mean is $u_\lambda^n$.
Thus, the two selected bounds are precisely
\eqref{eqn:small_ce} and \eqref{eqn:sublinear_div}.  

Gaussian entropy contraction \cite[Thm.~5.2.1 and Prop.~5.5.1]{BakryGentilLedoux2014} applied to each component gives
\begin{align}
D\big(P_{Y_2^n|J=\lambda}\big\|\calN(u_\lambda^n,s_2I_n)\big) \le \frac{s_1}{s_2}d_\lambda.
\end{align}
Thus, the corresponding $p_\lambda$-weighted average relative entropy at receiver~2 is also $O(n^{5/6})$.

{\em 3. The component means:}  For each component, put $v_\lambda= n^{-1}\bbE [ \|X^n-u_\lambda^n\|^2 \mid J=\lambda]$. Since $\|X^n\|^2=n$ almost surely and $u_\lambda^n=\bbE[X^n\mid J=\lambda]$, the conditional variance decomposition gives
\begin{align}
    \frac{\|u_\lambda^n\|^2}{n}=1-v_\lambda\quad\mbox{and}\quad 0\le v_\lambda\le 1. 
\end{align}
Recall that $d_\lambda=D\big( P_{Y_1^n|J=\lambda}\big\|\calN(u_\lambda^n,s_1 I_n)  \big)$. Gaussian maximum entropy at a fixed second moment (i.e., $h(Y_1^n | J=\lambda) \le \frac n2\log(2\pi e(v_\lambda+\nu_1))$) gives
\begin{align}
    d_\lambda\ge\frac{n}{2}\left[\frac{v_\lambda+\nu_1}{s_1}-1-\log \frac{v_\lambda+\nu_1}{s_1}\right]\ge\frac{n(v_\lambda-\alpha)^2}{4(1+\nu_1)^2}.\label{eqn:bound_d_lambda}
\end{align}
The second inequality follows from Taylor's theorem: the bracketed function and its first derivative vanish at $v=\alpha$, while its second derivative satisfies $(v+\nu_1)^{-2}\ge(1+\nu_1)^{-2}$ for $v\in[0,1]$.  Using the quadratic lower bound in~\eqref{eqn:bound_d_lambda}, we obtain
\begin{align}
    \left| \frac{\|u_\lambda^n\|^2 }{n}-\bar{\alpha}\right|= |v_\lambda-\alpha|\le 2(1+\nu_1)\sqrt{\frac{d_\lambda}{n}}. \label{eqn:u_lambda_bd}
\end{align}
Averaging over $p_\lambda$, using Jensen's inequality and the bound $\bard=O(n^{5/6})$ from~\eqref{eqn:sublinear_div} proves~\eqref{eqn:comp_means}.
\end{proof} 

\subsection{Auxiliary Densities and Preparation for Normal Approximation}  \label{sec:aux}

We use the partition from Proposition \ref{prop:partition} to construct auxiliary output measures for use in a Verd\'u–Han-type converse~\cite{VH94}. Receiver 1 uses a separate measure for each component, whereas the receiver 2 statistic uses the output distribution of the entire selected subcode. We first analyze the component distributions; Bayes' rule and the conditional entropy bound \eqref{eqn:small_ce} will control the passage to the full output distribution.  We choose the receiver-1 auxiliary measure using the receiver-2 component likelihood ratio, as specified in~\eqref{eqn:tilq1}. Lemma~\ref{lem:output} establishes the exact cancellation of the codeword-dependent terms in the weighted sum of conditional means, given the transmitted codeword, while Lemmas~\ref{lem:geom}--\ref{lem:ll_ratio} control the geometry and log likelihood-ratio fluctuations needed for the normal approximation.

Fix a component $\calI_\lambda$ and condition on $W\in\calI_\lambda$. In the following three lemmas, for brevity, we suppress the component subscript $\lambda$: $W$ is uniform on $\calI_\lambda$, $X^n=x^n(W,k)$, and $P_j$, with density $p_j$, is the output distribution under this conditional input distribution. Put $u^n:=\bbE[X^n]$ and let $\frakG_j:=\calN(u^n,s_jI_n)$ have density $g_j$, $j=1,2$. In particular, $J_2$ below uses the component output density. All implicit constants in these lemmas depend only on $\alpha$, $\nu_1$, and $\nu_2$, and are independent of the component, the code, the blocklength, and $\rho\in[-1,1]$.

\begin{lemma}[Auxiliary measures and cancellation] \label{lem:output} Let $f:= p_2/g_2$ and define the positive function 
\begin{align}
    \tilq_1(y^n):=g_1(y^n) \exp \big(\mu_\alpha (\Lambda_{1\rightarrow 2}\log f)(y^n) \big) ,\label{eqn:tilq1}
\end{align}
    where  $\Lambda_{1\rightarrow 2} $ is the    convolution operator with $\calN(0, (\nu_2-\nu_1)I_n)$. Then $\int \tilq_1\le 1$. Write $\tilQ_1$ for the resulting subprobability measure. Define  the log-likelihood ratios
    \begin{align}
        J_1(m_1,y^n):= \log\frac{\rvW_1^n(y^n|x^n(m_1,k))}
              {\tilq_1(y^n)}\quad \mbox{and}\quad J_2(y^n) :=\log \frac{p_2(y^n)}{q_0(y^n)}  \label{eqn:ll_ratios}
    \end{align}
    Let $J_1 :=J_1(W,Y_1^n)$ and $J_2 := J_2(Y_2^n)$ where $W$ is uniform on $\calI_\lambda$. For $m_1 \in \calI_\lambda$, denote the conditional means as  $\barJ_j(x^n(m_1, k)):=\bbE[J_j \mid W=m_1], j = 1,2$. These  satisfy
    \begin{align}
        \barJ_1(x^n)+\mu_\alpha\barJ_2(x^n) = nC_\alpha \label{eqn:cancellation}
    \end{align}
    for every codeword $x^n = x^n(m_1, k)$.
\end{lemma}
\begin{remark} The choice of $\tilQ_1$ ensures that the terms involving $\log(p_2/g_2)$ cancel exactly in the $(1,\mu_\alpha)$-weighted sum of the conditional means. Together with the exact energy constraint, this yields \eqref{eqn:cancellation} for every codeword.
\end{remark}

\begin{proof}
    First, we prove that $\int \tilq_1\le 1$.   Since $s_2-s_1=\nu_2-\nu_1$, we have $\frakG_1\Lambda_{1\rightarrow2}=\frakG_2$. As usual, Gaussian entropy contraction  \cite[Thm.~5.2.1 and Prop.~5.5.1]{BakryGentilLedoux2014} gives
\begin{align}
    \mu_\alpha D(Q\Lambda_{1\rightarrow2}\|\frakG_2) \le D(Q\|\frakG_1) \label{eqn:contracti}
\end{align}
for every probability measure $Q$ with finite relative entropy with respect to $\frakG_1$. Consequently, for every bounded measurable function $h:\bbR^n\to\bbR$, the variational formula for relative entropy (Donsker--Varadhan) gives
\begin{align}
    \log\int
    \exp\big(\mu_\alpha\Lambda_{1\rightarrow2}h\big)
    \,\rmd \frakG_1 &=\sup_Q\left\{
        \mu_\alpha\int h\,\rmd(Q\Lambda_{1\rightarrow2})
        -D(Q\|\frakG_1)
      \right\}\\
    &\le\mu_\alpha\sup_{Q'}\left\{
        \int h\,\rmd Q'-D(Q'\|\frakG_2)
      \right\}
    =\mu_\alpha\log\int \rme^h\,\rmd \frakG_2, \label{eqn:dk}
\end{align}
where both suprema range over probability measures on $\bbR^n$ and the inequality follows from  \eqref{eqn:contracti}.

Apply \eqref{eqn:dk} to a clipped version of $\log f$, i.e.,  $h_b:=\max\{-b,\min\{\log f,b\}\}$. Since $p_2$ is a finite Gaussian mixture and $g_2$ is a Gaussian density, for the fixed code there is a finite constant $A$ such that
$|\log f(y^n)|\le A(1+\|y^n\|^2)$. The dominated convergence theorem therefore gives
$\Lambda_{1\rightarrow2}h_b\to \Lambda_{1\rightarrow2}\log f$  pointwise as $b\to\infty$. Moreover, $\rme^{h_b}\le1+f$, so another application of dominated convergence yields
\begin{align}
    \int\rme^{h_b}\,\rmd \frakG_2 \rightarrow\int f\,\rmd \frakG_2=\int p_2(y^n)\,\rmd y^n=1.
\end{align}
Fatou's lemma now gives
\begin{align}
    \int\tilq_1(y^n)\,\rmd y^n &\le\liminf_{b\to\infty}
       \int\exp\big(\mu_\alpha
           \Lambda_{1\rightarrow2}h_b\big)\,\rmd \frakG_1
       \le\lim_{b\to\infty}
       \left(\int\rme^{h_b}\,\rmd \frakG_2\right)^{\mu_\alpha}
     =1.
\end{align}
This proves that $\int\tilq_1\le 1$.

    Write $v^n=x^n-u^n$ and 
    \begin{align}
        \varrho(x^n):=\bbE \big[\log f(x^n+Z_2^n)\big]. \label{eqn:varrho}
    \end{align}
    An expansion of the two log-likelihood ratios gives
    \begin{align}
        \barJ_1(x^n) &= nC_1(\alpha)+ \frac{\|v^n\|^2-n\alpha}{2s_1}-\mu_\alpha \varrho(x^n),\label{eqn:first_id}\\
        \barJ_2(x^n) &= nC_2(\alpha) - \frac{\|v^n\|^2-n\alpha}{2s_2}+\varrho(x^n). \label{eqn:sec_id}
    \end{align}
    For \eqref{eqn:first_id}, $\bbE[ (\Lambda_{1\rightarrow 2}\log f)(x^n+Z_1^n)]=\varrho(x^n)$ because convolution adds the missing Gaussian covariance $(\nu_2-\nu_1)I_n$. Equation~\eqref{eqn:sec_id} also uses $\|x^n\|^2=n$. Adding \eqref{eqn:first_id} to $\mu_\alpha$ times \eqref{eqn:sec_id} gives \eqref{eqn:cancellation} as desired. Notice that the second and third terms in \eqref{eqn:first_id} and \eqref{eqn:sec_id} cancel exactly. This is a key step in the converse.
\end{proof}
Our goal is to recover the dispersion matrix $\Sigma_\alpha(\rho)$ in the normal approximation. The next lemma uses relative entropy to bound the deviations of the relevant norms and inner products from their values in the spherical superposition construction (cf.\ Subsection~\ref{sec:spherical}). Define
\begin{align}
    d:=D(P_1\|\frakG_1), \label{eqn:comp_and_div}
\end{align}
the relative entropy of the component output distribution $P_1$ with respect to $\frakG_1$.

\begin{lemma}[Geometry from Gaussian relative entropy] \label{lem:geom} For the component under consideration, with $d$ defined in~\eqref{eqn:comp_and_div},
\begin{align}
    \bbE\left[ \, \bigg|\frac{\|X^n-u^n\|^2}{n}-\alpha \bigg| \, \right] +\bbE\left[  \frac{|(u^n)^\top(X^n-u^n)|}{n} \right] = O\bigg(\sqrt{\frac{d+1}{n}} \bigg) \label{eqn:geometry}
\end{align}
\end{lemma}
   \begin{proof}
Put $V^n:=X^n-u^n$. Since $u^n=\bbE[X^n]$ and
$\|X^n\|^2=n$ almost surely,
$\bbE[\|V^n\|^2]=n-\|u^n\|^2\le n$.
Talagrand's Gaussian transport inequality~\cite[Theorem~1.1]{Talagrand1996} provides a coupling of $Y_1^n-u^n$ and $G^n\sim\calN(0,s_1I_n)$ such that
\begin{align}
    \bbE\big[\|Y_1^n-u^n-G^n\|^2\big]\le2s_1d.
\end{align}
Moreover, $\bbE[\|Y_1^n-u^n\|^2] = \bbE[ \| (X^n-u^n )+ Z_1^n\|^2 ] = \bbE[ \|X^n-u^n\|^2 + \|Z_1^n\|^2] = n-\|u^n\|^2+n\nu_1 \le n(1+\nu_1)$ and $\bbE[\|G^n\|^2]=ns_1$.
Cauchy--Schwarz and $(a+b)^2\le2a^2+2b^2$ therefore give
\begin{align}
    \bbE\big|\|Y_1^n-u^n\|^2-\|G^n\|^2\big| &\le\sqrt{2s_1d}\sqrt{2n(1+\nu_1+s_1)} =O(\sqrt{nd}).
\end{align}
Since $\mathrm{Var}(\|G^n\|^2)=2ns_1^2$, it follows that
\begin{align}
    \bbE\big|\|Y_1^n-u^n\|^2-ns_1\big| =O(\sqrt{nd}+\sqrt n).
\end{align}
Now $\bbE[\|Y_1^n-u^n\|^2-ns_1\mid X^n] =\|V^n\|^2-n\alpha$. Conditional Jensen's inequality thus yields
\begin{align}
    \bbE\left|\frac{\|V^n\|^2}{n}-\alpha\right| \le\frac1n\bbE\big|\|Y_1^n-u^n\|^2-ns_1\big| =O\left(\sqrt{\frac{d+1}{n}}\right). \label{eqn:cond_jens}
\end{align}

Finally, the exact-energy constraint gives $2(u^n)^\top V^n=\bbE[\|V^n\|^2]-\|V^n\|^2$. The triangle inequality and Jensen's inequality imply
\begin{align}
    \bbE\left|\frac{(u^n)^\top V^n}{n}\right|
     \le\frac12\left(
        \bbE\left|\frac{\|V^n\|^2}{n}-\alpha\right|
        +\left|\frac{\bbE[\|V^n\|^2]}n-\alpha\right|
    \right)\le\bbE\left|\frac{\|V^n\|^2}{n}-\alpha\right|. \label{eqn:inn_prod}
\end{align}
Combining the bounds in \eqref{eqn:cond_jens} and \eqref{eqn:inn_prod} proves the lemma.
\end{proof}

\begin{lemma}[Log-likelihood ratio decomposition and conditional covariance bounds]\label{lem:ll_ratio} There are conditionally zero-mean random variables $\xi_{1,x^n}$ and  $\xi_{2,x^n}$ (i.e., $\bbE[ \xi_{j,X^n} |X^n=x^n]=0$), called centered remainder terms, such that  for every transmitted codeword $x^n$, conditionally on $X^n=x^n$, we have
\begin{align}
    J_1 - \barJ_1(x^n) = G_{1,x^n}- \xi_{1,x^n} \quad\mbox{and}\quad J_2 - \barJ_2(x^n) = G_{2,x^n}+\xi_{2,x^n} , \label{eqn:JminusJbar}
\end{align}
where
\begin{align}
    G_{1,x^n} &:=\frac{(x^n-u^n)^\top Z_1^n}{s_1}-\frac{\alpha}{2\nu_1s_1}  (\|Z_1^n\|^2-n\nu_1),  \label{eqn:defG1}\\
    G_{2,x^n} &:= \left( \frac{u^n}{T}- \frac{\bar{\alpha}(x^n-u^n)}{Ts_2}\right)^\top Z_2^n-\frac{\bar{\alpha}}{2Ts_2} (\|Z_2^n\|^2-n\nu_2), \label{eqn:defG2}
\end{align}
and 
\begin{align}
    \bbE_{X^n}\big[  \mathrm{Var}(\xi_{j,X^n} | X^n)\big] = O(d),\qquad j =1,2. \label{eqn:var_bd}
\end{align}
    Put $\Gamma(x^n):=n^{-1}\mathrm{Cov}( (G_{1,x^n}, G_{2,x^n})^\top )$, where the covariance is conditioned on the transmitted codeword, and write $v^n=x^n-u^n$. The  entries of $\Gamma(x^n)$   are 
    \begin{align}
        \Gamma_{11}(x^n) &=  \frac{\nu_1 \|v^n\|^2}{ns_1^2}+\frac{\alpha^2}{2s_1^2 } , \label{eqn:Gamma11}\\
        \Gamma_{22}(x^n) &= \frac{\nu_2}{n}\left\| \frac{u^n}{T}-\frac{\bar{\alpha}v^n}{Ts_2} \right\|^2 + \frac{\bar{\alpha}^2\nu_2^2}{2T^2s_2^2}, \label{eqn:Gamma22}\\
        \Gamma_{12}(x^n) &=\frac{\rho\sqrt{ \nu_1 \nu_2} }{ns_1T}\left( (u^n)^\top v^n- \frac{\bar{\alpha}}{s_2}\|v^n\|^2 \right)+ \frac{\alpha\bar{\alpha}\rho^2\nu_2}{2s_1 T s_2}. \label{eqn:Gamma12}
    \end{align}
    Consequently,
    \begin{align}
        \bbE\big[ \|\Gamma(X^n)-\Sigma_\alpha(\rho)\|_{\mathrm{HS}}  \big]  =O\left(\sqrt{\frac{ d+1}{n}}\right).  \label{eqn:Gamma_Sigma}
    \end{align}
\end{lemma}
\begin{remark}
The quantities $G_{1,x^n}$ and $G_{2,x^n}$ are the zero-mean fluctuations of the auxiliary information densities $\log(\rvW_1^n(Y_1^n|x^n)/g_1(Y_1^n))$ and $\log(g_2(Y_2^n)/q_0(Y_2^n))$, respectively, obtained by subtracting their conditional means given the transmitted codeword $X^n=x^n$. Under the spherical superposition conditions $\|x^n-u^n\|^2=n\alpha$ and $(u^n)^\top(x^n-u^n)=0$, we have $\Gamma(x^n)=\Sigma_\alpha(\rho)$. Thus, $G_{1,x^n}$ and $G_{2,x^n}$ recover the information density fluctuations appearing in the achievability proof.
\end{remark}
\begin{proof}
    Recall from \eqref{eqn:varrho} that $\varrho(x^n):=\bbE \big[\log f(x^n+Z_2^n)\big]$ where $f = p_2/g_2$. Now set
    \begin{align}
        \xi_{2,x^n}:=\log f(x^n+Z_2^n)-\varrho(x^n)\quad\mbox{and}\quad \xi_{1,x^n}:=\mu_\alpha (\Lambda_{1\rightarrow 2}\log f)(x^n+Z_1^n) -\mu_\alpha\varrho(x^n).
    \end{align}
    Expanding the Gaussian density ratios proves \eqref{eqn:JminusJbar}--\eqref{eqn:defG2}. The reverse local logarithmic Sobolev inequality
\cite[Thm.~5.5.2(v), Eqn.~(5.5.6)]{BakryGentilLedoux2014}, applied to the normalized OU transition at time $t=\frac12\log\mu_\alpha$ and integrated against the standard Gaussian distribution, gives
    \begin{align}
        \bbE_{P_2}\big[  \|\nabla \log f(Y_2^n)\|^2 \big]\le\frac{2d}{s_2(\mu_\alpha-1)}. \label{eqn:rev_Sobolev}
    \end{align}
    The conditional   Gaussian Poincar\'e inequality \cite{BLM2013}, together with
$\|\nabla\Lambda_{1\rightarrow2}\log f\|^2
\le\Lambda_{1\rightarrow2}\|\nabla\log f\|^2$
and $P_1\Lambda_{1\rightarrow2}=P_2$, gives
\begin{align}
\bbE_{X^n}\big[\mathrm{Var}(\xi_{2,X^n}| X^n)\big]
&\le \nu_2\bbE_{P_2}\big[\|\nabla\log f(Y_2^n)\|^2\big],\\
\bbE_{X^n}\big[\mathrm{Var}(\xi_{1,X^n}| X^n)\big]
&\le \mu_\alpha^2\nu_1
\bbE_{P_2}\big[\|\nabla\log f(Y_2^n)\|^2\big].
\end{align}
Equation~\eqref{eqn:rev_Sobolev} therefore proves
\eqref{eqn:var_bd}.
    

    Gaussian third moments vanish, so each linear term is uncorrelated with each centered quadratic term. The identities $\mathrm{Cov}(Z_{1,i}, Z_{2,i})=\rho\sqrt{\nu_1\nu_2}$ and $\mathrm{Cov}(Z_{1,i}^2, Z_{2,i}^2)=2\rho^2 \nu_1\nu_2$ give \eqref{eqn:Gamma11}--\eqref{eqn:Gamma12}. Substituting $\|v^n\|^2/n=\alpha$, $\|u^n\|^2/n=\bar{\alpha}$ and $((u^n)^\top  v^n)/n=0$ gives the dispersion matrix $\Sigma_\alpha(\rho)$.  Since $\|x^n\|^2=n$ and $x^n=u^n+v^n$,
\begin{align}
    \frac{\|u^n\|^2}{n}-\bar\alpha =-\left(\frac{\|v^n\|^2}{n}-\alpha\right) -\frac{2(u^n)^\top v^n}{n}.
\end{align}
Consequently, every entry of $\Gamma(x^n)-\Sigma_\alpha(\rho)$ is bounded in absolute value by a constant times $\big|\|v^n\|^2/n-\alpha\big| +|(u^n)^\top v^n|/n$. Taking expectations and applying Lemma~\ref{lem:geom} proves \eqref{eqn:Gamma_Sigma}.
\end{proof}

\subsection{Combining the Component Bounds and Normal Approximation} \label{sec:combining}

We now combine the component bounds from Subsection~\ref{sec:aux} and restore the component subscripts
$\lambda$. For the deterministic partition
$\calI=\sqcup_\lambda\calI_\lambda$, let $J=J(W)$ denote
the component label, so that $J=\lambda$ when
$W\in\calI_\lambda$. Write $M_\lambda=|\calI_\lambda|$
and $p_\lambda=M_\lambda/M$.  
For $j=1,2$, let $P_{j,\lambda}:=P_{Y_j^n|J=\lambda}$ have density $p_{j,\lambda}$. The receiver-2 output distribution of the entire selected subcode is
\begin{align}
    P_2:=\frac1M\sum_{w\in\calI}\rvW_2^n(\fndot| x^n(w,k)) =\sum_\lambda p_\lambda P_{2,\lambda},
\end{align}
with density $p_2=\sum_\lambda p_\lambda p_{2,\lambda}$.

For each component, put $u_\lambda^n:=\bbE[X^n\mid J=\lambda]$ and use $d_\lambda$, $J_{j,\lambda}$, $G_{j,x^n,\lambda}$, and $\xi_{j,x^n,\lambda}$ for the quantities defined in Subsection~\ref{sec:aux}, evaluated with $W$ uniform on $\calI_\lambda$ and the outputs generated according to the original joint Gaussian channel.

Recall from \eqref{eqn:success_sel} that the selected subcode's joint probability of successful decoding is
\begin{align}
    \frakP_{\mathrm{succ}}
    :=\Pr(Y_1^n\in\calA_W,\;Y_2^n\in\calB) =\frac{1}{M}\sum_{m_1 \in\calI}s(m_1, k).
\end{align}
\begin{lemma}[Information spectrum-type converse bound] \label{lem:info_spec}
    For each component $\calI_\lambda$, let $\tilq_{1,\lambda}$ be a positive subprobability density and define  the modified information densities
    \begin{align}
        \jmath_1 := \log\frac{\rvW_1^n(Y_1^n|X^n)}{\tilq_{1,J}(Y_1^n)}\quad\mbox{and}\quad \jmath_2:=\log\frac{p_2(Y_2^n)}{q_0(Y_2^n)}.
    \end{align}
    Then, for every $\gamma>0$,
    \begin{align}
         \frakP_{\mathrm{succ}}\le\Pr\left( \begin{bmatrix}
             \jmath_1 \\ \jmath_2 
         \end{bmatrix}\ge\begin{bmatrix}
             \log M_J-\gamma \\ -\log q-\gamma
         \end{bmatrix} \right)+  2\rme^{-\gamma}, \label{eqn:fb_conv}
    \end{align}
    where the vector inequality is to be interpreted in a coordinate-wise manner. 
\end{lemma}

\begin{proof}[Proof of Lemma~\ref{lem:info_spec}]
    Conditioned on $J=\lambda$, message 1 is uniform on $\calI_\lambda$. Since the decoding sets $\calA_w$ are disjoint,  using the definition of $\jmath_1$, 
    \begin{align}
        \Pr(Y_1^n\in\calA_W,~\jmath_1<\log M_\lambda-\gamma \mid J=\lambda) \le\frac{1}{M_\lambda}\sum_{w\in\calI_\lambda}M_\lambda \rme^{-\gamma}\tilQ_{1,\lambda}(\calA_w)\le \rme^{-\gamma}.
    \end{align}
    Averaging over $J$ preserves the upper bound. At receiver 2, using the definition of $\jmath_2$,
    \begin{align}
        \Pr(Y_2^n\in\calB,~\jmath_2 < -\log q-\gamma)\le \rme^{-\log q-\gamma}Q_0(\calB) = \rme^{-\gamma},
    \end{align}
    where the last step follows from $q=Q_0(\calB)$ (cf.\ \eqref{eqn:sel3}). Putting these two bounds together and using the definition of $\frakP_{\mathrm{succ}}$ proves the non-asymptotic converse in \eqref{eqn:fb_conv}.
\end{proof}

\begin{remark}[Relation to  existing information spectrum converses]
Lemma \ref{lem:info_spec} uses the change-of-measure and decoding-set arguments underlying the converses of Verd\'u and Han~\cite{VH94}, Hayashi and Nagaoka~\cite{Hayashi03}, and Boucheron and Salamatian~\cite{bouch00}. Its form reflects the selected subcode: omitting the common
slack $\gamma$, receiver 1 has threshold \(\log M_J\), whereas receiver 2 has threshold \(-\log Q_0(\mathcal B)\ge\log M_{2,n}-O(1)\), with its message fixed and \(\mathcal B\) a common decoding set. The receiver-2 statistic compares the full-subcode output density \(p_2\) with \(q_0\).

Averaged output densities~\cite{bouch00} and arbitrary auxiliary measures~\cite{Hayashi03} already appear in these earlier  information spectrum converses. To evaluate the present bound on the second-order scale, we use the partition's quantitative Gaussian comparisons (in \eqref{eqn:sublinear_div}) and conditional entropy bound (in \eqref{eqn:small_ce}), together with the construction of $\tilde q_{1,\lambda}$ from $p_{2,\lambda}/g_{2,\lambda}$. Lemma~\ref{lem:output} shows that the conditional means of the terms involving $\log(p_{2,\lambda}/g_{2,\lambda})$ cancel in the weighted sum, while Lemma~\ref{lem:ll_ratio} controls their fluctuations.
\end{remark}

We now use Lemma~\ref{lem:info_spec} and the component estimates from Subsection~\ref{sec:aux} to obtain a Gaussian upper bound on the joint probability of successful decoding.

\begin{proposition}[Normal approximation for selected subcode]
\label{prop:conv_normal}
Fix $\alpha\in[0,1]$, $\tau\in(0,1/2)$, and $0\le\kappa<\infty$. There exist constants $0<A<\infty$ and $n_0\in\bbN$, depending only on $\alpha$, $\nu_1$, $\nu_2$, $\tau$, and $\kappa$, such that the following holds for every $n\ge n_0$ and $\rho\in[-1,1]$. Consider a selected subcode whose codewords have energy $n$, which satisfies \eqref{eqn:images} with $e\in[0,1-\tau]$, and for which $\Xi\ge nC_\alpha-\kappa\sqrt n$ (cf.~Lemma~\ref{lem:sel}, Remark~\ref{rmk:Xi} and \eqref{eqn:lower_bd_Xi}).
 Then
\begin{align}
    \frakP_{\mathrm{succ}}
    \le
    \rvQ_{\alpha,\rho}\left(\frac{\Xi-nC_\alpha}{\sqrt n}-r_n \right)+r_n,
    \qquad r_n:=A n^{-1/48},
\end{align}
where $\rvQ_{\alpha,\rho} (\fndot)$ was defined in \eqref{eqn:Falpharho}.
\end{proposition}

\begin{proof}
{\em 1. The partition and the decomposition of the log-likelihood ratios $\jmath_1$ and $\jmath_2$:}

Apply Lemma \ref{lem:defi}, using the marginal error upper bounds in~\eqref{eqn:images} with $e\le 1-\tau$. It gives that the entropy deficit is $\Delta=O(\sqrt{n})$. Recall that Proposition \ref{prop:partition} then supplies, via the SFRL, a partition of the retained indices $\calI = \sqcup_\lambda\calI_\lambda$ such that $H(J|Y_2^n) = O(n^{1/3})$ and $\bard = \sum_\lambda p_\lambda d_\lambda=O(n^{5/6}).$ For each component, choose the auxiliary density $\tilq_{1,\lambda}$ as in Lemma~\ref{lem:output}.

For each component $\lambda$, define its posterior probability
\begin{align}
    \pi_\lambda(y^n)
    &:=\Pr(J=\lambda| Y_2^n=y^n)
      =\frac{p_\lambda p_{2,\lambda}(y^n)}
             {\sum_{\lambda'}p_{\lambda'}p_{2,\lambda'}(y^n)}
      =\frac{p_\lambda p_{2,\lambda}(y^n)}{p_2(y^n)}. \label{eqn:bayes}
\end{align}
Rearranging this identity and taking logarithms gives
\begin{align}
    \log\frac{p_2(y^n)}{q_0(y^n)}
    &=
    \log\frac{p_{2,\lambda}(y^n)}{q_0(y^n)}
    +\log p_\lambda-\log\pi_\lambda(y^n). \label{eqn:after_bayes}
\end{align}
Define the nonnegative random variable
$\Upsilon :=-\log\pi_J(Y_2^n)$. Conditionally on $W=w$ with $J(w)=\lambda$, using~\eqref{eqn:after_bayes}, we therefore have
\begin{align}
    \jmath_2
    =J_{2,\lambda}(Y_2^n)+\log p_\lambda+\Upsilon,
\end{align}
and $\bbE[\Upsilon]=H(J|Y_2^n)$. Abbreviate $m_j(w)=\barJ_{j,J(w)}(x^n(w,k))$ and also write    $G_j:=G_{j,X^n,J}$  and $\xi_j:=\xi_{j,X^n,J}$.
 (see Lemma~\ref{lem:ll_ratio} to recall definitions). The decomposition in  Lemma \ref{lem:ll_ratio} can be written as
\begin{align}
    \jmath_1=m_1(W)+G_1-\xi_1\quad\mbox{and}\quad\jmath_2= m_2(W)+G_2+\xi_2+\log p_J+\Upsilon. 
\end{align}
Since $\bbE[\xi_j|W]=0$,  from \eqref{eqn:var_bd}, averaging   gives $\bbE[\xi_j^2]=O(\bard)$. Take 
\begin{align}
    \gamma=\log n,\qquad \tau_n=n^{-1/48},\qquad\ell_n = \sqrt{n}\tau_n=n^{23/48}.\label{eqn:sequences}
\end{align}
Chebyshev's and Markov's inequalities yield
    \begin{align}
        \Pr\Big( \max\{|\xi_1|,|\xi_2|,\Upsilon\}>\ell_n \Big) = O\bigg( \frac{\bard}{\ell_n^2}   + \frac{H(J|Y_2^n)}{\ell_n}\bigg)= O(n^{-1/8}). \label{eqn:cheby_markov}
    \end{align}
    
    On the complement of the   event in~\eqref{eqn:cheby_markov}, the event in the probability in~\eqref{eqn:fb_conv}  implies $G_1\ge t_1(W)$ and $G_2\ge t_2(W)$ where 
    \begin{align}
        t_1(w) &:=\log M+\log p_{J(w)}-\gamma-m_1(w)-\ell_n,\\ 
        t_2(w) &:=-\log q-\log p_{J(w)}-\gamma-m_2(w)-2\ell_n.
    \end{align}
    The cancellation  in \eqref{eqn:cancellation} and $\mu_\alpha>1$ imply that, for every message, 
    \begin{align}
        t_1(w)+\mu_\alpha t_2(w)&=\Xi-m_1(w)-\mu_\alpha m_2(w)+ (1-\mu_\alpha)\log p_{J(w)}-(1+\mu_\alpha)\gamma-(1+2\mu_\alpha)\ell_n\\
        &\ge \Xi-nC_\alpha-(1+\mu_\alpha)\gamma-(1+2\mu_\alpha)\ell_n. \label{eqn:reduction}
    \end{align}
    Define the sequences
    \begin{align}
        w_n := \frac{\Xi-nC_\alpha}{\sqrt{n}}\quad\mbox{and}\quad h_n := \frac{ (1+\mu_\alpha)\gamma+(1+2\mu_\alpha)\ell_n}{\sqrt{n}}=O(n^{-1/48}).
    \end{align}
    Thus, \eqref{eqn:fb_conv} and \eqref{eqn:cheby_markov} reduce the proof to bounding $\Pr(G_1\ge t_1(W), G_2\ge t_2(W))$, where the normalized thresholds have weighted sum at least $w_n-h_n$.

{\em 2. Boundary points with $0<\alpha<1$:} 
    First, suppose that the first-order rate pair is on the interior of the boundary of $\calC$, i.e., $0<\alpha<1$ so $\Sigma:=\Sigma_{\alpha} (\rho)$ is positive definite by Proposition~\ref{prop:Fproperty}. Let $\Gamma_w$ be the covariance matrix of the random vector $( G_1, G_2)^\top/\sqrt{n}$ conditioned on $W=w$. Averaging  \eqref{eqn:Gamma_Sigma} over the components and applying Jensen's inequality yields
    \begin{align}
        \bbE\big[ \|\Gamma_W-\Sigma_\alpha(\rho)\|_{\mathrm{HS}}  \big]  =O\left(\sqrt{\frac{\bard+1}{n}}\right)=O(n^{-1/12}). \label{eqn:cov_average}
    \end{align}
    Call a message  $w\in\calI$ ``good'' if $\|\Gamma_w-\Sigma\|_{\mathrm{HS}} \le\lambda_{\min}(\Sigma)/2$. By Markov's inequality and~\eqref{eqn:cov_average}, the probability that~$W$ is not good is $O(n^{-1/12})$.  For good messages, the covariance eigenvalues are bounded above and away from zero by fixed constants.

    For each good message $w$, let $\rvZ_w=(\rvZ_{1,w},\rvZ_{2,w})^\top \sim\calN(0,\Gamma_w)$. The multivariate Berry--Esseen theorem \cite[Theorem~1.1]{Ben05} gives 
    \begin{align}
    \sup_{z_1,z_2\in\bbR}
    \bigg|
        \Pr\bigg(
            \frac{G_1}{\sqrt n}\ge z_1,\,
            \frac{G_2}{\sqrt n}\ge z_2
            \,\bigg|\,W=w
        \bigg)
        -\Pr(\rvZ_{1,w}\ge z_1,\rvZ_{2,w}\ge z_2)
    \bigg|
    = O\bigg(\frac{1}{\sqrt n}\bigg).
    \label{eqn:be_good}
    \end{align}
    The linear coefficient vectors in $G_1$ and $G_2$ have norms $O(\sqrt n)$, while their quadratic coefficients are bounded. A common orthogonal change of coordinates makes the coordinates of both linear coefficient vectors uniformly bounded: map an orthonormal basis of their span to orthonormal vectors with entries $O(n^{-1/2})$. By joint rotational invariance of the Gaussian noises, conditionally on $W=w$, the vector $(G_1,G_2)^\top$ is therefore distributed as a sum of $n$ independent centered random vectors with uniformly bounded third absolute moments. Together with the lower bound on $\lambda_{\min}(\Gamma_w)$ for good messages, this justifies the uniform $O(n^{-1/2})$ bound in~\eqref{eqn:be_good}.
 
     We next approximate $\Gamma_w$ with $\Sigma$. For good messages, the Gaussian relative entropy formula gives 
     \begin{align}
    D\big(\calN(0,\Gamma_w)\,\big\|\,\calN(0,\Sigma)\big)
    &=
    \frac12\left[
        \tr(\Sigma^{-1}\Gamma_w)-2
        -\log\det(\Sigma^{-1}\Gamma_w)
    \right] = O\big(\|\Gamma_w-\Sigma\|_{\mathrm{HS}}^2\big).
\end{align}
Indeed, the eigenvalues of $\Sigma^{-1/2}\Gamma_w\Sigma^{-1/2}$ lie in $[1/2,3/2]$, on which $x-1-\log x\le  (x-1)^2$. Pinsker's inequality therefore bounds the total variation distance between these two Gaussian distributions by $O(\|\Gamma_w-\Sigma\|_{\mathrm{HS}})$.

Let $\rvZ=(\rvZ_1,\rvZ_2)^\top\sim\calN(0,\Sigma)$. Combining this comparison with \eqref{eqn:be_good}, averaging over good messages, and bounding the probability difference by one for the remaining messages yields
\begin{align}
    &\bbE_W\left[
        \sup_{z_1,z_2\in\bbR}
        \bigg|
            \Pr\bigg(
                \frac{G_1}{\sqrt n}\ge z_1,\,
                \frac{G_2}{\sqrt n}\ge z_2
                \,\bigg|\,W
            \bigg)
            -\Pr(\rvZ_1\ge z_1,\rvZ_2\ge z_2)
        \bigg|
    \right] \notag\\
    &\qquad\le
    O\bigg(\frac{1}{\sqrt n}\bigg)
    +O\Big(\bbE\big[\|\Gamma_W-\Sigma\|_{\mathrm{HS}}\big]\Big)
    +\Pr(W\text{ is not good})
    =O(n^{-1/12}).
    \label{eqn:averaged_normal}
\end{align}
All constants are uniform over the good messages and $\rho\in[-1,1]$, for fixed $\alpha\in(0,1)$. Since the supremum is inside the expectation, \eqref{eqn:averaged_normal} applies also to thresholds depending on the transmitted message. The definition of $\rvQ_{\alpha,\rho}$ in \eqref{eqn:Falpharho} gives 
\begin{align}
    \frakP_{\mathrm{succ}}\le\rvQ_{\alpha,\rho}(w_n-h_n) +O(n^{-1/12}). \label{eqn:psucc1}
\end{align}

{\em 3. Endpoints of $\calC$:} The same reduction through \eqref{eqn:reduction} applies. At $\alpha=0$ and $\alpha=1$, respectively, the first and second coordinates of the Gaussian vector with covariance $\Sigma_\alpha(\rho)$ are zero almost surely. We therefore use the scalar Berry--Esseen theorem for the coordinate with positive limiting variance and Chebyshev's inequality for the other coordinate. 

We first control the coordinate with zero limiting variance. Since the codewords have exact energy $n$, 
\begin{align}
    \alpha=0~&: \qquad\bbE [G_1^2]=\frac{1}{\nu_1}\bbE\big[ \|X^n-u_J^n\|^2 \big]=\frac{n}{\nu_1}\sum_\lambda p_\lambda\left|  \frac{\|u_\lambda^n\|^2}{n}-1 \right|,\\
    \alpha=1~&:\qquad \bbE[G_2^2] =\frac{\nu_2}{T^2}\bbE\big[\|u_J^n\|^2\big].
\end{align}
Both quantities grow as $O(n^{11/12})$ by~\eqref{eqn:comp_means}.  Using the same $\tau_n$ as defined in \eqref{eqn:sequences}, the coordinate with zero limiting variance, denoted $G_{j_0}$, satisfies
\begin{align}
    \Pr( |G_{j_0}|>\sqrt{n}\tau_n) = O(n^{-1/12}/\tau_n^2)= O(n^{-1/24}). \label{eqn:chebyG}
\end{align}
The averaged covariance estimate \eqref{eqn:cov_average} holds at the endpoints as well. For the other coordinate, this estimate and the univariate Berry–Esseen bound give the corresponding averaged uniform scalar Gaussian approximation with error $O(n^{-1/12})$ (similar to \eqref{eqn:averaged_normal}). This follows by the same good-message and Gaussian total variation argument as above, now in a single dimension  and with the positive limiting variance $V_2(0)=\rvV(S_2)$ at $\alpha=0$
and $V_1(1)=\rvV(S_1)$ at $\alpha=1$ (recall definitions in \eqref{eqn:disp1} and \eqref{eq:dispersion-one-two}).

At $\alpha=0$, suppose that both threshold tests $G_1\ge t_1(W)$ and $G_2\ge t_2(W)$ hold and that $|G_1|\le\sqrt n\,\tau_n$. Then, clearly, $t_1(W)/\sqrt{n}\le\tau_n$. Hence, $G_2/\sqrt{n}\ge (w_n-h_n-\tau_n)/\mu_0$ by \eqref{eqn:reduction}. By the definition of $\rvQ_{0,\rho}$  in~\eqref{eqn:F_endpoints}, \eqref{eqn:chebyG} and the univariate Berry--Esseen bound,
\begin{align}
    \frakP_{\mathrm{succ}} \le \rvQ_{0,\rho}(w_n-h_n-\tau_n)+O(n^{-1/24}). \label{eqn:psucc2}
\end{align}
In an analogous manner, at $\alpha=1$,
\begin{align}
    \frakP_{\mathrm{succ}} \le \rvQ_{1,\rho}(w_n-h_n-\mu_\alpha\tau_n)+O(n^{-1/24}). \label{eqn:psucc3}
\end{align}

The bounds in \eqref{eqn:psucc1}, \eqref{eqn:psucc2} and \eqref{eqn:psucc3} prove Proposition \ref{prop:conv_normal}.
\end{proof}

\subsection{Completion of the Second-Order Converse }\label{sec:completion_conv}

Let a sequence of codes satisfy condition \eqref{eqn:second-order_def} with $\alpha\in (0,1)$ and with second-order coding rates $(L_1, L_2)$. Fix $\xi,\zeta>0$ with $\eps+\zeta<1$ as in Subsection~\ref{sec:reduction}.  For every retained message, both conditional marginal success probabilities are at least $1-\eps-\zeta$ (cf.~\eqref{eqn:images}). Lemma \ref{lem:sel} provides a subcode with 
\begin{align}
    \frakP_{\mathrm{succ}} \ge 1-\eps-\zeta\quad\mbox{and}\quad \Xi\ge \log M_{1,n}+\mu_\alpha\log M_{2,n}-O(1).
\end{align}
Both marginal success probabilities are uniformly positive (cf.~\eqref{eqn:images}), and the rate assumptions in~\eqref{eqn:rate_assump}, together with   the lower bound on $\Xi$ in~\eqref{eqn:sel_final}, imply that $\Xi\ge nC_\alpha-O(\sqrt{n})$. Proposition~\ref{prop:conv_normal}, together with the monotonicity of $\rvQ_{\alpha,\rho}$, then gives
\begin{align}
    1-\eps-\zeta\le\rvQ_{\alpha,\rho}\big(L_1 + \mu_\alpha L_2-(1+\mu_\alpha)\xi-o(1) \big) + o(1).
\end{align}
First let $n\to\infty$, then $\xi\downarrow0$. Continuity and strict monotonicity of $\rvQ_{\alpha,\rho}$ (cf.\ Proposition~\ref{prop:Fproperty}) yield $L_1 +\mu_\alpha L_2\le \rvG_{\alpha,\rho}(\eps+\zeta)$. Let $\zeta\downarrow0$. The continuity of $\rvG_{\alpha,\rho}$ proves the  converse for $\alpha\in (0,1)$.

At $\alpha=0$, the rate assumption and $\log M_{1,n}\ge0$ give $\log M_{1,n}/\sqrt n\ge L_1^+-\xi$ for all sufficiently large $n$. Applying the same argument with
$\rvQ_{0,\rho}$ as defined in~\eqref{eqn:F_endpoints} gives  
\begin{align}
    L_1^++\frac{S_1}{S_2}L_2\le\frac{S_1}{S_2}\sqrt{\rvV(S_2)}\Phi^{-1}(\eps).
\end{align}
The conclusion at $\alpha=1$ is analogous. This completes the proof of the second-order converse.

\section{Conclusion and Future Work}
Understanding how to derive second-order asymptotics for network information theory problems whose first-order characterizations involve auxiliary random variables has been a long-standing quest of the author since he began working in this area in the early 2010s. The present work joins recent advances that establish matching second-order bounds for problems with auxiliaries, including source coding with side information~\cite{Cao2026WynerAhlswedeKorner,Zhou2026WynerZiv}. The author finds the resolution of the Gaussian broadcast problem particularly satisfying.   Early work with Kosut established  second-order achievability bounds for the asymmetric broadcast channel~\cite{TK12}, while subsequent work with Scarlett characterized the second-order region of the Gaussian multiple-access channel with degraded message sets~\cite{ST2015}, using spherical superposition coding techniques that also underlie the achievability proof in the present work.  Subsection~\ref{sec:noise} also shows how noise correlation affects the second-order region. In particular, it is interesting that, at each interior point of the capacity boundary, the physically degraded channel has the smallest second-order region among Gaussian broadcast channels with the same receiver SNRs.

The main message of this paper is that the SFRL~\cite{LiElGamal2018} offers another potentially useful ingredient in strong and second-order converses, alongside tools such as the blowing-up lemma~\cite{AhlswedeGacsKorner1976}, the Gaussian Poincar\'e inequality~\cite{FongTan2017GaussianBC}, reverse hypercontractivity~\cite{LiuVanHandelVerdu2019}, exponent methods~\cite{Oohama2015DBC}, edge removal~\cite{KosutKliewer2019}, and change of measure~\cite{TyagiWatanabe2020}. Its role here is to convert a suitably constructed auxiliary channel into a deterministic partition of the retained message set. The component label plays the role of the auxiliary random variable, with quantitative control of its conditional entropy given the receiver-2 output and the average relative entropy of the component output distributions from the corresponding Gaussian distributions.  

Several key arguments in this paper are highly specific to the Gaussian setting, particularly the Ornstein--Uhlenbeck entropy dissipation and Gaussian functional inequalities~\cite{BakryGentilLedoux2014} used to construct and analyze the auxiliary channel and observation. It is not  clear how to adapt these arguments to other degraded broadcast channels, such as the binary symmetric broadcast channel. Finding suitable replacements, and extending the approach to two-receiver discrete memoryless broadcast channels that are degraded, less noisy, or more capable~\cite[Sec.~5.6]{elgamal}, would be a fruitful endeavor.  Their   capacity regions are   known, and representations in terms of upper concave envelopes~\cite{Nair2013ConcaveEnvelopes} provide promising starting points for studying the optimizing auxiliaries. The challenge is to turn these first-order descriptions into quantitative distributional estimates that preserve the relevant second-order fluctuations.

\section*{Acknowledgments}
\emph{Generative-AI use disclosure.}
 The author used OpenAI Codex (GPT-6 Astra, Ultra mode) and Anthropic's Claude (Opus 5) during the development and preparation of this manuscript. These systems assisted in exploring and stress-testing proof strategies, formulating and refining intermediate lemmas, checking algebraic and probabilistic calculations, identifying possible logical gaps, and refining portions of the exposition. The author wrote the manuscript, independently verified every mathematical statement, proof, and reference,  and assumes full responsibility for its contents.

 The author thanks Recep Can Yavas and Jonathan Scarlett for their constructive comments, which helped improve the manuscript.

\bibliographystyle{ieeetr} 
\bibliography{isitbib_uniform}

@STRING{ISIT = {Proc. Int. Symp. Inform. Theory}}

@STRING{TIT = {IEEE Trans. Inform. Theory}}

@article{Bergmans1973,
  author       = {Patrick P. Bergmans},
  title        = {Random coding theorem for broadcast channels with degraded components},
  journal      = TIT,
  volume       = {19},
  number       = {2},
  pages        = {197--207},
  month        = mar,
  year         = {1973},
  doi          = {10.1109/TIT.1973.1054980}
}

@article{TuninettiEtal2023,
  author       = {Daniela Tuninetti and Paul Sheldon and Besma Smida and Natasha Devroye},
  title        = {On second order rate regions for the static scalar {Gaussian} broadcast channel},
  journal      = {IEEE J. Sel. Areas Commun.},
  volume       = {41},
  number       = {7},
  pages        = {1982--1999},
  month        = jul,
  year         = {2023},
  doi          = {10.1109/JSAC.2023.3280991}
}

@article{Bergmans1974,
  author       = {Patrick P. Bergmans},
  title        = {A simple converse for broadcast channels with additive white {Gaussian} noise},
  journal      = TIT,
  volume       = {20},
  number       = {2},
  pages        = {279--280},
  month        = mar,
  year         = {1974},
  doi          = {10.1109/TIT.1974.1055184}
}

@article{Talagrand1996,
  author       = {Michel Talagrand},
  title        = {Transportation cost for {Gaussian} and other product measures},
  journal      = {Geom. Funct. Anal.},
  volume       = {6},
  number       = {3},
  pages        = {587--600},
  month        = may,
  year         = {1996},
  doi          = {10.1007/BF02249265}
}

@misc{HuangLiuLiu2026,
  author       = {Huang, Mian and Liu, Yanxiao and Liu, Yi},
  title        = {Sub-optimality of {Marton}'s Inner Bound for the Two-Receiver Broadcast Channel},
  month        = aug,
  year         = {2026},
  howpublished = {arXiv:2608.19869},
  doi          = {10.48550/arXiv.2608.19869},
  url          = {https://arxiv.org/abs/2608.19869},
  eprint       = {2608.19869},
  archiveprefix = {arXiv},
  primaryclass = {cs.IT}
}

@article{Gallager1974,
  author       = {Gallager, R. G.},
  title        = {Capacity and Coding for Degraded Broadcast Channels},
  journal      = {Probl. Inf. Transm.},
  volume       = {10},
  number       = {3},
  pages        = {185--193},
  year         = {1974}
}

@article{Liu2021WAK,
  author       = {Liu, Jingbo},
  title        = {Dispersion Bound for the {Wyner--Ahlswede--K{\"o}rner} Network via a Semigroup Method on Types},
  journal      = TIT,
  volume       = {67},
  number       = {2},
  pages        = {869--885},
  month        = feb,
  year         = {2021},
  doi          = {10.1109/TIT.2020.3041791}
}

@misc{Zhou2026WynerZiv,
  author       = {Zhou, Lin},
  title        = {Exact Second-Order Asymptotics for the Discrete {Wyner--Ziv} Problem},
  month        = oct,
  year         = {2026},
  howpublished = {arXiv:2610.11998},
  doi          = {10.48550/arXiv.2610.11998},
  url          = {https://arxiv.org/abs/2610.11998},
  eprint       = {2610.11998},
  archiveprefix = {arXiv},
  primaryclass = {cs.IT}
}

@article{Nair2013ConcaveEnvelopes,
  author       = {Nair, Chandra},
  title        = {Upper Concave Envelopes and Auxiliary Random Variables},
  journal      = {Int. J. Adv. Eng. Sci. Appl. Math.},
  volume       = {5},
  number       = {1},
  pages        = {12--20},
  month        = mar,
  year         = {2013},
  doi          = {10.1007/s12572-013-0081-7}
}

@misc{Cao2026WynerAhlswedeKorner,
  author       = {Cao, Daming},
  title        = {Exact Second-Order Asymptotics for the {Wyner--Ahlswede--K{\"o}rner} Problem},
  month        = aug,
  year         = {2026},
  howpublished = {arXiv:2608.21991},
  doi          = {10.48550/arXiv.2608.21991},
  url          = {https://arxiv.org/abs/2608.21991},
  eprint       = {2608.21991},
  archiveprefix = {arXiv},
  primaryclass = {cs.IT}
}

@article{Ben05,
  author       = {V. Bentkus},
  title        = {A {Lyapunov}-type bound in {$\mathbb{R}^d$}},
  journal      = {Theory Probab. Appl.},
  volume       = {49},
  number       = {2},
  pages        = {311--323},
  month        = jan,
  year         = {2005},
  doi          = {10.1137/S0040585X97981123}
}

@article{LiElGamal2018,
  author       = {Li, Cheuk Ting and El Gamal, Abbas},
  title        = {Strong functional representation lemma and applications to coding theorems},
  journal      = TIT,
  volume       = {64},
  number       = {11},
  pages        = {6967--6978},
  month        = nov,
  year         = {2018},
  doi          = {10.1109/TIT.2018.2865570}
}

@article{GuoShamaiVerdu2005,
  author       = {Dongning Guo and Shlomo Shamai and Sergio Verd{\'u}},
  title        = {Mutual information and minimum mean-square error in {Gaussian} channels},
  journal      = TIT,
  volume       = {51},
  number       = {4},
  pages        = {1261--1282},
  month        = apr,
  year         = {2005},
  doi          = {10.1109/TIT.2005.844072}
}

@book{BakryGentilLedoux2014,
  author       = {Bakry, Dominique and Gentil, Ivan and Ledoux, Michel},
  title        = {Analysis and Geometry of {Markov} Diffusion Operators},
  series       = {Grundlehren der mathematischen Wissenschaften},
  volume       = {348},
  publisher    = {Springer},
  address      = {Cham},
  year         = {2014},
  doi          = {10.1007/978-3-319-00227-9}
}

@book{BLM2013,
  author       = {Boucheron, St{\'e}phane and Lugosi, G{\'a}bor and Massart, Pascal},
  title        = {Concentration Inequalities: A Nonasymptotic Theory of Independence},
  publisher    = {Oxford University Press},
  address      = {Oxford},
  year         = {2013},
  doi          = {10.1093/acprof:oso/9780199535255.001.0001}
}

@article{DytsoPoorShamai2023,
  author       = {Dytso, Alex and Poor, H. Vincent and {Shamai (Shitz)}, Shlomo},
  title        = {Conditional Mean Estimation in {Gaussian} Noise: A Meta Derivative Identity With Applications},
  journal      = TIT,
  volume       = {69},
  number       = {3},
  pages        = {1883--1898},
  month        = mar,
  year         = {2023},
  doi          = {10.1109/TIT.2022.3216012}
}

@article{bouch00,
  author       = {S. Boucheron and M. R. Salamatian},
  title        = {About priority encoding transmission},
  journal      = TIT,
  volume       = {46},
  number       = {2},
  pages        = {699--705},
  month        = mar,
  year         = {2000},
  doi          = {10.1109/18.825846}
}

@article{MolavianJaziLaneman2015,
  author       = {E. MolavianJazi and J. N. Laneman},
  title        = {A Second-Order Achievable Rate Region for {Gaussian} Multi-Access Channels via a Central Limit Theorem for Functions},
  journal      = TIT,
  volume       = {61},
  number       = {12},
  pages        = {6719--6733},
  month        = dec,
  year         = {2015},
  doi          = {10.1109/TIT.2015.2492547}
}

@article{ScarlettMartinezGuillen2015,
  author       = {J. Scarlett and A. Martinez and A. {Guill{\'e}n i F{\`a}bregas}},
  title        = {Second-Order Rate Region of Constant-Composition Codes for the Multiple-Access Channel},
  journal      = TIT,
  volume       = {61},
  number       = {1},
  pages        = {157--172},
  month        = jan,
  year         = {2015},
  doi          = {10.1109/TIT.2014.2371026}
}

@book{Petrov,
  author       = {Valentin V. Petrov},
  title        = {Sums of Independent Random Variables},
  series       = {Ergebnisse der Mathematik und ihrer Grenzgebiete},
  volume       = {82},
  publisher    = {Springer-Verlag},
  address      = {Berlin, Heidelberg},
  year         = {1975},
  doi          = {10.1007/978-3-642-65809-9}
}

@article{ST2015,
  author       = {J. Scarlett and V. Y. F. Tan},
  title        = {Second-order asymptotics for the {Gaussian MAC} with degraded message sets},
  journal      = TIT,
  volume       = {61},
  number       = {12},
  pages        = {6700--6718},
  month        = dec,
  year         = {2015},
  doi          = {10.1109/TIT.2015.2487340}
}

@article{yavas21,
  author       = {R. C. Yavas and V. Kostina and M. Effros},
  title        = {{Gaussian} multiple and random access channels: Finite-blocklength analysis},
  journal      = TIT,
  volume       = {67},
  number       = {11},
  pages        = {6983--7009},
  month        = nov,
  year         = {2021},
  doi          = {10.1109/TIT.2021.3111676}
}

@article{cover72,
  author       = {Cover, Thomas M.},
  title        = {Broadcast channels},
  journal      = TIT,
  volume       = {18},
  number       = {1},
  pages        = {2--14},
  month        = jan,
  year         = {1972}
}

@book{Csi97,
  author       = {Csisz{\'a}r, Imre and K{\"o}rner, J{\'a}nos},
  title        = {Information Theory: Coding Theorems for Discrete Memoryless Systems},
  edition      = {2nd},
  publisher    = {Cambridge University Press},
  address      = {Cambridge, U.K.},
  year         = {2011}
}

@article{FongTan2017GaussianBC,
  author       = {Silas L. Fong and Vincent Y. F. Tan},
  title        = {A Proof of the Strong Converse Theorem for {Gaussian} Broadcast Channels via the {Gaussian} {Poincar\'e} Inequality},
  journal      = TIT,
  volume       = {63},
  number       = {12},
  pages        = {7737--7746},
  month        = dec,
  year         = {2017},
  doi          = {10.1109/TIT.2017.2753224}
}

@book{elgamal,
  author       = {El Gamal, Abbas and Kim, Young-Han},
  title        = {Network Information Theory},
  publisher    = {Cambridge University Press},
  address      = {Cambridge, U.K.},
  year         = {2012}
}

@article{ZamaniOechteringSkoglund2024,
  author       = {Zamani, Amirreza and Oechtering, Tobias J. and Skoglund, Mikael},
  title        = {On the Privacy-Utility Trade-Off With and Without Direct Access to the Private Data},
  journal      = TIT,
  volume       = {70},
  number       = {3},
  pages        = {2177--2200},
  month        = mar,
  year         = {2024},
  doi          = {10.1109/TIT.2023.3326070}
}

@article{TyagiWatanabe2020,
  author       = {Tyagi, Himanshu and Watanabe, Shun},
  title        = {Strong Converse Using Change of Measure Arguments},
  journal      = TIT,
  volume       = {66},
  number       = {2},
  pages        = {689--703},
  month        = feb,
  year         = {2020},
  doi          = {10.1109/TIT.2019.2953877}
}

@article{KosutKliewer2019,
  author       = {Kosut, Oliver and Kliewer, J{\"o}rg},
  title        = {Strong Converses Are Just Edge Removal Properties},
  journal      = TIT,
  volume       = {65},
  number       = {6},
  pages        = {3315--3339},
  month        = jun,
  year         = {2019},
  doi          = {10.1109/TIT.2018.2889687}
}

@article{LiuVanHandelVerdu2019,
  author       = {Liu, Jingbo and van Handel, Ramon and Verd{\'u}, Sergio},
  title        = {Second-Order Converses via Reverse Hypercontractivity},
  journal      = {Math. Stat. Learn.},
  volume       = {2},
  number       = {2},
  pages        = {103--163},
  year         = {2019},
  doi          = {10.4171/MSL/13}
}

@misc{OohamaAsymmetricBC,
  author       = {Oohama, Yasutada},
  title        = {New Strong Converse for Asymmetric Broadcast Channels},
  month        = apr,
  year         = {2016},
  howpublished = {arXiv:1604.02901},
  note         = {Version 4, revised August 2020},
  url          = {https://arxiv.org/abs/1604.02901},
  eprint       = {1604.02901},
  archiveprefix = {arXiv},
  primaryclass = {cs.IT}
}

@inproceedings{Oohama2015DBC,
  author       = {Oohama, Yasutada},
  title        = {Strong Converse Exponent for Degraded Broadcast Channels at Rates outside the Capacity Region},
  booktitle    = {Proc. IEEE Int. Symp. Inf. Theory (ISIT)},
  pages        = {939--943},
  address      = {Hong Kong, China},
  month        = jun,
  year         = {2015},
  doi          = {10.1109/ISIT.2015.7282593}
}

@inproceedings{ZhangQianChenKhisti2021,
  author       = {Zhang, George and Qian, Jingjing and Chen, Jun and Khisti, Ashish},
  title        = {Universal Rate-Distortion-Perception Representations for Lossy Compression},
  booktitle    = {Advances in Neural Information Processing Systems},
  volume       = {34},
  pages        = {11517--11529},
  publisher    = {Curran Associates, Inc.},
  year         = {2021},
  url          = {https://papers.nips.cc/paper_files/paper/2021/hash/5fde40544cff0001484ecae2466ce96e-Abstract.html}
}

@inproceedings{TheisWagner2021,
  author       = {Theis, Lucas and Wagner, Aaron B.},
  title        = {A Coding Theorem for the Rate-Distortion-Perception Function},
  booktitle    = {ICLR Neural Compression Workshop},
  year         = {2021},
  note         = {arXiv:2104.13662},
  url          = {https://arxiv.org/abs/2104.13662}
}

@article{AhlswedeGacsKorner1976,
  author       = {R. Ahlswede and P. G{\'a}cs and J. K{\"o}rner},
  title        = {Bounds on Conditional Probabilities with Applications in Multi-User Communication},
  journal      = {Z. Wahrscheinlichkeitstheor. Verw. Geb.},
  volume       = {34},
  number       = {2},
  pages        = {157--177},
  month        = jun,
  year         = {1976},
  doi          = {10.1007/BF00535682}
}

@article{TanTom15,
  author       = {V. Y. F. Tan and M. Tomamichel},
  title        = {The Third-Order Term in the Normal Approximation for the {AWGN} Channel},
  journal      = TIT,
  volume       = {61},
  number       = {5},
  pages        = {2430--2438},
  month        = may,
  year         = {2015},
  doi          = {10.1109/TIT.2015.2411256}
}

@article{Tan14,
  author       = {Tan, Vincent Y. F.},
  title        = {Asymptotic estimates in information theory with non-vanishing error probabilities},
  journal      = {Found. Trends Commun. Inf. Theory},
  volume       = {11},
  number       = {1--2},
  pages        = {1--184},
  year         = {2014},
  doi          = {10.1561/0100000086}
}

@article{Hayashi09,
  author       = {M. Hayashi},
  title        = {Information spectrum approach to second-order coding rate in channel coding},
  journal      = TIT,
  volume       = {55},
  number       = {11},
  pages        = {4947--4966},
  month        = nov,
  year         = {2009}
}

@article{Hayashi03,
  author       = {Hayashi, Masahito and Nagaoka, Hiroshi},
  title        = {General formulas for capacity of classical-quantum channels},
  journal      = TIT,
  volume       = {49},
  number       = {7},
  pages        = {1753--1768},
  month        = jul,
  year         = {2003},
  doi          = {10.1109/TIT.2003.813556}
}

@article{LiShao2002,
  author       = {Li, Wenbo V. and Shao, Qi-Man},
  title        = {A Normal Comparison Inequality and Its Applications},
  journal      = {Probab. Theory Related Fields},
  volume       = {122},
  number       = {4},
  pages        = {494--508},
  month        = apr,
  year         = {2002},
  doi          = {10.1007/s004400100176}
}

@article{Marton79,
  author       = {K. Marton},
  title        = {A coding theorem for the discrete memoryless broadcast channel},
  journal      = TIT,
  volume       = {25},
  number       = {3},
  pages        = {306--311},
  month        = may,
  year         = {1979}
}

@article{PPV10,
  author       = {Y. Polyanskiy and H. V. Poor and S. Verd\'{u}},
  title        = {Channel coding rate in the finite blocklength regime},
  journal      = TIT,
  volume       = {56},
  number       = {5},
  pages        = {2307--2359},
  month        = may,
  year         = {2010},
  doi          = {10.1109/TIT.2010.2043769}
}

@BOOK{Poor,
  title = {An Introduction to Signal Detection and Estimation},
  publisher = {Springer},
  year = {1998},
  author = {H. V. Poor},
  address = {2nd}
}

@article{TK12,
  author       = {V. Y. F. Tan and O. Kosut},
  title        = {On the Dispersions of Three Network Information Theory Problems},
  journal      = TIT,
  volume       = {60},
  number       = {2},
  pages        = {881--903},
  month        = feb,
  year         = {2014}
}

@article{VH94,
  author       = {S. Verd\'{u} and T. S. Han},
  title        = {A general formula for channel capacity},
  journal      = TIT,
  volume       = {40},
  number       = {4},
  pages        = {1147--1157},
  month        = jul,
  year         = {1994},
  doi          = {10.1109/18.335960}
}
\end{document}